\documentclass[aps,prb,twocolumn,10pt,a4paper,showpacs]{revtex4-1}
\usepackage{graphicx}
\usepackage{amsmath}
\usepackage{amssymb}
\usepackage{bbm}

\begin{document}



\newcommand{\dt}{\tilde \delta}
\newcommand{\ua}{\uparrow}
\newcommand{\da}{\downarrow}

\newcommand{\comment}[1]{}

\newcommand{\w}{\omega}

\newcommand{\pd}{\phantom\dagger}

\newcommand{\dmp}{\Delta_{\mathrm{mp}}(\w)}
\newcommand{\Xtiltyp}{\tilde{X}_{\mathrm{t}}(\wtil)}
\newcommand{\Xmp}{X_{\mathrm{mp}}}
\newcommand{\tildeltyp}{\tilde{\Delta}_{\mathrm{t}}(\tilde{\w})}
\newcommand{\tildeltypo}{\tilde{\Delta}_{\mathrm{t}}}
\newcommand{\etatil}{\tilde{\eta}}
\newcommand{\wtil}{\tilde{\w}}
\newcommand{\wtilp}{\tilde{\w}^{\prime}}
\newcommand{\dI}{\Delta_{I}^{\pd}(\w)}
\newcommand{\dIno}{\Delta_{I}(\w)}
\newcommand{\dIo}{\Delta_{I}^{\pd}}
\newcommand{\dIono}{\Delta_{I}}
\newcommand{\dJ}{\Delta_{J}^{\pd}(\w)}
\newcommand{\e}{{\cal{E}}}
\newcommand{\etilI}{{\tilde{\cal{E}}}_{I}^{\pd}}
\newcommand{\etilIo}{{\tilde{\cal{E}}}_{I}}
\newcommand{\etilJ}{{\tilde{\cal{E}}}_{J}^{\pd}}
\newcommand{\etilJo}{{\tilde{\cal{E}}}_{J}}
\newcommand{\eI}{{\cal{E}}_{I}^{\pd}}
\newcommand{\eIo}{{\cal{E}}_{I}}
\newcommand{\eJ}{{\cal{E}}_{J}^{\pd}}
\newcommand{\eJo}{{\cal{E}}_{J}}
\newcommand{\ecal}{{\cal{E}}}
\newcommand{\gtilI}{\tilde{G}_{I}^{\pd}(\wtil)}
\newcommand{\gtilIno}{\tilde{G}_{I}(\wtil)}
\newcommand{\gtilIo}{\tilde{G}_{I}^{\pd}}
\newcommand{\gtilIono}{\tilde{G}_{I}}
\newcommand{\gtilJ}{\tilde{G}_{J}^{\pd}(\wtil)}
\newcommand{\gtilJo}{\tilde{G}_{J}^{\pd}}
\newcommand{\gtilJono}{\tilde{G}_{J}^{\pd}}
\newcommand{\dostilI}{\tilde{D}_{I}^{\pd}(\wtil)}
\newcommand{\dostilIno}{\tilde{D}_{I}(\wtil)}
\newcommand{\stilI}{\tilde{S}_{I}^{\pd}(\wtil)}
\newcommand{\stilIno}{\tilde{S}_{I}(\wtil)}
\newcommand{\stilIo}{\tilde{S}_{I}^{\pd}}
\newcommand{\stilIono}{\tilde{S}_{I}}
\newcommand{\stilJ}{\tilde{S}_{J}^{\pd}(\wtil)}
\newcommand{\stilJo}{\tilde{S}_{J}^{\pd}}
\newcommand{\stilJono}{\tilde{S}_{J}}
\newcommand{\xtilI}{\tilde{X}_{I}^{\pd}(\wtil)}
\newcommand{\dtilI}{\tilde{\Delta}_{I}^{\pd}(\wtil)}
\newcommand{\dtilIno}{\tilde{\Delta}_{I}(\wtil)}
\newcommand{\dtilIo}{\tilde{\Delta}_{I}^{\pd}}
\newcommand{\dtilIono}{\tilde{\Delta}_{I}}
\newcommand{\nh}{N_{\mathcal{H}}}

\newcommand{\n}[2][I]{n^{\scriptscriptstyle(#1)}_{#2}}
\newcommand{\Ntot}[2][I]{N^{\scriptscriptstyle(#1)}_{#2}}


\title{Many-body localization in Fock-space: a local perspective}

\author{David E. Logan and Staszek Welsh}
\affiliation{Department of Chemistry, Physical and Theoretical Chemistry, Oxford University, South Parks Road, Oxford, OX1 3QZ, United Kingdom}

\date{\today}

\begin{abstract}
A canonical model for many-body localization (MBL) is studied, of interacting spinless fermions on a lattice with 
uncorrelated quenched site-disorder. The model maps onto a tight-binding model on a `Fock-space (FS) lattice' of many-body states, with an extensive local connectivity. We seek to understand some aspects of MBL from this perspective,
via local propagators for the FS lattice and their self-energies (SE's); focusing on the SE probability distributions, over disorder and FS sites. A  probabilistic mean-field theory (MFT) is first developed, centered on self-consistent determination of the geometric mean of the distribution. Despite its simplicity this captures some key features of the problem, including recovery of an MBL transition, and predictions for the forms of the SE distributions.
The problem is then studied numerically in $1d$ by exact diagonalization, free from MFT assumptions. The geometric mean indeed appears to act as a suitable order parameter for the transition. Throughout the MBL phase the appropriate SE distribution is 
confirmed to have a universal form, with  long-tailed L\'evy behavior as predicted by MFT. In the delocalized phase for weak disorder, SE distributions are clearly log-normal; while on approaching the transition they 
acquire an intermediate L\'evy-tail regime,  indicative of the incipient MBL phase.
\end{abstract}

\pacs{71.23.-k, 71.10.-w, 05.30.-d }


\maketitle

\section{Introduction}
\label{section:intro}

Sixty years ago Anderson~\cite{PWA1958} famously discovered  the phenomenon of localization, specifically in single-particle systems. While the importance of disorder in interacting systems has long been 
appreciated,~\cite{PWA1958,MottMIT} traditional study of it has focused on ground state quantum phases and their transitions. More recently, stimulated in part by basic issues relating to thermalization  (or its absence) in isolated quantum systems, attention has  turned to study of highly excited quantum states, and the phenomenon of many-body 
localization~\cite{MirlinPolyakovlPRL2005,BAAAnnPhys2006,*AltshuleretalPRB2007,Oganesyan+HusePRB2007} (MBL)
occurring at finite energy density above the ground state.

Over the last decade or so the MBL problem has attracted 
great interest,~\cite{PeterP-PRB2008,Pal+HusePRB2010,ReichmanPRB2010,Monthus+GarelPRB2010,BardarsonPollmannMoorePRL2012,SerbynPapicAbaninPRL2013,DeLucaScardicchioEPL2013,IyerHusePRB2013,SerbynPapicAbaninLIOMPRL2013,VoskAltmanPRL2013,Huse+Pal+SondhiPRB2013,VoskAltmanPRL2014,AltmanDemlerOganesyanPRX2014,BardarsonPollmannPRL2014,LaumannSondhiPRB2014,Huse+Nand+OganPRB2014,HuseReview2015,DasMoessner2015,PapicAbaninPRL2015,GopalKDemlerPRL2015,ReichmanPRL2015,AbaninLIOMs2015,BeraFHMBardarsonPRL2015,SubrotoPRL2015,PixleyDasSarmaPRL2015,Mondragon-ShemME_PRB2015,LuitzAletPRB2015,Devakul15,NandkishoreBhattPRL2015,VoskHuseAltmanPRX2015,SIDVaasseurPotterPRL2015,SIDPotterVasseurPRX2015,Serbyn+P+AbaninPRX2015,ImbriePRL2016,*Imbrie16, ScardicchioFAppPRB2016,DeRoeckNoMEsPRB2016,Geraedts16,Khemani2016,Serbyn16,Serbyn16b,BaldwinQREM2016,PeterPHubbard2016,DeRoeckNoMBLinHigherDimPRB2017,RademakerPRL2016,Parameswaran17,Lezama17,Nag17,MoessnerSondhi2017,Dumitrescu17,DeTomasi17,Imbrie17,Wahl17, MarcinM+PeterPPTB2018}
and a rich understanding of it has begun to emerge (for a review see e.g.\ [\onlinecite{HuseReview2015}]).
Powerful diagnostics have been deployed to understand properties of both the ergodic and MBL phases, 
including level statistics and related measures,~\cite{Oganesyan+HusePRB2007,Pal+HusePRB2010,ReichmanPRL2015,LuitzAletPRB2015,NandkishoreBhattPRL2015,Serbyn16} and entanglement entropies and spectra.~\cite{PeterP-PRB2008,BardarsonPollmannMoorePRL2012,SerbynPapicAbaninPRL2013,VoskAltmanPRL2013,VoskAltmanPRL2014,BardarsonPollmannPRL2014,Geraedts16,Serbyn16b}  
The MBL phase itself has naturally attracted particular attention, highlights including an existence proof for the 
phase,~\cite{ImbriePRL2016,Imbrie16} the description of it in terms of local integrals of motion,\cite{SerbynPapicAbaninLIOMPRL2013,VoskAltmanPRL2013,Huse+Nand+OganPRB2014,AbaninLIOMs2015,ImbriePRL2016,*Imbrie16,RademakerPRL2016,Imbrie17,MarcinM+PeterPPTB2018} and an emerging understanding of how MBL states  exhibit broken symmetries and topological 
order.~\cite{Huse+Pal+SondhiPRB2013,VoskAltmanPRL2014,AltmanDemlerOganesyanPRX2014,BardarsonPollmannPRL2014,LaumannSondhiPRB2014}  Numerical methods have of course played a key role, from more or less ubiquitous exact diagonalization studies, to a 
variety of RG-based methods.~\cite{Monthus+GarelPRB2010,Huse+Pal+SondhiPRB2013,VoskAltmanPRL2013,VoskAltmanPRL2014,AltmanDemlerOganesyanPRX2014,VoskHuseAltmanPRX2015,SIDVaasseurPotterPRL2015,SIDPotterVasseurPRX2015,Parameswaran17,Dumitrescu17} 
A range of relevant models have likewise been studied, notably the `standard' spinless fermion or random XXZ models,
but including also quasiperiodic models,~\cite{IyerHusePRB2013,SubrotoPRL2015,PixleyDasSarmaPRL2015,Nag17,DeTomasi17} and MBL 
systems subject to periodic driving.~\cite{DasMoessner2015,PapicAbaninPRL2015,Geraedts16,Khemani2016,MoessnerSondhi2017}
An understanding of MBL is nevertheless still in relative infancy, with abundant potential for new insights to emerge.

Here  we take a rather different approach to the problem; focusing for specificity on the standard model of interacting 
spinless fermions~\cite{Oganesyan+HusePRB2007} with quenched site-disorder, for a lattice of $N \equiv L^{d}$ sites 
occupied by  $N_{e}$ fermions at non-zero filling $\nu =N_{e}/N$. As reprised in sec.\ \ref{section:model}, 
the model can be  mapped exactly onto a tight-binding model~\cite{SWDEL1} (TBM) on a `Fock-space (FS) lattice' of dimension
$N_{{\cal{H}}} ={^{N}}C_{N_{e}}$ ($\equiv \binom{N}{N_{e}}$), each site $I$ of which corresponds to a disordered, interacting many-body state with specified fermion occupancy of  all real-space sites, and is thus associated with a FS `site-energy';
while FS sites are connected by the one-electron hopping matrix element of the original Hamiltonian $H$.
The FS lattice is of course a complex network,\cite{SWDEL1,Monthus+GarelPRB2010,LuitzAletPRB2015,BaldwinQREM2016} 
with the associated TBM  very different from  that typically encountered in one-body localization. 
The local connectivity of a FS lattice site for example  -- the number of FS sites to which it is connected under hopping -- is typically extensive  ($\propto N$); and while the $N$ site-energies of the underlying real-space lattice sites are independent random variables, the FS site energies are certainly not (there being exponentially many of them, 
$N_{{\cal{H}}} ={^{N}}C_{N_{e}}$).

Our aim here is to exploit the above mapping, seeking to obtain some insight into aspects of MBL. 
To that end we consider the local (i.e.\ site-diagonal) propagators for the FS lattice,  appropriately 
rescaled in terms of the standard deviation of the eigenvalue spectrum 
(secs.\ \ref{subsection:energy rescaling},\ref{section:propagatorsformal}); focusing specifically on their
associated local (or Feenberg) self-energies.~\cite{Feenberg1948,Economoubook} 
An approach of this general ilk has a long history in studies
of one-body localization,~\cite{EconomousCohen1972,AAT1973,ThoulessReview1974,LicciardelloEconomou1975,Heinrichs1977,Stein1979,DELPGWPRB1987,*DELPGWdipolar1987,DELPGWJCP1986,*DELPGWPRB1985,*DELPGWPRB1984} and in an early approach to MBL in the context of vibrational energy flow,~\cite{DELPGWJCP1990,LeitnerReview2015}  but has not to our knowledge been considered in the recent era of MBL. The imaginary part  $\dtilIono$ of the local self-energy is of central importance, as its behavior discriminates cleanly  between  delocalized and MBL phases, being finite with probability unity for the former and vanishingly small for an MBL phase. We thus study its probability distribution, over disorder realizations and FS sites; noting at the outset that 
its mean value $\langle \dtilIono\rangle$ -- which amounts to Fermi's golden rule -- is non-zero throughout both phases 
(sec.\ \ref{section:propagatorsformal}), and as such is irrelevant as a diagnostic for the MBL transition.  

A probabilistic mean-field theory is first developed (secs.\ \ref{section:MFT},\ref{section:ProbDists1}), 
treated at the level of the second-order renormalized perturbation series~\cite{Feenberg1948,Economoubook}  for the self-energy; and centering on a self-consistent determination of the geometric mean of the distribution, which acts as an effective order parameter for the transition. Despite its simplicity this appears to capture  some key features of the problem, including recovery of an MBL transition in the thermodynamic limit  -- with some physical understanding of how and why this arises, given the extensive local connectivity of the FS lattice -- and  predictions for the self-energy distributions in each phase, and their evolution with disorder, interaction and filling fraction.

 Free from the assumptions entering the mean-field theory, in sec.\ \ref{section:EDresults} the problem is 
then studied numerically in $1d$ via exact diagonalization. Results arising broadly concur well with those from the 
mean-field approach, particularly in the MBL phase. The geometric mean indeed appears to act as a suitable order 
parameter for the transition. Throughout the MBL phase the appropriate self-energy distribution is likewise confirmed to 
have a universal form, with the long-tailed L\'evy behavior 
as  predicted by the mean-field theory; and with further understanding provided by considering the single-particle limit, where much larger system sizes can be studied, and for which exact results can be obtained.~\cite{Altshuler+Prigodin1989} In the delocalized phase for weak disorder, self-energy distributions are found to be log-normal; and while remaining unimodal 
with increasing disorder, on approaching the transition they appear to acquire an intermediate L\'evy tail regime,  indicative
of the incipient MBL phase, before crossing over to an exponentially damped inverse Gaussian form.


\section{Model and background}
\label{section:model}

The spinless fermion model considered is the standard one,~\cite{Oganesyan+HusePRB2007}
\begin{subequations}
\label{eq:1}
\begin{align}
H~=&~ H_{W}^{\pd}+H_{t}^{\pd}+H_{V}^{\pd}
\label{eq:1a}
\\
=& ~ \sum_{i} \epsilon_{i}^{\pd} \hat{n}_{i}^{\pd} +
\sum_{\langle ij \rangle} t ~(c_{i}^{\dagger}c_{j}^{\pd}+c_{j}^{\dagger}c_{i}^{\pd})
+\sum_{\langle ij \rangle} V~\hat{n}_{i}^{\pd}\hat{n}_{j}^{\pd}
\label{eq:1b}
\end{align}
\end{subequations}
(where $\hat{n}_{i} =c_{i}^{\dagger}c_{i}^{\pd}$);  here considered on a $d$-dimensional hypercubic lattice of coordination number $Z_{\mathrm{d}}=2d$. The hoppings ($t$) and interactions ($V$) are nearest neighbor (NN), and
$\langle ij \rangle$ denotes distinct NN pairs. In the disordered $H_{W}$, the site energies $\{\epsilon_{i}\}$ 
are independent random variables, with a common distribution  $P(\epsilon_{i})$ (chosen to have zero mean). 
The lattice has $N$ sites and contains $N_{e}$ fermions; and we are  
interested in the thermodynamic limit where both $N \equiv L^{d}$ and $N_{e} \rightarrow \infty$, holding the filling  $\nu = N_{e}/N$ fixed and non-vanishing. 
For the particular case of $d=1$ the model is equivalent under a Jordan-Wigner
transformation to the random XXZ model (with total spin $S^{Z}_{\mathrm{tot}} \equiv (\nu -\tfrac{1}{2})N$).

The dimension  of the associated Fock-space (FS) is $N_{{\cal{H}}} = {^{N}}C_{N_{e}}$,
and is exponentially large in the number 
of sites 
($\nh \propto \exp[s_{\infty}N]$ with
$s_{\infty}=-[\nu\ln\nu +(1-\nu)\ln(1-\nu)]$ the configurational entropy per site).
The Hamiltonian  may be recast equivalently as an effective tight-binding model on a `FS-lattice' of 
$N_{{\cal{H}}}$ sites $\{I\}$,~\cite{SWDEL1} 
\begin{equation}
\label{eq:2}
H~=~ \sum_{I} \eI |I\rangle\langle I|~+~
{\sum_{I,J (J\neq I)}}
T_{IJ}^{\pd}|I\rangle\langle J| .
\end{equation}
The $N_{{\cal{H}}}$ basis states $\{|I\rangle\}$ are the eigenstates of $H_{0}=H_{W}+H_{V}$, viz.\ 
$|I\rangle =$ with occupation number $\n{i} =0$ or $1$ only for each real-space site $i$ 
(such that $\sum_{i}\n{i} =N_{e}$ for any $|I\rangle$). Corresponding FS site-energies follow from 
$H_{0}|I\rangle =\eI |I\rangle$ as
\begin{equation}
\label{eq:3}
{\cal{E}}_{I}^{\pd} = ~\sum_{i} \epsilon_{i}^{\pd} \n{i} ~+~V r_{I}^{\pd} ~~~~~
:~ r_{I}^{\pd}~=~\sum_{\langle ij \rangle} ~\n{i}\n{j}
\end{equation}
where $r_{I}$ is thus defined (and simply counts the total number of occupied NN real-space pairs  in $|I\rangle$). 
FS sites are connected  under the hopping, $T_{IJ}=\langle I|H_{t}|J\rangle$, with $|T_{IJ}| =t$ when non-zero
($T_{IJ}$ is either $+t$ or $-t$ for general $d\geq 2$, its sign depending
on the configuration of fermions in $|I\rangle$ and $|J\rangle$; while $T_{IJ} =t$ for all connected FS sites 
in the $d=1$ open chain).
The number of FS sites to which any given $I$ is connected defines the local coordination number, $Z_{I}$. 
This is readily shown~\cite{SWDEL1} to be given by $Z_{I} = 2(dN_{e}-r_{I})$, so follows directly 
from $r_{I}$.

Eigenvalues and eigenfunctions of $H$ are denoted by $E_{n}$ and $|\Psi_{n}\rangle$. 
The eigenvalue spectrum $D(\w)=\nh^{-1}\sum_{n}\delta(\w -E_{n})$ is normally 
distributed,~\cite{SWDEL1} with an extensive mean $\overline{E}\propto N$ and a standard deviation 
$\mu_{E} \propto \sqrt{N}$; and it is states prescribed by this energy regime on which we focus
throughout. Such states correspond in the  commonly used sense to infinite-temperature states (and
include all but an exponentially small fraction of eigenstates).

 Given the mapping to an effective TBM on a FS lattice, the same questions can be asked as arise for a 
single-particle disordered  TBM, including the central one of whether eigenstates of some given energy 
are  localized or extended. As for one-body localization (1BL), the answer to this question resides 
in principle in the distribution, over  FS sites and disorder realizations, 
of the squared 
amplitudes $|A_{nI}|^{2} = |\langle I|\Psi_{n}\rangle|^{2}$ in the eigenfunction expansion
\begin{equation}
\label{eq:4}
|\Psi_{n}\rangle ~=~\sum_{I} A_{nI}^{\pd} |I\rangle
\end{equation}
 (with $\sum_{I}|A_{nI}|^{2}=1=\sum_{n}|A_{nI}|^{2}$, normalization).
This merits elaboration, as there are both differences and similarities between the  MBL and 1BL cases.

In a one-body problem the site label $I$ refers of course to a single real-space site 
($|I\rangle \equiv |i\rangle = c_{i}^{\dagger}|\mathrm{vac}\rangle$, with the Fock-space dimension 
simply $\nh \equiv {^{N}}C_{1} =N$). Here, an extended state has support on $\mathcal{O}(N)$ sites, 
with a typical $|A_{nI}|^{2}$ on the  order of $\sim 1/N$ (recall $\sum_{I}|A_{nI}|^{2} =1$); while
a  localized state has support on a finite number $n_{s}$ of sites,
with a typical $|A_{nI}|^{2}$ thus $\mathcal{O}(1/n_{s})$, likewise finite. 
The \emph{fraction} of sites on which a 
extended state has support is thus $\mathcal{O}(1)$ in the thermodynamic limit $N\rightarrow \infty$, 
while the fraction on which a localized state has support vanishes. In the many-body 
case,~\cite{LuitzAletPRB2015,DeLucaScardicchioEPL2013}  an extended state analogously has support on  
$\mathcal{O}(\nh)$ sites, with a typical $|A_{nI}|^{2} \sim 1/\nh$. A many-body localized state has 
however support on $\mathcal{O}(\nh^{\alpha})$ sites with $\alpha <1$, and correspondingly a typical 
$|A_{nI}|^{2} \sim \nh^{-\alpha}$; in other words it has support on a number of FS sites 
of order $\nh^{\alpha}$ $\propto e^{\alpha s_{\infty}N}$
which is exponentially large in $N$, in contrast to a finite number in the corresponding 1BL case.
Nevertheless, the \emph{fraction} $\sim \nh^{\alpha -1}$ of FS sites on which an MBL state has support again
vanishes in the thermodynamic limit, while the corresponding fraction for an extended many-body state
remains $\mathcal{O}(1)$; which is in evident parallel to the 1BL case. Bounded as they are in the thermodynamic 
limit, whether these fractions are finite or vanish is one reflection of the distinction between extended 
and localized states, for both MBL and 1BL problems.

These simple qualitative considerations naturally have implications for e.g.\ the first participation entropy,
one diagnostic of the amplitude distributions,~\cite{LuitzAletPRB2015,DeLucaScardicchioEPL2013}
defined  for a state $|\Psi_{n}\rangle$ by
$S_{1}^{\mathrm{pe}}=-\sum_{I}|A_{nI}|^{2}\ln|A_{nI}|^{2}$. For an extended many-body state, 
taking $|A_{nI}|^{2} \sim 1/\nh$  on all $\nh$ sites gives $S_{1}^{\mathrm{pe}}=\ln \nh$; while for an MBL state, 
taking $|A_{nI}|^{2} \sim \nh^{-\alpha}$ on $\nh^{\alpha}$ sites, gives 
$S_{1}^{\mathrm{pe}} =\alpha \ln \nh$ ($\alpha <1$). This behavior is indeed as found by numerical study of the participation  entropy.~\cite{LuitzAletPRB2015,DeLucaScardicchioEPL2013} 
Note further that, since $S_{1}^{\mathrm{pe}} \propto \ln\nh$ grows with system size, it is 
$S_{1}^{\mathrm{pe}}/\ln\nh \propto S_{1}^{\mathrm{pe}}/N$ which remains bounded in the thermodynamic limit.
This illustrates that relevant physical quantities are liable to require suitable rescaling with system size 
to obtain a well defined thermodynamic limit (further examples of which will arise below).


\subsection{Basic distributions}
\label{subsection:distributions0}

It proves helpful  in the following to have some understanding of the FS lattice in a statistical sense; specifically the distributions, over both FS sites and (where appropriate) disorder realizations, of: $r_{I}$ (eq.\ \ref{eq:3}) and hence the FS coordination number $Z_{I}= 2(dN_{e}-r_{I})$ (which are disorder-independent), as well as the distributions of FS site energies $\{\eI\}$, and of the eigenvalues $\{E_{n}\}$ of $H$. These are considered in [\onlinecite{SWDEL1}], from which we now recap required results.

In the thermodynamic limit of interest all such distributions are normal, with extensive means~\cite{FN1} 
$\overline{O} \propto N$ and standard deviations $\mu_{O}^{\pd} \propto \sqrt{N}$. The mean of $r_{I}$ is 
$\overline{r} = \nu dN_{e} =\nu^{2}dN$, so  the average coordination number
\begin{equation}
\label{eq:5}
\overline{Z}~=~ Z_{\mathrm{d}} (1-\nu) N_{e} ~=~ 2\nu (1-\nu) d N .
\end{equation}
That $\overline{Z} \propto N$  is physically obvious, but begs the question of how Fock-space localization survives the thermodynamic limit, itself considered in secs.\  \ref{section:MFT} ff.  As for the mean FS site energy ($\overline{\e}$) 
and eigenvalue ($\overline{E}$), they coincide and are given by
\begin{equation}
\label{eq:6}
\overline{E}~=~ \overline{\e}~=~V\overline{r} ~=~ V \nu^{2} dN .
\end{equation}

The variance of $r_{I}$ is $\mu_{r}^{2} = [\nu(1-\nu)]^{2}dN$ (whence $\mu_{Z}^{2}=4\mu_{r}^{2}$ for the coordination number).
That for the FS site energies (eq.\ \ref{eq:3}) is $\mu_{\e}^{2}=\mu_{W}^{2}+\mu_{r}^{2}$, with $\mu_{W}$ the standard deviation of the disorder term in eq.\ \ref{eq:3} ($\equiv H_{W}$), given by
\begin{equation}
\label{eq:7}
\mu_{W}^{\pd}~=~\sqrt{N}~ \big[\nu(1-\nu)\langle\epsilon^{2}\rangle\big]^{\tfrac{1}{2}}
\end{equation}
($\langle\epsilon^{2}\rangle = \int d\epsilon P(\epsilon)\epsilon^{2}$); such that 
\begin{equation}
\label{eq:8}
\mu_{\e}^{\pd}~=~\sqrt{N}~ \big[\nu(1-\nu)\langle\epsilon^{2}\rangle+V^{2}[\nu(1-\nu)]^{2}d\big]^{\tfrac{1}{2}} .
\end{equation}
Finally, the eigenvalue variance is given simply by~\cite{SWDEL1} 
\begin{equation}
\label{eq:9}
\mu_{E}^{2}~=~\mu_{\e}^{2}~+~t^{2}\overline{Z}.
\end{equation}


\subsection{Energy rescaling}
\label{subsection:energy rescaling}

The normalized eigenvalue distribution is just the density of states (DoS), 
$D(\w) =N_{{\cal{H}}}^{-1}\sum_{n}\delta(\w -E_{n})$
($\equiv N_{{\cal{H}}}^{-1}\langle\sum_{n}\delta(\w -E_{n})\rangle_{\epsilon}$ by self-averaging).
As above, it is the Gaussian
\begin{equation}
\label{eq:10}
D(\w) ~=~\frac{1}{\sqrt{2\pi}\mu_{E}}\exp \Big(- \frac{[\w -\overline{E}]^{2}}{2\mu_{E}^{2}}\Big),
\end{equation}
which we add is very well captured~\cite{SWDEL1} by exact diagonalization on the modest system sizes amenable to numerics.
$D(\w)$ is obviously $N$-dependent, from both $\overline{E} \propto N$ and because its standard deviation 
$\mu_{E} \propto \sqrt{N}$. As such it is natural to refer energies relative to the mean, and to rescale them in terms 
of the spectral width $\mu_{E}$ via 
\begin{equation}
\label{eq:11}
\wtil = (\w -\overline{E})/\mu_{E}~\equiv~(\w -\overline{\e})/\mu_{E} .
\end{equation}
With this the  DoS $\tilde{D}(\wtil)$ is a standard normal distribution (zero mean and unit variance),
\begin{equation}
\label{eq:12}
\tilde{D}(\wtil)~=~\frac{1}{\sqrt{2\pi}}\exp\big(-\tfrac{1}{2}\wtil^{2}\big)
\end{equation}
with no explicit $N$-dependence.  This rescaling proves central to our perspective on MBL, as will be seen in the following;
and  while so far motivated physically is in fact required on general grounds.


\section{Local propagators and self-energies}
\label{section:propagatorsformal}

We turn now the FS site propagators, and in particular the local (site-diagonal) propagator, the
associated self-energy of which is our primary focus.

The site-dependent propagators for the FS lattice, 
$G_{JI}^{\pd}(\w)$ ($\leftrightarrow G_{JI}^{\pd}(t) =-i\theta (t)\langle J|e^{-iHt}|I\rangle$),
are given by $G_{JI}^{\pd}(\w) =\langle J |(\w^{+}-H)^{-1}|I\rangle$ with $\w^{+}=\w +i\eta$ and $\eta \equiv 0+$.
Since $H$ has the TBM form eq.\ \ref{eq:2}, it follows directly that
\begin{equation}
\label{eq:13}
(\w^{+}-\eI) G_{IJ}^{\pd}(\w) ~=~\delta_{IJ}^{\pd} ~+~\sum_{K} T_{IK}^{\pd}G_{KJ}^{\pd}(\w),
\end{equation}
which generates the familiar locator-series expansion for the propagators.~\cite{Feenberg1948,Economoubook}
In particular, the local propagator  $G_{I}^{\pd}(\w) \equiv G_{II}^{\pd}(\w)$ is 
$G_{I}(\w) = \sum_{n}|A_{nI}^{\pd}|^{2}/(\w^{+} -E_{n})$; from which follows the \emph{local} DoS,
$D_{I}(\w) =-\tfrac{1}{\pi}\mathrm{Im}G_{I}(\w) = \sum_{n}|A_{nI}^{\pd}|^{2}\delta(\w -E_{n})$, 
i.e.\  the local density of eigenvalues which overlap site $I$. These in turn are related to the 
\emph{total} DoS -- the Gaussian eq.\ \ref{eq:10} -- by $D(\w) =\tfrac{1}{N_{{\cal{H}}}}\sum_{I}D_{I}(\w)$.

The local propagator has been known since Anderson's original work~\cite{PWA1958} to be of particular importance in 
one-body localization; and that is also true in the MBL context, given the mapping to the effective TBM eq.\ \ref{eq:2}. 
It is most effectively analyzed in terms of the Feenberg self-energy~\cite{Feenberg1948,Economoubook} 
$S_{I}(\w)=X_{I}(\w) -i\Delta_{I}(\w)$, and in particular the renormalized perturbation series (RPS) for it; where 
$S_{I}(\w)$ is defined as usual by~\cite{Feenberg1948,Economoubook} 
\begin{equation}
\label{eq:14}
G_{I}^{\pd}(\w) ~=~ \big[\w^{+} -\eI -S_{I}^{\pd}(\w)\big]^{-1} 
~=~\big[g_{I}^{-1}(\w) -S_{I}^{\pd}(\w)\big]^{-1} 
\end{equation}
with $g_{I}^{\pd}(\w)=(\w^{+}-\eI)^{-1}$ the (purely) site-diagonal propagator for the extreme MBL limit of
$T_{IJ}=0$ ($H_{t} \equiv 0$).

 We return to this below but first, as above, simply rescale the energy according to eq.\ \ref{eq:11}.
This obviously requires rescaling the propagator as $\gtilIno =\mu_{E}G_{I}(\w)$, given from eq.\ \ref{eq:14} by
\begin{equation}
\label{eq:15}
\mu_{E}^{\pd}G_{I}^{\pd}(\w) ~= ~
\gtilI~=~ \Big[\wtil^{+} -\etilI -\tilde{S}_{I}^{\pd}(\wtil)\Big]^{-1} 
\end{equation}
where $\wtil^{+}=\wtil +i\etatil$ and $\etatil =\eta/\mu_{E}$; with the rescaled FS site-energy (cf.\ eq.\ \ref{eq:11})
\begin{equation}
\label{eq:16}
\etilI ~=~ (\eI -\overline{\e})/\mu_{E}^{\pd},
\end{equation}
and where the self-energy in consequence rescales as
\begin{equation}
\label{eq:17}
\tilde{S}_{I}^{\pd}(\wtil)~=~S_{I}^{\pd}(\w)/\mu_{E}^{\pd} ~=~\xtilI -i\dtilI .
\end{equation}
The local propagator $\gtilI = \sum_{n}|A_{nI}^{\pd}|^{2}/(\wtil^{+} -\tilde{E}_{n})$
(with $\tilde{E}_{n} :=(E_{n}-\overline{E})/\mu_{E}$), so the local DoS 
\begin{equation}
\label{eq:18}
\dostilI =
\sum_{n}|A_{nI}^{\pd}|^{2}\frac{\etatil/\pi}{(\wtil -\tilde{E}_{n})^{2}+\etatil^{2}}
\equiv 
\sum_{n}|A_{nI}^{\pd}|^{2}\delta(\wtil -\tilde{E}_{n}),
\end{equation}
such that the total DoS, $\tilde{D}(\wtil) =\tfrac{1}{N_{{\cal{H}}}}\sum_{I}\dostilI$, is 
the standard normal form eq.\ \ref{eq:12}.

In the following we focus directly on the Feenberg self-energy $\stilIno$, and in particular its imaginary 
part $\dtilIno$ (long used in the context of 1BL~\cite{EconomousCohen1972,AAT1973,ThoulessReview1974,LicciardelloEconomou1975,Heinrichs1977,Stein1979,DELPGWPRB1987,*DELPGWdipolar1987,DELPGWJCP1986,*DELPGWPRB1985,*DELPGWPRB1984}),
which corresponds physically to the inverse lifetime for FS site/state $I$ participating in states of energy $\wtil$.
Our analysis rests on the contention that
its behavior discriminates between localized and extended states, namely that for extended states $\dtilIno$ is 
non-vanishing with probability unity over an ensemble of disorder realizations; while for localized states by 
contrast $\dtilIno$ is vanishingly small with probability one,
specifically is proportional to the imaginary part of the energy ($\etatil \rightarrow 0+$), so that 
$y=\dtilIono/\etatil$ itself is finite. To support this we sketch in Appendix \ref{section:App1}
some simple arguments indicating, whether 1BL or MBL is considered, that for extended states 
$\Delta_{I}(\w)$ is non-zero and is $\propto \mu_{E}$ in the many-body case, while 
$\Delta_{I}(\w) \propto \eta$ is vanishingly small for localized states. We add that the behavior 
described is also consistent with our numerical results.

Since $\dtilIono$ is distributed, it is in effect a probabilistic order parameter. One should thus study its 
distribution -- for extended states the probability density $F(\dtilIono)$ over an ensemble of disorder realizations that 
any FS site has a particular $\dtilIno$ (physically, the distribution of local inverse lifetimes/rates for FS states $I$).
Likewise for localized states one considers the corresponding density $\tilde{F}(y)$ for $y=\dtilIono/\etatil$,
given by $\tilde{F}(y)=\etatil F(\etatil y)$. It is these distributions we will consider in the following sections.

Before proceeding, we draw attention to two significant points. First, note that the disorder \emph{averaged} 
$\langle \dtilIno\rangle$ -- which amounts to Fermi's golden rule rate in leading order (see below) -- is as well 
known always non-zero, regardless of whether states are localized or extended (as too are the higher moments 
$\langle[\dtilIno]^{p}\rangle$, $p \geq 2$). As such, $\langle \dtilIno\rangle$ is not an order parameter for the MBL 
transition (although one anticipates it will be  adequately `typical' of $F(\dtilIono)$ sufficiently deep in a regime of
extended states). This has immediate implications for  the distribution $\tilde{F}(y)$ of $y=\dtilIono/\etatil$ 
characteristic of the localized regime, because, since $\etatil \equiv 0+$, all of \emph{its} moments -- other than 
$p=0$ (normalization) -- must in consequence diverge. The latter in turn suggests that the large-$y$ asymptotic behavior 
of $\tilde{F}(y)$ may be a power-law, $\tilde{F}(y) \sim y^{-\xi}$ with the exponent in the range $1<\xi <2$. This is indeed the case, as will be seen in subsequent sections.

\begin{figure}
\includegraphics{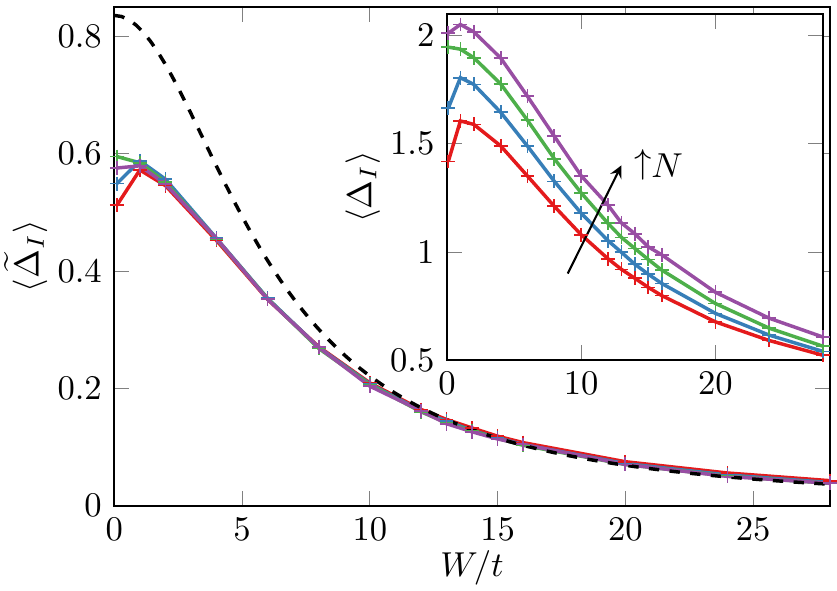}
\caption{\label{fig:fig1} 
The averaged $\langle \tilde{\Delta}_{I} \rangle$ \emph{vs} disorder $W/t$ for band center states ($\wtil =0$) in $d=1$; 
shown as a function of system size for $N=10$ (red), $12$ (blue), $14$ (green), $16$ (purple), as obtained by exact diagonalization. Results are for a box  $P(\epsilon) =\theta(\tfrac{W}{2}-|\epsilon|)$,
at half-filling, and for $V/t =2$. $\langle \tilde{\Delta}_{I} \rangle$ converges rapidly with increasing $N$, and is well converged by $N \simeq 12$.  The Fermi golden rule result (eq.\ \ref{eq:19}) is shown for comparison (dashed line).
\emph{Inset}: Corresponding system size evolution of $\langle \Delta_{I}(\w) \rangle$
($=\mu_{E}\tilde{\Delta}_{I}(\wtil)$). Discussion in text.
}
\end{figure}

Second, consider the averaged $\langle \dtilIno\rangle$ itself. Fig.\ \ref{fig:fig1} shows numerical results for 
$\langle \dtilIno\rangle$ \emph{vs} disorder strength $W/t$, for the $d=1$ case with a standard box site-energy distribution $P(\epsilon) =\theta(\tfrac{W}{2}-|\epsilon|)$ of full width $W$ ($\theta(x)$ is the unit step function).
We consider half-filling $\nu =1/2$ with interaction $V/t =2$ (corresponding for the random XXZ model to 
$S^{z}_{\mathrm{tot}} =0$ with spin-isotropic  Heisenberg exchange), and states at the band center ($\wtil =0$). 
With these parameters, a range of previous  exact diagonalization studies estimate the MBL transition to occur for 
$W/t$ in the range  
$\sim 12$$-$$16$~\cite{Pal+HusePRB2010,ReichmanPRB2010,DeLucaScardicchioEPL2013,LuitzAletPRB2015,BeraFHMBardarsonPRL2015}
(for $W$ and $t$ as we have defined them), although a numerical linked-cluster expansion method~\cite{Devakul15}
suggests these may underestimate somewhat the critical $W/t$. Results here are obtained by exact diagonalization, for 
system sizes of $N=10$$-$$16$ real-space sites (further discussion will be given in sec.\ \ref{section:EDresults}).
As seen clearly in fig.\ \ref{fig:fig1}, $\langle \dtilI\rangle$ converges rapidly as a function of system size $N$.
It is in fact already well converged by $N \simeq 12$ for essentially all $W/t$ save the lowest.
By contrast, the inset to fig.\ \ref{fig:fig1} shows the corresponding behavior for $\langle\dIno \rangle$, which for any given $W/t$ is seen to increase ($\propto \sqrt{N}$) with system size. The latter is expected, since $\dI =\mu_{E}\dtilIno$
(eq.\ \ref{eq:17}) and $\mu_{E} \propto \sqrt{N}$ (eqs.\ \ref{eq:9},\ref{eq:8},\ref{eq:5}). It is also readily 
understood physically, for $\langle\dIno\rangle$ is given to leading order (in the Feenberg RPS)
by the Fermi golden rule result $\langle \dIno\rangle \simeq \pi \overline{Z}t^{2}D(\w)$, with 
$D(\w)\propto 1/\mu_{E}$ the Gaussian DoS  eq.\ \ref{eq:10}. Since the mean coordination number $\overline{Z} \propto N$ 
(eq.\ \ref{eq:5}), $\langle \dIno\rangle \propto \sqrt{N}$; whence 
\begin{equation}
\label{eq:19}
\langle\dtilI\rangle ~\simeq ~\frac{\pi t^{2}\overline{Z}}{\mu_{E}} D(\w)
~=~\frac{\pi t^{2}\overline{Z}}{\mu_{E}^{2}} \tilde{D}(\wtil)
\end{equation}
(with $\tilde{D}(\wtil)$ the standard normal DoS eq.\ \ref{eq:12}), and indeed remains finite as the thermodynamic 
limit $N\rightarrow \infty$ is approached. As seen in fig.\ \ref{fig:fig1} (dashed line), the golden rule result 
eq.\ \ref{eq:19} captures rather  well the evolution of $\langle\dtilIno\rangle$ with disorder (and in fact becomes quantitatively accurate with increasing disorder).

Although thus far illustrated primarily for the average $\langle \dtilIno\rangle$, the latter considerations underscore 
the importance of the energy rescaling eq.\ \ref{eq:11}, and in particular the consequent rescaling  (eq.\ \ref{eq:17}) of 
the self-energy, $\dtilIno =\dIno/\mu_{E}$ (see also Appendix \ref{section:App1} for a further argument for this): 
it is $\dtilIno$ which in the thermodynamic limit is non-vanishing and finite  with probability unity in an extended 
regime. It is thus its probability distribution $F(\dtilIono)$ which must be considered, rather than that for $\Delta_{I}$.


\section{A mean-field approach}
\label{section:MFT}

Here and in sec.\ \ref{section:ProbDists1} we develop a self-consistent mean-field approach to MBL which, while 
undoubtedly simple, appears to capture at least some key features of the problem. The approach has its 
antecedents in a `typical medium' theory introduced many years ago,~\cite{DELPGWJCP1990,LeitnerReview2015}  in work on localization 
and vibrational energy flow in a many-body quantum state space, following a similar earlier approach to one-body 
localization.~\cite{DELPGWPRB1987,DELPGWdipolar1987}

As mentioned above we consider the Feenberg self-energy $\stilIono$ for the local FS propagator $\gtilIono$, 
in particular the renormalized perturbation series (RPS) for it~\cite{Economoubook,Feenberg1948} (which follows 
from analysis of eq.\ \ref{eq:13}). With this, the self-energy is expressed as a function of the local propagators 
$\{\gtilJono\}$, such that determining $\stilIono$ becomes a question of self-consistency (in a probabilistic sense). 
For the many-body system of interest the RPS has exactly the same structure as for a one-body TBM, because
the Fock-space Hamiltonian eq.\ \ref{eq:2} is of TBM form. The RPS for $S_{I}(\w)$ is given~\cite{Economoubook} as a sum 
of all closed, self-avoiding hopping paths on the FS lattice which begin and end on FS site $I$, contain $n \geq 2$ powers of the hopping $t$ (with only even $n$ possible since the FS lattice is bipartite~\cite{SWDEL1}), and where intermediate 
vertices corresponding to sites $J \neq I$ contain the local propagator for that site. Specifically
\begin{equation}
\label{eq:20}
\begin{split}
S_{I}^{\pd}(\w) ~=&~ \sum_{J} T_{IJ}^{2} G_{J}^{\pd}(\w)~+~....
\\~=&~t^{2}\sum_{J} \frac{1}{\w^{+}-\eJ -S_{J}^{\pd}(\w)}~+~....
\end{split}
\end{equation}
with the second-order ($n=2$) term shown explicitly, the sum being over the $Z_{I}$ Fock-space sites $J$ which are 
connected to site $I$ under the hopping $t$; and with $+...$ referring to higher-order ($n\geq 4$) RPS terms.
With $\gtilIono =\mu_{E}G_{I}$ (eq.\ \ref{eq:15}), $\stilIono =S_{I}/\mu_{E}$ (eq.\ \ref{eq:17}) follows as~\cite{FN2}
\begin{equation}
\label{eq:21}
\stilI ~=~ \frac{t^{2}}{\mu_{E}^{2}} \sum_{J} \gtilJ
~=~\frac{t^{2}}{\mu_{E}^{2}} \sum_{J} \frac{1}{\wtil^{+}-\etilJ -\stilJ}
\end{equation}
We have dropped the higher-order RPS terms here, because in practice we consider the problem only at second-order 
level. This is well known to be exact if the Fock-space lattice has the topology of a Bethe 
lattice,~\cite{Economoubook,AAT1973} which it does not. Here we employ it as a natural leading-order approximation,
conjecturing that higher-order  RPS contributions will produce only a quantitative modification of results arising at second-order level~\cite{FN3} (for which the numerical results of sec.\ \ref{section:EDresults}, which do not make this approximation, provide support).

Granted this, we then analyze the problem simply and approximately in the spirit of a probabilistic mean-field theory, via the following strategy: \\ 
(a) The self-energy $\stilJono$ in eq.\ \ref{eq:21} for sites $J$ connected to $I$ is first replaced by a typical value, denoted $\tilde{S}_{\mathrm{t}}(\wtil) = \tilde{X}_{\mathrm{t}}(\wtil) -i\tildeltyp$. \\
(b) With this, the distribution $F(\dtilIo)$ of $\dtilIo$ is obtained.\\
(c) Self-consistency is then imposed by requiring that a typical value of $\dtilIo$ arising from this distribution coincides with the input $\tildeltyp$.
In practice, `typicality' is measured by the geometric mean,
\begin{equation}
\label{eq:22}
\ln\tildeltyp ~=~\langle \ln\dtilI\rangle
\end{equation}
with the average $\langle ...\rangle = \int d\wtil ~ ... F(\dtilIo)$ over the distribution $F(\dtilIo)$.
$\tildeltyp$ acts in effect as an order parameter, being non-zero if states of energy $\wtil$ are extended; and vanishingly small in the MBL phase, with $\tildeltyp \propto \etatil \rightarrow 0+$. With this, equation \ref{eq:21} becomes
\begin{equation}
\label{eq:23}
\stilI ~=~\frac{t^{2}}{\mu_{E}^{2}} \sum_{J} \frac{1}{\overline{\w}^{+}-\etilJ +i\tildeltyp}
\end{equation}
where for convenience we have subsumed $\mathrm{Re}\tilde{S}_{\mathrm{t}}(\wtil) = \tilde{X}_{\mathrm{t}}(\wtil)$ 
into the energy, by defining $\overline{\w} \equiv \overline{\w}(\wtil)$ as
\begin{equation}
\label{eq:24}
\overline{\w}~=~\wtil ~-~ \tilde{X}_{\mathrm{t}}(\wtil).
\end{equation}

Now consider the Fock-space site-energy $\eJ$ for any site $J$ connected to $I$ under hopping. Since the hopping is 
between NN real-space sites, $\eJ$ differs from $\eI$ simply by the difference in site-energies of sites between which 
the fermion hops, and by the resultant change in the NN interaction contribution; e.g.\ for $d=1$,
\begin{equation}
\nonumber
{\cal{E}}_{J}^{\pd} ~=~{\cal{E}}_{I}^{\pd} +\epsilon_{i+1}^{\pd}-\epsilon_{i}^{\pd}~+~
V [n_{i+2}^{\pd}-n_{i-1}^{\pd}]
\end{equation}
with $\epsilon_{i+1}$ the site-energy of the occupied real-space site in $J$ ($\equiv |J\rangle$) to which a fermion 
hops under $T_{IJ}$, and  $\epsilon_{i}$ that for the NN real-space site occupied in $I$ from whence it came 
[the occupation numbers $n_{i+2}$ and $n_{i-1}$ are the same for both $I$ and $J$]. The $\eJ$ are thus highly correlated 
with $\eI$, and differ from it by an amount of ${\cal{O}}(W,V)$, with $W$ the disorder scale on which the 
real-space site-energies fluctuate. Hence $\tilde{{\cal{E}}}_{J} ={\cal{E}}_{J}/\mu_{E}$ entering eq.\ \ref{eq:23} is 
\begin{equation}
\label{eq:25}
\tilde{{\cal{E}}}_{J}^{\pd} ~=~\tilde{{\cal{E}}}_{I}^{\pd} ~+~
{\cal{O}}\Big(\tfrac{W}{\mu_{E}},\tfrac{V}{\mu_{E}}\Big)
~ \equiv ~ \tilde{{\cal{E}}}_{I}^{\pd},
\end{equation}
since $\mu_{E} \propto \sqrt{N}$. The key point here is that the rescaled FS site-energies $\tilde{{\cal{E}}}_{J}$ 
entering eq.\ \ref{eq:23} for $\tilde{S}_{I}(\wtil)$ are the same as that for site $I$ in the thermodynamic limit, 
i.e.\ are effectively resonant with it.

Eq.\ \ref{eq:23} thus reduces to
\begin{subequations}
\label{eq:26}
\begin{align}
\stilI ~=&~ \frac{t^{2}}{\mu_{E}^{2}} \sum_{J} \frac{1}{\overline{\w}^{+}-\etilI
+i\tildeltyp}
\label{eq:26a}
\\
~=&~\frac{Z_{I}t^{2}}{\mu_{E}^{2}} ~\frac{1}{\overline{\w}^{+}-\etilI
+i\tildeltyp} .
\label{eq:26b}
\end{align}
\end{subequations}
All terms in the sum in eq.\ \ref{eq:26a} are the same, and their number is the coordination number $Z_{I}$ of site $I$.
This has a mean of $\overline{Z} \propto N $ (eq.\ \ref{eq:5}) and a standard deviation $\mu_{Z}^{\pd}\propto \sqrt{N}$.
Since $\mu_{E}^{2} \propto N$, only the mean $\overline{Z}$ is  relevant in the thermodynamic limit, so we replace 
$Z_{I} \equiv \overline{Z}$ in eq.\ \ref{eq:26b}. $\dtilI = -\mathrm{Im}\tilde{S}_{I}(\wtil)$  then follows as
\begin{equation}
\label{eq:27}
\dtilI  ~=~\frac{\Gamma~[\etatil+\tildeltyp]}{[\overline{\w}
-\etilI]^{2}+[\etatil+\tildeltyp]^{2}} ,
\end{equation}
where $\Gamma =\overline{Z}t^{2}/\mu_{E}^{2}$ is thus defined and is finite as $N\rightarrow \infty$, 
being given by (eqs.\ \ref{eq:5},\ref{eq:8},\ref{eq:9})
\begin{equation}
\label{eq:28}
\Gamma ~=~\frac{\overline{Z}t^{2}}{\mu_{E}^{2}}
~=~\frac{t^{2}}{\left[t^{2} + \frac{1}{2d}\langle\epsilon^{2}\rangle +\frac{1}{2}\nu(1-\nu)V^{2}\right]}
~\leq ~1 .
\end{equation} 
Eq.\ \ref{eq:27} will be used in the following section to determine the probability distribution of 
$\tilde{\Delta}_{I}(\wtil)$. Before that it is however useful to have some insight into
(a) the physical origin of the factor $Z_{I}t^{2}/\mu_{E}$ in the basic expression eq.\ \ref{eq:26b} for $\stilIono$; 
and (b) some key differences between the present problem and that arising in 1-body localization.


\subsubsection{Physical digression}
\label{subsubsection:digression1}

There are two distinct elements contributing to the factor of $Z_{I}t^{2}/\mu_{E}^{2}$ in eq.\ \ref{eq:26}.
First, an effective rescaling of the hopping, $t \rightarrow t/\mu_{E}^{\pd}$. This is a general consequence of 
rescaling the energy as in eqs.\ \ref{eq:11},\ref{eq:16}, and hence the local propagator as 
$\gtilIono =\mu_{E}G_{I}$  (eq.\ \ref{eq:15}). To see this directly define $\tilde{G}_{IJ} =\mu_{E}G_{IJ}$ generally, 
so the `equation of motion' eq.\ \ref{eq:13} reads
\begin{equation}
\nonumber
(\wtil^{+}-\etilI) \tilde{G}_{IJ}^{\pd}(\wtil) =\delta_{IJ}^{\pd} +\sum_{K} \tilde{T}_{IK}^{\pd}\tilde{G}_{KJ}^{\pd}(\wtil)
\end{equation}
with $\tilde{T}_{IK}^{\pd} =T_{IK}^{\pd}/\mu_{E}^{\pd}$ 
(and $|\tilde{T}_{IK}^{\pd}| = t/\mu_{E}^{\pd}$). 
Comparison to eq.\ \ref{eq:13}
shows that the local $\gtilIono$ in particular, and hence $\stilIono$, are the same functions of 
$\{ \tilde{T}_{JK}^{\pd}\}$ (and  $\{(\wtil  -\etilJo)\}$) that $G_{I}$ and $S_{I}$ are of 
$T_{IK}^{\pd}$ (and $\{(\w -\eJo)\}$); or equivalently for the RPS itself, 
that $\stilIono$ is the same function of $t/\mu_{E}$ and the $\{\gtilJono\}$ that $S_{I}$ is of $t$ and the $\{ G_{J}\}$.
This effective rescaling of $t \rightarrow t/\mu_{E} \propto t/\sqrt{N}$ is analogous to that required in
dynamical mean-field theory~\cite{dmftgeorgeskotliar} to ensure that the limit of infinite spatial dimensions is 
well-defined.

 To illustrate the physical origin of the coordination number factor in $Z_{I}t^{2}/\mu_{E}^{2}$, consider a particularly
simple limit: of `one shell', where a given FS site $I$ is coupled under the hopping $H_{t}$ to its $Z_{I} \propto N$ 
neighbors $\{ J\}$, themselves uncoupled from each other (eq.\ \ref{eq:26} above, obviously with $\tilde{S}_{\mathrm{t}} =0$,
in fact captures this limit  exactly). This is formally equivalent to a non-interacting Anderson impurity~\cite{hewsonbook} coupled to a `conduction band' containing  $Z_{I} \propto N$ states (the `impurity' being $I$, band states 
the $\{ J\}$). In this case the hopping contribution $H_{t}$ is given by
\begin{equation}
\label{eq:29}
H_{t}^{\pd} ~=~ |I\rangle ~ \sum_{J} T_{IJ}^{\pd} \langle J| +\mathrm{h.c.} ,
\end{equation}
showing that $|I\rangle$ couples directly \emph{only} to the particular linear combination of states denoted 
by $|0\rangle \propto \sum_{J} T_{IJ}^{\pd}|J\rangle$. This corresponds to the so-called $0$-orbital in the Wilson 
chain representation of the Anderson model,~\cite{hewsonbook} and is given in normalized form by
\begin{equation}
\nonumber
|0\rangle ~=~ \frac{1}{\sqrt{Z_{I}}t} \sum_{J} T_{IJ}^{\pd}|J\rangle 
\end{equation}
(since $\sum_{J}T_{IJ}^{2} =Z_{I}t^{2}$).
In terms of it, $H_{t}$ (eq.\ \ref{eq:29}) is thus
\begin{equation}
\label{eq:30}
H_{t}^{\pd}~=~\sqrt{Z_{I}}t~|I\rangle\langle 0| +\mathrm{h.c.}
\end{equation}
so that $|I\rangle$ and $|0\rangle$ are coupled by an effective hopping of $\sqrt{Z_{I}}t$.~\cite{FN4}
 Combined with the $t \rightarrow t/\mu_{E}$ rescaling above, the effective hopping entering $\stilI$ is thus 
$t_{\mathrm{eff}} =\sqrt{Z_{I}}t/\mu_{E}$ (the square of which naturally appears in the second-order RPS expression 
eq.\ \ref{eq:26b}, or eq.\ \ref{eq:27}); with $t_{\mathrm{eff}}$, and hence
$\Gamma =\overline{Z} t^{2}/\mu_{E}^{2}$ (eq.\ \ref{eq:28}), thus remaining finite as $N\rightarrow \infty$.

It may also be helpful to contrast the situation arising above for MBL with that occurring for 1-body localization. In the latter case the counterpart of eq.\ \ref{eq:20} for the self-energy $S_{i}(\w)$ of real-space site $i$ is
\begin{equation}
\nonumber
S_{i}^{\pd}(\w) ~=~ 
 t^{2}\sum_{j} \frac{1}{\w^{+}-\epsilon_{j}^{\pd} -S_{j}^{\pd}(\w)}
\end{equation}
with the sum over the  NN sites to $i$ (of which there are, say, $K_{c}$). Since the site-energies $\{\epsilon_{k}\}$ are independent random variables, the right hand side of this expression is a sum of $K_{c}$ independent random 
terms; in contrast e.g.\ to eq.\ \ref{eq:26a} for MBL where, since the $Z_{I}$ terms in the sum coincide,  there is only a single random term. In addition, the probability distribution for $S_{i}(\w)$ in this case is \emph{independent} of the site energy of site $i$; while in the MBL case the distribution of $\stilIo$ both depends on the Fock-space site-energy $\etilIo$ for that site and, at the level of eq.\ \ref{eq:26}, is in fact entirely determined by it.


\section{Self-consistent distributions and MBL transition}
\label{section:ProbDists1}

Using eq.\ \ref{eq:27} for $\dtilIo$, its distribution (for any given energy $\wtil$) is obtained by integrating over the distribution ${\cal{P}}(\etilIo)$ of $\etilIo=(\eIo-\overline{\e})/\mu_{E}$ (eq.\ \ref{eq:16}),
\begin{equation}
\label{eq:31}
F(\dtilIo) =\int^{\infty}_{-\infty} d\etilI~{\cal{P}}(\etilI)~\delta\Big(
\dtilIo - \frac{\Gamma~[\etatil+\tildeltyp]}{[\etilI-\overline{\w}]^{2}+[\etatil+\tildeltyp]^{2}}\Big) .
\end{equation}
From sec.\ \ref{subsection:distributions0}, ${\cal{P}}(\etilI)$ is a Gaussian with vanishing mean,
\begin{equation}
\label{eq:32}
{\cal{P}}(\etilI)~=~\frac{1}{\sqrt{2\pi}\lambda} \exp\left(
-\frac{\tilde{{\cal{E}}}_{I}^{2}}{2\lambda^{2}}
\right),
\end{equation}
with $\lambda = \mu_{\e}/\mu_{E}$ independent of $N$ and given (from eqs.\ \ref{eq:8},\ref{eq:9},\ref{eq:5}) by
\begin{equation}
\label{eq:33}
\lambda^{2} ~=~\frac{\mu_{{\cal{E}}}^{2}}{\mu_{E}^{2}} ~=~
\frac{\langle \epsilon^{2}\rangle +\nu(1-\nu)dV^{2}}{\langle\epsilon^{2}\rangle 
+\nu(1-\nu) dV^{2} +2dt^{2}}.
\end{equation}

From now on in this section we focus explicitly on the band center $\wtil =0$, eqs.\ \ref{eq:11},\ref{eq:12}
(equivalently $\overline{\w} =0$, eq.\ \ref{eq:24}, since $\tilde{X}_{\mathrm{t}}(\wtil =0)$ vanishes by symmetry).
With the form eq.\ \ref{eq:32} for ${\cal{P}}(\etilIo)$, eq.\ \ref{eq:31} is readily evaluated to give
\begin{equation}
\label{eq:34}
\begin{split}
&F(\dtilIo) ~=~
\\
&\frac{1}{\sqrt{1- \dtilIo \frac{[\etatil+\tildeltypo]}{\Gamma}}}~\sqrt{\frac{\kappa}{\pi}}~
\frac{1}{\tilde{\Delta}_{I}^{\frac{3}{2}}}~
\exp\Big[-\kappa \Big(\frac{1}{\dtilIo} - \frac{[\etatil +\tildeltypo]}{\Gamma}\Big)\Big]
\end{split}
\end{equation}
for $0 \leq \tilde{\Delta}_{I} <\Gamma/(\tilde{\eta}+\tilde{\Delta}_{\mathrm{t}})$ (and zero otherwise), where
\begin{equation}
\label{eq:35}
\kappa ~=~\frac{\Gamma (\etatil+\tildeltypo)}{2\lambda^{2}}
\end{equation}
is thus defined. Eq.\ \ref{eq:34} encompasses both the MBL and delocalized regimes. We now consider them separately.


\subsection{MBL regime}
\label{subsection:MBL1}

In the MBL regime $y=\dtilIo/\etatil$ is finite, so one considers its distribution $\tilde{F}(y) =\etatil F(\etatil y)$ 
(sec.\ \ref{section:propagatorsformal}). Since $\tilde{\eta} =0+$, eq.\ \ref{eq:34} then gives 
\begin{equation}
\label{eq:36}
\tilde{F}(y)~=~\sqrt{\frac{\tilde{\kappa}}{\pi}}~\frac{1}{y^{\frac{3}{2}}}~
\exp\Big(-\frac{\tilde{\kappa}}{y}\Big)~~~~~~:~ y =\frac{\tilde{\Delta}_{I}^{\pd}}{\tilde{\eta}}
\end{equation}
(holding for all $y \in (0,\infty)$), where (eq.\ \ref{eq:35})
\begin{equation}
\label{eq:37}
\tilde{\kappa}~=~\frac{\kappa}{\tilde{\eta}} ~=~\frac{\Gamma}{2\lambda^{2}}~
\Big(
1+\frac{\tilde{\Delta}_{\mathrm{t}}}{\tilde{\eta}}
\Big).
\end{equation}
Eq.\ \ref{eq:37} is precisely a L\'evy distribution, with a characteristic long tail $\propto y^{-3/2}$, such that all 
moments of $y$ diverge. For the reasons explained in sec.\ \ref{section:propagatorsformal}, this is a physically natural 
form for the distribution of $\tilde{\Delta}_{I}/\tilde{\eta}$ in the MBL phase. We will compare this behavior
to results obtained by exact diagonalization in sec.\ \ref{section:EDresults} (where none of the approximations entering 
the mean-field theory are made).  

Next we impose self-consistency on the geometric mean (eq.\ \ref{eq:22}). From the normalized $\tilde{F}(y)$ 
eq.\ \ref{eq:36}, the average $\langle \ln(\dtilIo/\etatil)\rangle = \int_{0}^{\infty}dy~\tilde{F}(y) \ln y$ follows, 
\begin{subequations}
\label{eq:38}
\begin{align}
\Big\langle\ln\Big(\frac{\tilde{\Delta}_{I}}{\tilde{\eta}}\Big)\Big\rangle ~=&~
\ln \tilde{\kappa} ~-~\frac{4}{\sqrt{\pi}}\int_{0}^{\infty}dx ~ \ln x ~ e^{-x^{2}}
\label{eq:38a}
\\
=&~\ln (4\tilde{\kappa})+\gamma   ~~~~~~:~\gamma = 0.577216..
\label{eq:38b}
\end{align}
\end{subequations}
with $\gamma$ the Euler-Mascheroni constant. Imposing  $\ln(\tildeltypo/\etatil) = \langle\ln(\dtilIo/\etatil)\rangle$
(eq.\ \ref{eq:22}) gives $\tildeltypo/\etatil =4e^{\gamma}\tilde{\kappa}$, with $\tilde{\kappa}$ given by eq.\ \ref{eq:37}.
Hence the self-consistency condition  $y_{\mathrm{t}}^{\pd} =[2e^{\gamma}\Gamma/\lambda^{2}](1+y_{\mathrm{t}}^{\pd})$
for $y_{\mathrm{t}}=\tildeltypo/\etatil$ ($\geq 0$); yielding
\begin{equation}
\label{eq:39}
\frac{\tildeltypo}{\etatil} ~=~ T \big[1-T\big]^{-1} ~~~~~~:~T \leq 1
\end{equation}
with $T (\geq 0)$ defined by 
\begin{equation}
\label{eq:40}
T~=~\frac{2e^{\gamma}\Gamma}{\lambda^{2}} ~=~ \frac{2e^{\gamma}\overline{Z}t^{2}}{(\lambda\mu_{E}^{\pd})^{2}} .
\end{equation}
Note that $\Gamma$ (eq.\ \ref{eq:28}) and $\lambda$ (eq.\ \ref{eq:33}) are finite as $N \rightarrow \infty$
(with $\overline{Z}$ and $\mu_{E}^{2}$ each $\propto N$), whence $T$ remains bounded in the thermodynamic limit.
Since $\dtilIo/\etatil$ is non-negative and finite with probability one throughout the MBL phase, eq.\ \ref{eq:39} 
shows that the phase is self-consistent only for $T<1$. The transition to delocalization from the MBL phase thus occurs as 
$T \rightarrow 1-$, where $\tildeltypo/\etatil \sim [1-T]^{-s}$ diverges with an exponent of unity (fig.\ \ref{fig:fig2}).

The mean-field transition criterion $T=1$ will be considered further below. Here we simply note that $\Gamma$ 
(eq.\ \ref{eq:28}), $\lambda$ (eq.\ \ref{eq:33}), and hence $T$, are invariant under $\nu \leftrightarrow (1-\nu)$.
Filling fractions $\nu$ $(=N_{e}/N)$ and $(1-\nu)$ are thus equivalent; as required physically, and reflecting the invariance of $H$ under a particle-hole transformation.

The self-consistent $\tilde{\kappa}$ entering the L\'evy distribution eq.\ \ref{eq:36} follows from 
eqs.\ \ref{eq:37},\ref{eq:39},\ref{eq:40} as
\begin{equation}
\label{eq:41}
\tilde{\kappa}~=~ \frac{1}{4e^{\gamma}} \frac{\tildeltypo}{\etatil} ~=~\frac{1}{4e^{\gamma}} \frac{T}{1-T}.
\end{equation}
It sets the scale for the emergence of the $\sim y^{-3/2}$ tails characteristic  of the distribution, which 
arise (eq.\ \ref{eq:36}) for $y \gg \tilde{\kappa}$, and get pushed to progressively larger $y$-values on 
approaching the transition $T \rightarrow 1-$. 

More importantly, note  that the L\'evy distribution for $y =\dtilIo/\etatil$ can be written in the one-parameter scaling form
\begin{subequations}
\label{eq:42}
\begin{align}
\tilde{F}(y) ~=&~\frac{1}{c\tilde{\kappa}} ~f_{L}^{\pd}\Big(\frac{y}{c\tilde{\kappa}}\Big) 
~=~ \frac{1}{\tilde{c\kappa}} ~f_{L}^{\pd}(x) 
\label{eq:42a}
\\
f_{L}^{\pd}(x) ~=&~\frac{1}{\sqrt{\pi c}~x^{\frac{3}{2}}}~
\exp\big(-\frac{1}{cx}\big) ~~~~:~ x =\frac{y}{c\tilde{\kappa}},
\label{eq:42b}
\end{align}
\end{subequations}
with $c$ an arbitrary constant (independent of the physical parameters); and with $f_{L}(x)$ -- the probability density of 
$x=y/(c\tilde{\kappa})$ -- dependent solely on $x$. If e.g.\ $c =4e^{\gamma}$ is chosen (with $\gamma$ again Euler's constant), 
then since the geometric mean of eq.\ \ref{eq:36} is $y_{\mathrm{t}} = 4e^{\gamma} \tilde{\kappa}$, 
eq.\ \ref{eq:42} reads $\tilde{F}(y) = y_{\mathrm{t}}^{-1} f_{L}(x)$ with $f_{L}(x)$ the probability density of 
$x \equiv y/y_{\mathrm{t}}$. The central point here is that all L\'evy distributions can be scaled onto each other, 
with $f_{L}(x)$ as such characteristic of the \emph{entire} MBL phase. We return to this important point when considering numerical results in sec.\ \ref{section:EDresults}.


\subsection{Delocalized regime}
\label{subsection:MBD1}

In a delocalized regime $\dtilIono$ (and hence $\tildeltypo$) is finite, and since $\etatil =0+$
eq.\ \ref{eq:34} for $F(\tilde{\Delta}_{I})$ thus reduces to
\begin{equation}
\label{eq:43}
F(\dtilIo) ~=~
~\sqrt{\frac{\kappa}{\pi}}~\frac{1}{\tilde{\Delta}_{I}^{\frac{3}{2}}}~
\exp\Big(-\frac{\kappa}{\dtilIo}\Big)~
\times \frac{1}{\sqrt{1- \dtilIo \frac{\tildeltypo}{\Gamma}}}
\exp\Big(\kappa\frac{\tildeltypo}{\Gamma}\Big)
\end{equation}
with $\kappa =\Gamma\tildeltypo/(2\lambda^{2})$. This is the product of a  L\'evy distribution
(in $\dtilIono$ itself), times a contribution that is integrably divergent as the upper limit 
$\dtilIono =\Gamma/\tildeltypo$ of the distribution is approached (that limit acting as a cutoff to the L\'evy tail).

To determine $\ln \tildeltypo =\langle \ln\dtilIono\rangle$ self-consistently it is more economical to work directly 
with eqs.\ \ref{eq:31},\ref{eq:32} for $F(\dtilIo)$; from which  (with $\etatil=0$ and $\overline{\w}=0$),
\begin{subequations}
\label{eq:44}
\begin{align}
\langle \ln\dtilIo &\rangle ~=~
\frac{1}{\sqrt{2\pi}\lambda}
\int^{\infty}_{-\infty}dx~e^{-\frac{x^{2}}{2\lambda^{2}}}~
\ln \Big[\frac{\Gamma\tildeltypo}{x^{2}+{\tildeltypo}^{2}}\Big]
\label{eq:44a}
\\
=&~\frac{2}{\sqrt{\pi}}\int^{\infty}_{0}dy ~ e^{-y^{2}}
\ln \left[
\frac{\Gamma\tildeltypo}{2\lambda^{2}y^{2}\left(1+\frac{{\tildeltypo}^{2}}{2\lambda^{2}y^{2}}\right)}
\right]
\label{eq:44b}
\end{align}
\end{subequations}
Hence  (noting eqs.\ \ref{eq:38})
\begin{equation}
\label{eq:45}
\begin{split}
\ln\tildeltypo~&=~\ln \left[\frac{2e^{\gamma}\Gamma}{\lambda^{2}} \tildeltypo\right]
\\
&
~-~ \frac{2}{\sqrt{\pi}}\int_{0}^{\infty}dy ~\exp(-y^{2})~
\ln\left(1+\frac{\tilde{\Delta}_{\mathrm{t}}^{2}}{2\lambda^{2}y^{2}}\right)
\end{split}
\end{equation}
where the first term is recognized  (eq.\ \ref{eq:40}) as being $\ln (T\tildeltypo)$. The low-$\tildeltypo$ 
behavior of the integral in eq.\ \ref{eq:45} is 
\begin{equation}
\int_{0}^{\infty}dy ~\exp(-y^{2})~\ln\Big(1+\frac{x}{y^{2}}\Big)
~\overset{x \rightarrow 0+}{\sim} ~\pi \sqrt{x} ~-~\sqrt{\pi}x ~+~{\cal{O}}\big(x^{\tfrac{3}{2}}\big).
\nonumber
\end{equation}
Eq.\ \ref{eq:45} thus yields the self-consistency condition for $\tildeltypo$ in the delocalized phase 
close to the transition,
\begin{equation}
\tildeltypo~\overset{\tildeltypo \rightarrow 0+}{\sim}~
T\tildeltypo~\Big(
1-\frac{\sqrt{2\pi}}{\lambda} \tildeltypo+ 
\frac{[1 +\pi]}{\lambda^{2}}{\tildeltypo}^{2}+{\cal{O}}({\tildeltypo}^{3})
\Big).
\nonumber
\end{equation}
Since $\tildeltypo\geq 0$ necessarily, this has a non-trivial solution only for $T \geq 1$, given to leading order by
\begin{equation}
\label{eq:46}
\tilde{\Delta}_{\mathrm{t}}^{\pd}~\overset{T \rightarrow 1+}{\sim}~
\frac{\lambda}{\sqrt{2\pi}T}~\left[ T-1\right]^{s} ~~~~~~:~s=1.
\end{equation}
The transition approached from the delocalized phase thus occurs (as it ought) at the same point $T=1$ as the approach to it from the MBL phase; and $\tildeltypo$ vanishes as $T \rightarrow 1+$ with the same exponent, $s=1$, with which 
$\tildeltypo/\etatil$ diverges as the transition is approached from the MBL side (eq.\ \ref{eq:39}).

\begin{figure}
\includegraphics{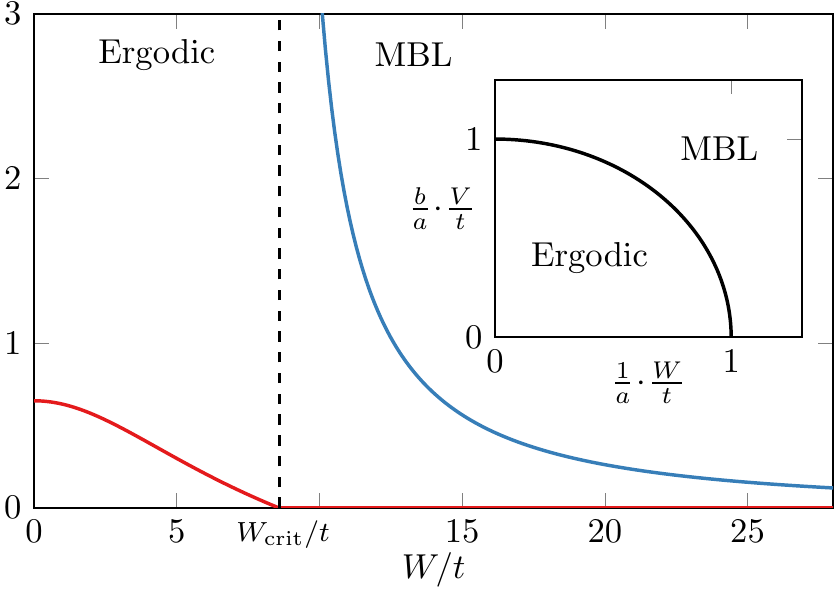}
\caption{\label{fig:fig2} 
Evolution with disorder ($W/t$) of the self-consistent mean-field $\tildeltypo$ (delocalized/ergodic phase) and 
$\tildeltypo/\etatil$ (MBL phase), for the band center $\wtil =0$. Shown for a box site-energy distribution, for $d=1$ at 
half-filling with $V/t=2$. The dashed line indicates the transition point.
\emph{Inset}: 
Mean-field phase diagram in the $(W/t, V/t)$-plane (with constants $a,b$ given by eq.\ \ref{eq:47}). Discussion in text.
}
\end{figure}

Throughout the delocalized phase more generally, the self-consistent $\tildeltypo$ is obtained numerically.
One further limit is however readily extracted. As the width, $\lambda$, of the distribution of 
$\etilIo$ (eq.\ \ref{eq:32}) vanishes, $F(\dtilIono)\rightarrow \delta(\dtilIono -\tildeltypo)$
tends to a $\delta$-distribution. In this case eq.\ \ref{eq:44a}  gives
$\ln\tildeltypo =\langle \ln \dtilIono\rangle =\ln(\Gamma/\tildeltypo)$, and hence
\begin{equation}
\nonumber
\tilde{\Delta}_{\mathrm{t}}^{\pd} ~ \overset{\lambda \rightarrow 0}\sim ~ 
\sqrt{\Gamma}
~=~\frac{t}{\sqrt{t^{2} + \frac{1}{2d}\langle\epsilon^{2}\rangle +\frac{1}{2}\nu(1-\nu)V^{2}}}
~\sim ~1
\end{equation}
(using eq.\ \ref{eq:28} for $\Gamma$). This is just the small $\lambda$ limit of the Fermi golden rule behavior 
(see eq.\ \ref{eq:19}).~\cite{FN5}

The behavior considered above and in sec.\ \ref{subsection:MBL1} is exemplified in fig.\ \ref{fig:fig2}, showing the 
evolution of $\tildeltypo$ and $\tildeltypo/\etatil$ (for the MBL phase)   with disorder, $W/t$.
Results are shown for the box site-energy distribution  $P(\epsilon) =\theta(\tfrac{W}{2}-|\epsilon|)$ (for which 
$\langle\epsilon^{2}\rangle =W^{2}/12$), and for $d=1$ at half-filling with interaction $V/t =2$.\\

Now consider further the simple mean-field transition criterion, viz.\ $T=1$ with $T$ given by eq.\ \ref{eq:40}. 
From eqs.\ \ref{eq:40},\ref{eq:33},\ref{eq:28} (with $\langle \epsilon^{2}\rangle =W^{2}/12$ as above),
\begin{equation}
\label{eq:47}
\begin{split}
T~=&~ \frac{a^{2}}{1+b^{2} \big(\frac{V}{W}\big)^{2}}~\Big[\frac{t}{W}\Big]^{2} 
\\
:~ a =& 4e^{\gamma/2} \sqrt{3d} \simeq 9.2 \sqrt{d}, ~~~~b = 2\sqrt{3\nu(1-\nu) d} 
\end{split}
\end{equation}
Defining $x =W/at$ and $y=bV/at$, the condition $T < 1$ for MBL states is $y^{2} > 1-x^{2}$. The resultant phase boundary is shown in fig.\ \ref{fig:fig2} (inset), and the following points should be noted.
\noindent (1) A transition arises for all space dimension $d$, including for the non-interacting limit $V=0$. The latter is of course wrong for $d=1,2$;  but is as  expected from a mean-field theory, which may handle adequately generic behavior above a lower critical dimension but not below it.
(2) As seen from the $V$-dependence of the phase boundary in fig.\ \ref{fig:fig2} inset, increasing the interaction
for given, sufficiently low disorder, eventually drives the system to an MBL phase, as one anticipates physically 
(interactions effectively self-generate disorder in the distribution of $\{\eIo\}$~\cite{FNlater}). This is indeed 
as found by exact diagonalization~\cite{ReichmanPRL2015} (although a transition occurring at a finite-$V/t$ as the disorder vanishes, rather than e.g.\ as $V/t \rightarrow \infty$, is presumably an artifact of the theory).
(3) The transition for $d=1$ at half-filling  and for $V/t =2$ occurs at $W/t \simeq 8.6$. As mentioned earlier, a range of exact diagonalization studies for these parameters estimate the MBL transition to occur for
$W/t$ in the range $\sim 12$$-$$16$;~\cite{Pal+HusePRB2010,ReichmanPRB2010,DeLucaScardicchioEPL2013,LuitzAletPRB2015,BeraFHMBardarsonPRL2015}
so the mean-field estimate appears not wildly out of line.

\begin{figure}
\includegraphics{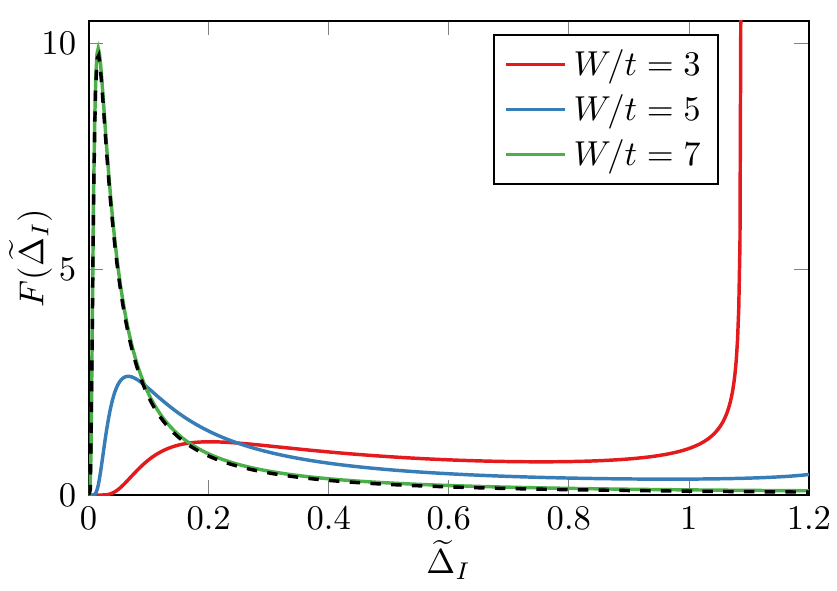}
\caption{\label{fig:fig3} 
Mean-field  $F(\tilde{\Delta}_{I}^{~})$ \emph{vs} $\tilde{\Delta}_{I}^{~}$ (eq.\ \ref{eq:43})
in the delocalized phase; for the same parameters as figs.\ \ref{fig:fig1},\ref{fig:fig2} 
and shown for $W/t = 3,5,7$ ($W_{\mathrm{crit}}/t \simeq 8.6$). With increasing disorder, while remaining 
in the delocalized phase, a L\'evy form for $F(\tilde{\Delta}_{I}^{~})$ emerges (shown for the $W/t =7$ case, 
dashed line). Discussion in text.
}
\end{figure}

Finally here, we comment briefly on the mean-field distribution $F(\dtilIono)$  in the delocalized phase.
From eq.\ \ref{eq:43}, this is the product of a L\'evy distribution (in $\dtilIono$ itself), times a factor 
that is integrably divergent as the upper limit $\dtilIono =\Gamma/\tildeltypo$ of the distribution is approached
(and which in passing we add can be shown to generate $F(\dtilIono) \rightarrow \delta(\dtilIono-\tildeltypo)$  as 
$\lambda \rightarrow 0$, mentioned above).
Here we simply note from eq.\ \ref{eq:43} that the L\'evy contribution itself (whose mode occurs at $\dtilIono =2\kappa/3$)
is `well formed' in an obvious sense if $\dtilIono \gg \kappa = (\Gamma/2\lambda^{2})\tildeltypo \propto \tildeltypo$, while the factor $[1-\dtilIono (\tildeltypo/\Gamma)]^{-1/2}$ is unity for practical purposes provided 
$\dtilIono \ll \Gamma/\tildeltypo \propto 1/\tildeltypo$.  On approaching the transition where 
$\tildeltypo \rightarrow 0$, one thus expects to see the emergence of L\'evy behavior in $F(\dtilIono)$ over an 
increasingly wide range of $\dtilIo$; as illustrated in fig.\ \ref{fig:fig3}, from which L\'evy-like behavior is seen to 
emerge reasonably far into the delocalized phase. We return to this when considering exact diagonalization results in 
sec.\ \ref{subsection:delocnumerics}.


\subsubsection{Energy dependence}
\label{subsubsection:finitew}

While we have focused above on the center of the eigenvalue spectrum, $\wtil =(\w-\overline{E})/\mu_{E} =0$,
the analysis can naturally be extended to $\wtil \neq 0$. We comment on it briefly.
In this case the real part $\Xtiltyp =\mathrm{Re}\tilde{S}_{\mathrm{t}}(\wtil)$ no longer vanishes by symmetry, 
and is related to $\tildeltyp =-\mathrm{Im}\tilde{S}_{\mathrm{t}}(\wtil)$ by the Hilbert transform 
$\pi \tilde{S}_{\mathrm{t}}(\wtil) = \int^{\infty}_{-\infty}d\tilde{\w}^{\prime} \tildeltypo(\tilde{\w}^{\prime})/(\wtil^{+}-\wtil^{\prime})$; thus determining the $\wtil$-dependence of  $\overline{\w} =\wtil -\Xtiltyp$ (eq.\ \ref{eq:24}), and 
hence of the general probability density eq.\ \ref{eq:31} for $F(\dtilIono) \equiv F(\dtilIono;\wtil)$. Results arising mirror those obtained above for $\wtil =0$. In the MBL phase for example, the self-consistent $\tildeltyp/\etatil$ 
has the same form as \ref{eq:39}, viz.\
\begin{equation}
\label{eq:48}
\frac{\tildeltyp}{\etatil} ~=~T(\wtil)  [1-T(\wtil)]^{-1} ~~~~~~:~T(\wtil) \leq 1
\end{equation}
with $T(\wtil) < 1$ for the MBL phase to be self-consistent, and $T(\wtil)=T(-\wtil)$. The transition for energy $\wtil$ 
thus occurs as $T(\wtil) \rightarrow 1-$. This condition determines the mobility edges, $\wtil_{\pm} =\pm \wtil_{\mathrm{m}}$, separating regions of localized and delocalized states. $T(\wtil)$ is readily shown to  be of form
\begin{equation}
\label{eq:49}
T(\wtil) ~=~ T(0) \exp\big(-I(\wtil)\big)
\end{equation}
with $T(0)=2e^{\gamma}\Gamma/\lambda^{2}$ the band center result (eq.\ \ref{eq:40}), 
$I(\wtil)\overset{\wtil \rightarrow 0}{\sim}  \alpha {\wtil}^{2} +{\cal{O}}(\wtil^{4})$ (with $\alpha >0$ a constant 
${\cal{O}}(1)$); and with $I(\wtil)=I(-\wtil)$ more generally an increasing function of $\wtil$ from the band center 
$\wtil =0$, whence $T(\wtil)$ decreases with $\wtil$. Band center states are thus the last to become MBL with increasing disorder, and resultant mobility-edge trajectories in the $(W/t,\wtil)$-plane accordingly have the expected characteristic 
`D-shape'.~\cite{LuitzAletPRB2015} Note that since the mobility edges open up continuously on decreasing disorder $W$ below the critical value for band center delocalization, they thus occur at finite values of 
$\wtil_{\mathrm{m}} \equiv (\w_{\mathrm{m}}-\overline{E})/\mu_{E}$;  i.e.\ for $\w_{\mathrm{m}}-\overline{E}$
$\propto \sqrt{N}$ (since $\mu_{E} \propto \sqrt{N}$), as pointed out on general grounds in [\onlinecite{SWDEL1}].


\section{Numerical results}
\label{section:EDresults}

Results obtained by exact diagonalization (ED) are now considered, for a $d=1$ open chain with site-energy 
distribution $P(\epsilon) =\theta(\tfrac{W}{2} -|\epsilon|)$. While other parameter regimes have been studied, 
here we consider explicitly half-filling ($\nu =1/2$) with interaction $V/t =2$, and for states 
at the band center $\wtil =0$.~\cite{FN6}; for which the MBL transition 
occurs for $W/t\sim 12$$-$$16$.~\cite{Pal+HusePRB2010,ReichmanPRB2010,DeLucaScardicchioEPL2013,LuitzAletPRB2015,BeraFHMBardarsonPRL2015}

Since our main aim is to determine the probability distributions of the $\dtilIono$ (or $\dtilIono/\etatil$), and their evolution with disorder $W/t$, we first describe how these are calculated for finite-size systems. The self-energy 
$\stilIno$ -- in its entirety (rather than e.g.\ at truncated RPS level) -- is defined via the inverse of the local 
propagator $\gtilIno$, eq.\ \ref{eq:15}; from which $\dtilIno =-\mathrm{Im}\stilIno$ follows,
\begin{equation}
\label{eq:50}
\begin{split}
\dtilI ~=&~\mathrm{Im}\Big(\frac{1}{\gtilI}\Big)~-~\etatil
\\
\gtilI ~=&~ \sum_{n} \frac{|A_{nI}^{\pd}|^{2}}{\wtil +i\etatil -\tilde{E}_{n}^{\pd}} 
\end{split}
\end{equation}
(where $\etatil =\eta/\mu_{E}$). To obtain $\dtilI$, in principle one considers~\cite{ThoulessReview1974} 
first the thermodynamic limit $N\rightarrow \infty$, followed by $\eta \rightarrow 0+$.
For any finite-size system, however -- where the eigenfunction amplitudes $A_{nI}$ (eq.\ \ref{eq:4}) are obtained 
from ED -- the thermodynamic limit obviously cannot be taken, whence in turn $\eta$ cannot be set to zero from the outset.
This just reflects the fact that any finite-size system, no matter how large, strictly speaking has a discrete eigenvalue spectrum. The smallest typical energy scale in that spectrum is however the eigenvalue spacing $[N_{{\cal{H}}}D(\w)]^{-1}$, and it is an $\eta$ of this order that should be considered in finite-$N$ calculations (which amounts simply to regularizing the $\delta$-functions in e.g.\ $D(\w) =N_{{\cal{H}}}^{-1}\sum_{n}\delta(\w -E_{n})$, replacing them by Lorentzians of halfwidth $\eta$). We thus take $\eta =[N_{{\cal{H}}}D(\w)]^{-1}$, and hence $\etatil =[N_{{\cal{H}}}\tilde{D}(\wtil)]^{-1}$ with $\tilde{D}(\wtil)$ the standard normal DoS eq.\ \ref{eq:12}, i.e.\
\begin{equation}
\label{eq:51}
\etatil ~=~ \frac{1}{N_{{\cal{H}}}\tilde{D}(\wtil)}~\overset{\wtil =0}{=}~ \frac{\sqrt{2\pi}}{N_{{\cal{H}}}}
\end{equation}
such that $\etatil \propto 1/N_{{\cal{H}}} \sim e^{-cN}$ is exponentially small in the number of sites $N$.
We have taken the $\sqrt{2\pi}$ prefactor indicated in eq.\ \ref{eq:51} (but have naturally confirmed that results are insensitive to this choice). 

With the procedure sketched, the distributions are determined by averaging over disorder realizations and FS sites; with
$F(\dtilIono)$ the probability density, over an ensemble of disorder realizations, that any site has a particular 
value of $\dtilIono$. Calculations are for system sizes $N=10 -16$ ($N_{{\cal{H}}} \sim 250 -13000$), with $5000$ 
disorder realizations typically sampled for all but the highest $N=16$.

\begin{figure}
\includegraphics{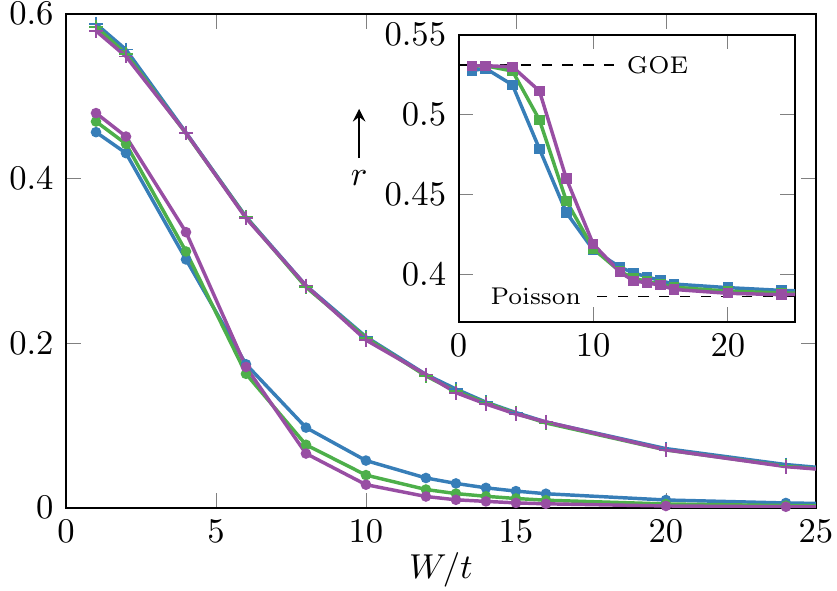}
\caption{\label{fig:fig4} 
Geometric mean  $\tildeltypo$ (circles) of $F(\dtilIono)$ \emph{vs} disorder $W/t$, for system sizes $N=12$ (blue), 
$14$ (green), $16$ (purple).  Arithmetic means $\langle \dtilIono\rangle$ (crosses) for the same sizes are also shown (as in
fig.\ \ref{fig:fig1}), and are well converged in $N$ by $N\simeq 12$.
\emph{Inset}: Mean ratio $r$ (squares) of consecutive eigenvalue spacings 
\emph{vs} $W/t$ for the same system sizes. Discussion in text.  
}
\end{figure}

These distributions \emph{per se} will be considered in the following sections. But  first (fig.\ \ref{fig:fig4}) we give 
an overview of the resultant geometric mean $\tildeltypo$ of $F(\dtilIo)$, and its evolution with disorder $W/t$ and 
system size $N$. As a relevant comparator, the inset to fig.\ \ref{fig:fig4} shows a measure often used
to distinguish localized from delocalized states;~\cite{Oganesyan+HusePRB2007,Pal+HusePRB2010,ReichmanPRL2015,NandkishoreBhattPRL2015,LuitzAletPRB2015,Serbyn16} viz.\ the ratio of consecutive eigenvalue spacings,
$r_n = \mathrm{min}(\delta_n, \delta_{n-1}) / \mathrm{max}(\delta_n, \delta_{n-1})$ where 
$\delta_n = E_{n+1} - E_n$ (with $E_n$ the ordered eigenvalues).
The mean ratio $r$  \emph{vs} $W/t$ is shown, for states in the immediate vicinity of the band center and $N=12$$-$$16$.
For the Gaussian orthogonal ensemble (GOE) characteristic of extended states $r_{\mathrm{GOE}} = 0.53..$,
while for Poissonian statistics appropriate to the MBL phase, 
$r_{\mathrm{Pois}} =0.38..$.~\cite{Oganesyan+HusePRB2007,AtasPRL2013}  
In the extended regime 
sufficiently below the critical disorder, $r$ increases with increasing $N$ for given $W/t$ (and for low enough disorder has 
effectively reached the GOE limit). In the MBL regime by contrast $r$ shows the reverse trend, decreasing with increasing  
$N$ towards the Poisson limit. In between is a continuous crossover, as expected for finite-size systems. The data are 
clearly consistent with the occurrence of the MBL transition (whose existence is not in doubt~\cite{ImbriePRL2016}),
though to gauge the critical $W/t$ with some confidence requires larger system sizes coupled with a finite-size scaling 
analysis. This has been done~\cite{LuitzAletPRB2015} for  system sizes up to an impressive $N=22$, leading to an
estimate of the critical $W/t \simeq 14.9$ (rather larger than naive inspection of the raw data might suggest).

The main part of fig.\ \ref{fig:fig4} shows corresponding results for the $W/t$- and $N$-dependence of the geometric mean 
$\tildeltypo =\exp(\langle \ln\dtilIo\rangle)$; together with those for the arithmetic mean 
$\langle \dtilIo\rangle$ ($=\int d\dtilIo ~\dtilIo F(\dtilIo)$) discussed in sec.\ \ref{section:propagatorsformal} 
(fig.\ \ref{fig:fig1}). The same qualitative characteristics are seen for $\tildeltypo$ as for $r$ above:  for low enough disorder $\tildeltypo$ increases with increasing system size and ultimately tends to a finite limit (a geometric mean 
cannot exceed its arithmetic counterpart); while for larger $W/t$ in the MBL phase the reverse behavior is seen, and it 
indeed appears likely that $\tildeltypo$ asymptotically vanishes with increasing $N$, as required for localized states. 
In between is once again an expected crossover behavior (which prevents us being credibly quantitative about the critical value of $W/t$). Overall, as for $r$, the data for $\tildeltypo$ are likewise  consistent with the occurrence of the MBL transition. Its behavior is however in marked contrast to that for the arithmetic mean $\langle \dtilIo\rangle$ which, as discussed in sec.\ \ref{section:propagatorsformal}, is well converged in $N$ even by $N=12$; and, being non-zero in both the ergodic and MBL phases, does not discriminate between them. 

We turn now to the numerically determined distributions of $\dtilIono$ and $\dtilIono/\etatil$, beginning with the MBL phase. 


\subsection{MBL regime}
\label{subsection:MBL2}

\begin{figure}
\includegraphics{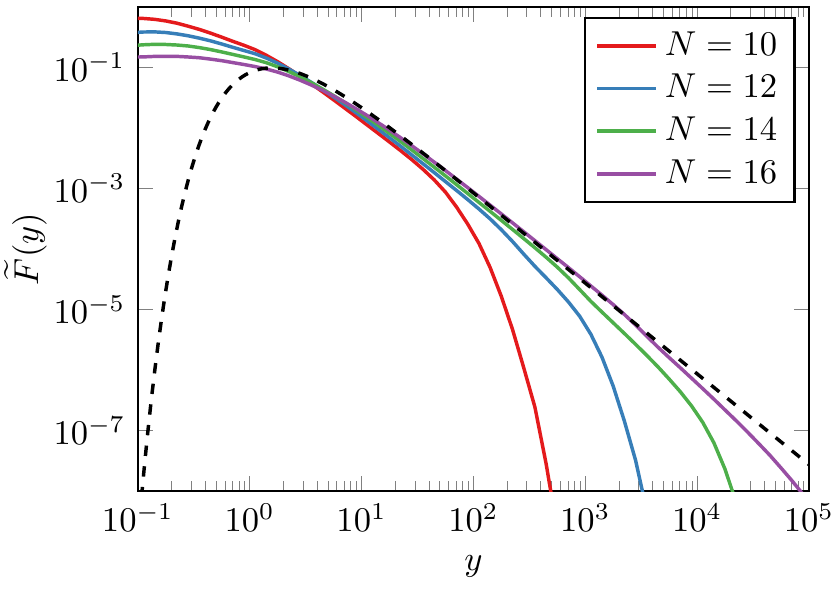}
\caption{\label{fig:fig5} 
Numerical results in the MBL phase for $W/t=20$, showing the distribution $\tilde{F}(y)$ 
\emph{vs} $y$ ($=\tilde{\Delta}_{I}/\etatil$),  for $N=10$ (red), $12$ (blue), $14$ (green) and $16$ (purple). 
The dashed line is  a fit of the $N=16$ results to a (full) L\'evy distribution. 
}
\end{figure}

Fig.\ \ref{fig:fig5} shows the distribution $\tilde{F}(y)$ of $y=\dtilIono/\etatil$, for fixed disorder strength
$W/t=20$ and system sizes $N=10$$-$$16$. For each $N$, a window of power-law tail behavior $\propto y^{-3/2}$ occurs; illustrated e.g.\ by a fit for the $N=16$ data to a full  L\'evy distribution,  which (eq.\ \ref{eq:36}) has 
tails $\propto y^{-3/2}$ (and captures the numerics rather well down to $y\sim 1$). This power-law window extends up to
an $N$-dependent cutoff  $y \sim {\cal{O}}(N_{{\cal{H}}})$ (i.e.\ $\dtilIono =\etatil y \sim {\cal{O}}(1)$ since 
$\etatil \propto 1/N_{{\cal{H}}}$, eq.\ \ref{eq:51}); beyond which the distribution falls of exponentially (itself 
considered in sec.\ \ref{subsubsection:MBL2a}). With increasing $N$ the power-law tails extend over an increasingly large 
$y$-range, which suggests $\tilde{F}(y)\propto y^{-3/2}$ as the leading large-$y$ asymptotic behavior of $\tilde{F}(y)$ 
in the thermodynamic limit $N\rightarrow \infty$, as arises from the mean-field approach of 
secs.\ \ref{section:MFT},\ref{section:ProbDists1}. To investigate this we consider a one-parameter scaling ansatz of form
\begin{equation}
\label{eq:52}
\tilde{F}(y) ~=~ \frac{1}{\alpha} f\Big(\frac{y}{\alpha}\Big),
\end{equation} 
with $f(x=y/\alpha)$ independent of system size (such that all the $N$-dependence resides in $\alpha$);
and which scaling form also arises for a pure L\'evy distribution (see eq.\ \ref{eq:42}).
With  this ansatz the geometric mean $y_{\mathrm{t}}$ follows as $y_{\mathrm{t}} =\alpha  x_{\mathrm{t}}$, where
$x_{\mathrm{t}} = \exp(\int_{0}^{\infty}dx~f(x)\ln x)$ is $N$-independent.  $\alpha$ is thus proportional to 
$y_{\mathrm{t}}$; and we choose $\alpha =y_{\mathrm{t}}$, so  $f(x=y/\alpha)$ is the distribution of 
$x \equiv y/y_{\mathrm{t}}$.

\begin{figure}
\includegraphics{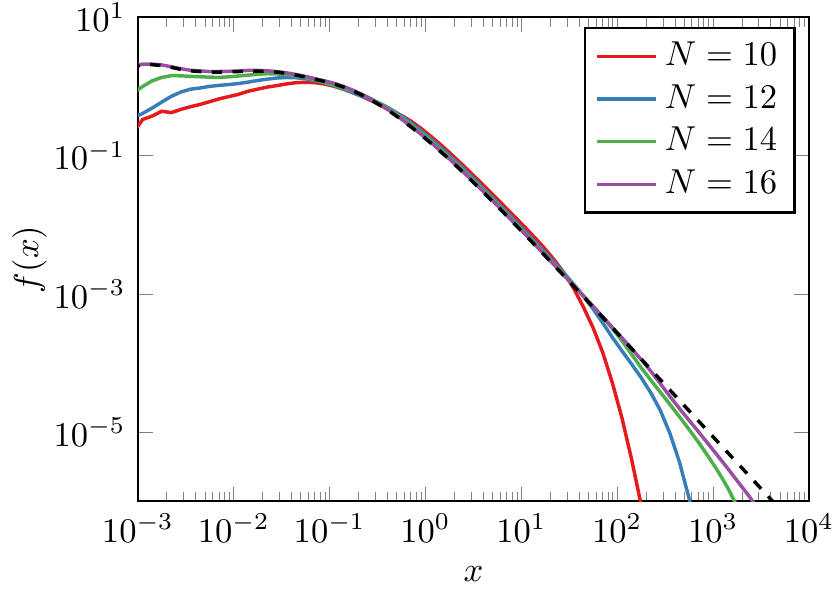}
\caption{\label{fig:fig6} 
The distribution $f(x)$ \emph{vs} $x$ of $x = y/y_{\mathrm{t}}$ for $W/t =20$, and $N=10$ (red), $12$ (blue), $14$ (green) 
and $16$ (purple). The dashed line is the estimated large-$N$ limiting form of $f(x)$.
}
\end{figure}

The resultant $f(x)$ is shown in fig.\ \ref{fig:fig6}, again for $W/t=20$. $f(x)$ now has $x^{-3/2}$ tails that are common 
for each $N$, and which encompass an increasingly large $x$-range with increasing $N$. The limiting distribution can be estimated by extrapolating the $N = 16$ data as an  $x^{-3/2}$ tail (and re-normalizing the distribution with the extrapolated tail in place). This is shown as the dashed line in fig.\ \ref{fig:fig6}. The range of validity of this distribution is not moreover confined to the power-law tail region, but extends down to $x \simeq 10^{-1}$; since for the system sizes shown in fig.\ \ref{fig:fig6} the $f(x)$ distributions are quite well converged in $N$ for $x \gtrsim 10^{-1}$, which comprises the great majority ($\sim 90\%$) of its weight. The dashed line shown is thus expected to capture the bulk of the limiting $f(x)$  (although is not converged in $N$ for $x \lesssim 10^{-1}$).

\begin{figure}
\includegraphics{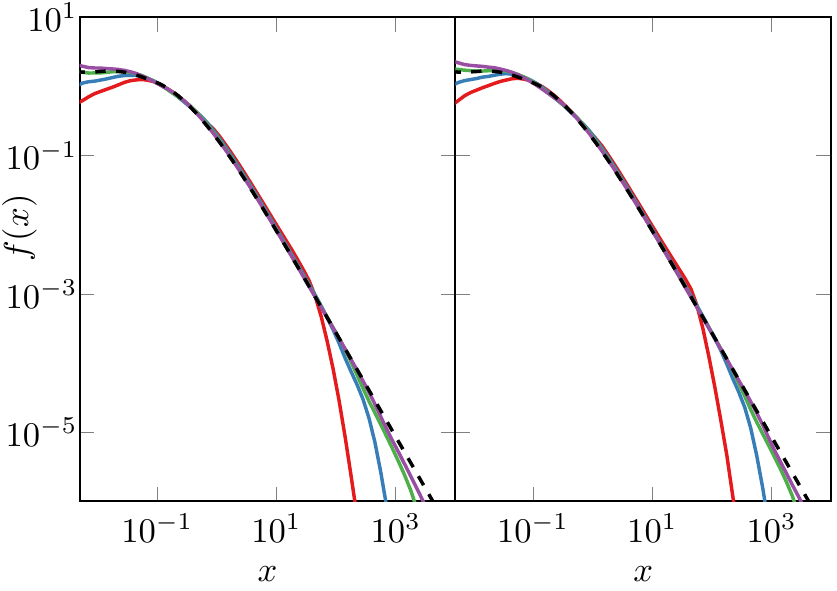}
\caption{\label{fig:fig7} 
$f(x)$ \emph{vs} $x = y/y_{\mathrm{t}}$ shown for $W/t =24$ (left panel), and
$28$ (right panel); again for $N=10$ (red), $12$ (blue), $14$ (green) and $16$ (purple). The dashed line is the 
estimated limiting form of $f(x)$ obtained from the $W/t =20$ results of fig.\ \ref{fig:fig6}.
Results shown are barely distinguishable from those for $W/t=20$ in fig.\ \ref{fig:fig6}. See text for discussion.
}
\end{figure}

The example above refers specifically to $W/t =20$, but the same analysis can obviously be performed for $W/t$ throughout the MBL phase. Results arising are illustrated in fig.\ \ref{fig:fig7}, which in direct parallel to fig.\ \ref{fig:fig6} shows the 
$f(x)$ determined  for $W/t =24$ and $28$; and includes also (dashed line) the estimated limiting $f(x)$ distribution obtained from the $W/t=20$ results of fig.\ \ref{fig:fig6}. As evident from figs.\ \ref{fig:fig7},\ref{fig:fig6}, the $f(x)$ for these different $W/t$ are scarcely distinguishable  (as we have confirmed holds over a wide range of 
$W/t$, down to $W/t \simeq 16$). In particular the limiting $f(x)$ appears to be the same in all cases,  
attesting to its universality as a function characteristic of the \emph{entire} MBL phase (as discussed in 
sec.\ \ref{subsection:MBL1}, eq.\ \ref{eq:42}, in regard to the pure L\'evy distribution arising within 
the mean-field approach).

This universality of $f(x)$ means (see  eq.\ \ref{eq:52}) that the \emph{entire} $y=\dtilIono/\etatil$-dependence of the distribution $\tilde{F}(y)$ is encoded solely in a single quantity -- its geometric mean $\alpha =y_{\mathrm{t}}$, which thus contains all system size and $W/t$ dependence. We have naturally investigated this, and find the following 
qualitative  behavior. For any given system size $N$, increasing disorder $W/t$ and thus moving further into the MBL phase leads to a progressive decrease in $y_{\mathrm{t}}$. This is just as expected physically. 

However on fixing $W/t$ and progressively increasing the system size, $\alpha = y_{\mathrm{t}}$ does not appear to saturate 
(which the considerations of sec.\ \ref{subsection:MBL1} imply it should); but grows progressively with $N$ 
for the system sizes up to $N=16$ that can realistically be studied [$\tildeltypo$ and $\etatil \propto 1/N_{{\cal{H}}}$ 
each decrease with increasing $N$ in the MBL phase, fig.\ \ref{fig:fig4}, but  $y_{\mathrm{t}}=\tildeltypo/\etatil$ itself 
increases over the accessible $N$-range]. The obvious question is why, and a plausible explanation would seem to 
be that the system sizes accessible in practice are not large enough to establish the convergence of  $y_{\mathrm{t}}$.
But it is clearly desirable to have evidence for that. To this end we now consider the case of one-body localization 
(1BL), which in turn provides some further insight into the behavior of the MBL distributions shown 
in figs.\ \ref{fig:fig5}-\ref{fig:fig7}.


\subsubsection{1-body localization and back to MBL}
\label{subsubsection:MBL2a}

It is natural to ask the same questions about 1BL ($N_{e}=1$, $\nu =1/N$), for in that case one can easily consider much larger system sizes $N$ than for MBL.

For $d$$=$$1$, where all states are localized for any disorder, we calculate  the distribution $\tilde{F}(y)$ 
of $y=\dIono/\eta$ ($\equiv \dtilIono/\etatil$) in direct parallel to the procedure above for the MBL case.
With~\cite{FN7} $\eta = [16t^{2}+W^{2}]^{1/2}/N$ ($:=\mu_{E}/N$ with $\mu_{E}$ here simply defined), 
the $\dIono/\eta$ are obtained via eq.\ \ref{eq:50} for states in the immediate vicinity of the band 
center. The scaling form eq.\ \ref{eq:52} holds in this case (as detailed below), and the value of $N$ for 
which the geometric mean $\alpha =y_{\mathrm{t}}$ in practice saturates can be determined.
\begin{figure}
\includegraphics{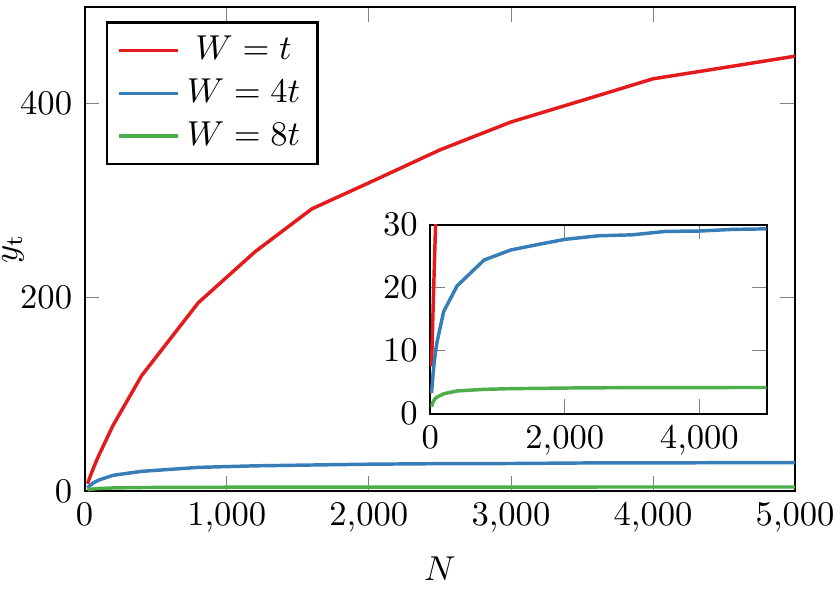}
\caption{\label{fig:fig8} 
For 1BL, geometric means $y_{\mathrm{t}}=\Delta_{\mathrm{t}}/\eta$ \emph{vs}
system size $N$ for $W/t =8$ (green), $4$ (blue) and $1$ (red).
}
\end{figure}
This is illustrated in fig.\ \ref{fig:fig8}, where the resultant $N$-dependence of $y_{\mathrm{t}}$ is shown for 
three different $W/t =1,4, 8$ (as examples of `weak', `moderate' and `strong' disorder). The results are striking.
For $W/t=8$ the geometric mean is converged only for $N \gtrsim 500$. On decreasing disorder the corresponding value 
increases further, to $N \gtrsim 4000$ for $W/t=4$; while for $W/t=1$, $y_{\mathrm{t}}$ converges for $N$ in excess of 
$10^{4}$. In particular, for what in this context are comparatively small $N \lesssim 100$, the $y_{\mathrm{t}}$ are sharply increasing with $N$ for all three disorder strengths. The geometric mean $y_{\mathrm{t}}$ (and hence the entire distribution 
$\tilde{F}(y)$) does then converge with increasing $N$, but the $N$-values for which this is reached in practice
are  far in excess of anything that can be handled in the MBL problem  ($N \lesssim 20$ or so); which seems consistent with the view that the system sizes that can be studied for MBL are not sufficient to reach the converged value of 
$y_{\mathrm{t}}$, especially for disorder strengths close to the transition.

The full $\tilde{F}(y)$ can in fact be obtained for the one-body $d$$=$$1$ problem. In this case the probability 
distribution of the local density of states (LDoS) is known exactly in the thermodynamic limit;~\cite{Altshuler+Prigodin1989}
specifically the distribution $W(\rho_{1})$ of $\rho_{1}=\rho/\langle\rho\rangle$, with $\rho$ the LDoS (at the given energy) and $\langle \rho\rangle$ its disorder average. This is given~\cite{Altshuler+Prigodin1989} for arbitrary non-zero $\eta$ 
(i.e.\ with Lorentzian broadening of width $\eta$ for associated $\delta$-functions in the LDoS, as employed in our numerics), and is an inverse Gaussian distribution.~\cite{Jorgensenbook} $\Delta_{I}$ is proportional to the LDoS, $\Delta_{I} =c\rho_{1}$(with the constant $c$ readily shown to be of order $t$), so the distribution $\tilde{F}(y)$ of $y=\Delta_{I}/\eta$ follows as
$\tilde{F}(y)= \tfrac{\eta}{c} W(\tfrac{\eta y}{c})$ and is 
\begin{equation}
\label{eq:53}
\tilde{F}(y)~=~ \sqrt{\frac{\xi}{\pi}}~y^{-3/2}
~\exp\Big[
-\frac{\xi~(\tfrac{\eta}{c}y-1)^{2}}{y} \Big]
\end{equation}
(specifically with $\xi =4\tau c$ in weak disorder, and $\tau$ the mean free time~\cite{Altshuler+Prigodin1989}).
Since eq.\ \ref{eq:53} holds in the thermodynamic limit $N\rightarrow \infty$, the limit of $\eta \rightarrow 0+$ may 
be taken with impunity, to give the desired limiting result for the full distribution $\tilde{F}(y)$.  
This is obviously
\begin{equation}
\label{eq:54}
\tilde{F}(y)~=~ \sqrt{\frac{\xi}{\pi}}~y^{-3/2}
~\exp\Big[
-\frac{\xi}{y} \Big]
~~~~:~ \eta \rightarrow 0+ ,
\end{equation}
which is precisely a L\'evy distribution again.
\begin{figure}
\includegraphics{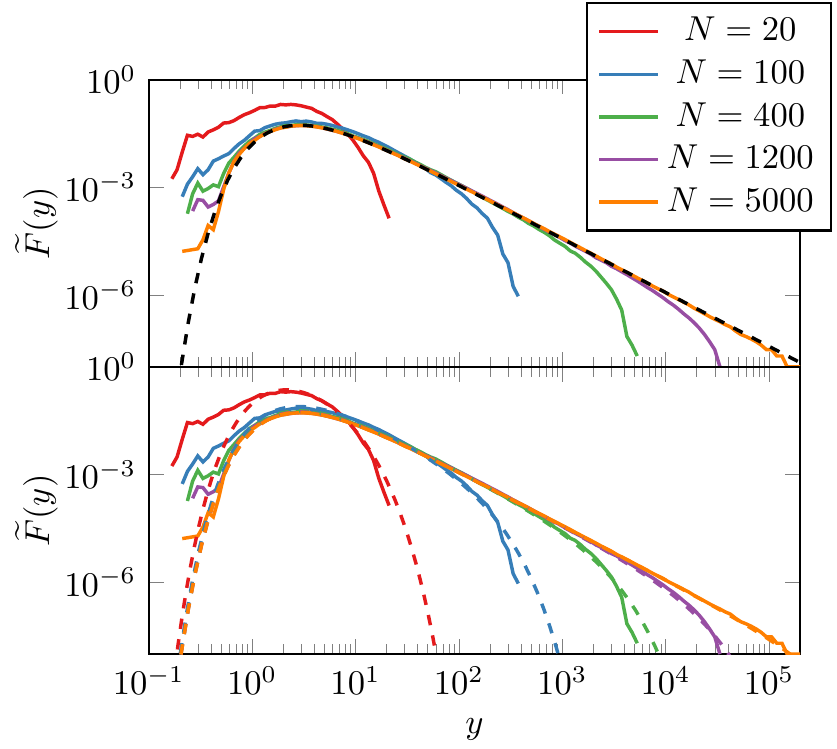}
\caption{\label{fig:fig9}  
\emph{Upper panel}: For 1BL with $W/t=4$, numerical distributions $\tilde{F}(y)$ 
\emph{vs} $y$ ($=\Delta_{I}/\eta$), for system sizes $N$ ranging from $N=20$ to $5000$ as indicated.
The dashed line is  a fit of the $N=5000$ results to a full L\'evy distribution eq.\ \ref{eq:54}.
\emph{Lower panel}: same data, showing fits (dashed lines) to the inverse Gaussian distribution eq.\ \ref{eq:53}
appropriate to non-zero $\eta$, which clearly capture the departures from the leading L\'evy tails.
The same fit parameters ($\xi$ and $c$) were used throughout, the only $N$-dependence arising in $\eta \propto 1/N$.
}
\end{figure}

Finite-$N$ numerics indeed show clear recovery  of this behavior. Fig.\ \ref{fig:fig9} (upper panel)
shows $\tilde{F}(y)$ \emph{vs} $y$  for a wide range of system sizes from $N=20$ to $5000$ (cf.\ fig.\ \ref{fig:fig5} 
for the corresponding MBL distributions), with the L\'evy distribution very well captured by $N \approx 5000$ across essentially the full six-decade $y$-range shown (save for the very lowest, which amounts to a negligible fraction of the distribution). As seen from the figure, the tails of the finite-$N$ distributions depart from L\'evy form 
($\propto y^{-3/2}$) for $y \sim {\cal{O}}(N)$ (i.e.\  $\Delta_{I} =\eta y \sim{\cal{O}}(1)$), in direct parallel to the 
MBL numerics (fig.\ \ref{fig:fig5}); beyond which the L\'evy tails are exponentially damped.
The latter $N$-dependent crossover should be captured by the inverse Gaussian form eq.\ \ref{eq:53}, since it pertains 
to finite-$\eta$. That it clearly does is shown in the lower panel of fig.\ \ref{fig:fig9};  with the value of $y$ at which the crossover begins moving to progressively larger values with increasing $N$, such that the L\'evy distribution is recovered in the thermodynamic limit.

With the above in mind, it is natural to ask whether the finite-$N$ MBL numerics are similarly captured by the 
form eq.\ \ref{eq:53}.
\begin{figure}
\includegraphics{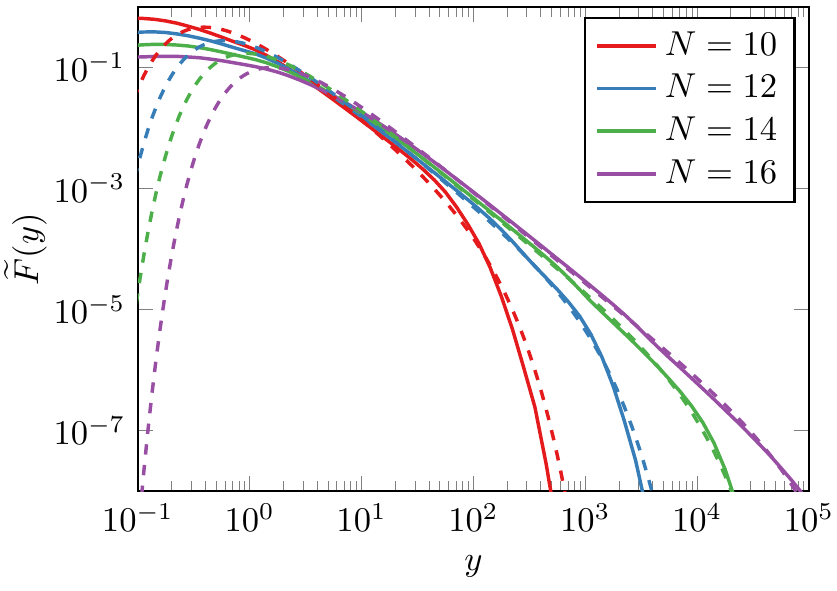}
\caption{\label{fig:fig10}   
Numerical MBL distributions $\tilde{F}(y)$ \emph{vs} $y$ ($=\tilde{\Delta}_{I}/\tilde{\eta}$) for $W/t =20$, and 
system sizes $N =10-16$ as indicated. Dashed lines show one-parameter fits to the inverse Gaussian distribution 
eq.\ \ref{eq:53}, discussed in text.
}
\end{figure}
This indeed appears to be the case, as shown in fig.\ \ref{fig:fig10} where the results of fig.\  \ref{fig:fig5} are 
again fit to an inverse Gaussian distribution (and which, though shown here for $W/t =20$, is found to be representative
of the MBL phase). While comparison to eq.\ \ref{eq:53} is at first sight a two-parameter fit (in $\xi$, and $c$ or equivalently $\tilde{c}=c/\mu_{E}$), note that if eq.\ \ref{eq:53} captures the data then the value of $\tilde{c}$ is 
in practice known, because the mean value of $y =\dIono/\eta =\dtilIono/\etatil$ for the inverse Gaussian eq.\ \ref{eq:53} 
is $\langle y\rangle =c/\eta =\tilde{c}/\etatil$, whence $\tilde{c}=\langle \dtilIono \rangle$.
As shown in sec.\ \ref{section:propagatorsformal} (fig.\ \ref{fig:fig1}), the arithmetic mean 
$\langle \dtilIono \rangle$ is well described by the Fermi golden rule result
 eq.\ \ref{eq:19} in the MBL regime; from which  (together with eqs.\ \ref{eq:5},\ref{eq:9} for $\overline{Z}$, $\mu_{E}$) 
\begin{equation}
\label{eq:55}
\tilde{c}~=~ \sqrt{\tfrac{\pi}{2}}~ 
\Big[1 +\tfrac{1}{2}\nu(1-\nu)\big(\tfrac{V}{t}\big)^{2} +\tfrac{1}{24}\big(\tfrac{W}{t}\big)^{2}\Big]^{-1}
\end{equation}
for the band center $\wtil =0$. In the results shown in fig.\ \ref{fig:fig10} we have used this $N$-independent 
$\tilde{c}$ (adding that it is found to be equally satisfactory for  $W/t \gtrsim 16$).

Despite the inevitably restricted range of $N$ available, fig.\ \ref{fig:fig10} shows that the same essential characteristics are  seen in the MBL data as for the 1BL case; giving further support to the result arising from the mean-field approach of secs.\ \ref{section:MFT},\ref{section:ProbDists1} that the limiting $\tilde{F}(y)$ in the MBL phase is a L\'evy distribution, at least for the bulk of $\tilde{F}(y)$ (the natural caveat  again being the lowest $y$-values, convergence of which we believe lies well beyond the accessible $N$-range).


\subsection{Delocalized regime}
\label{subsection:delocnumerics}

\begin{figure}
\includegraphics{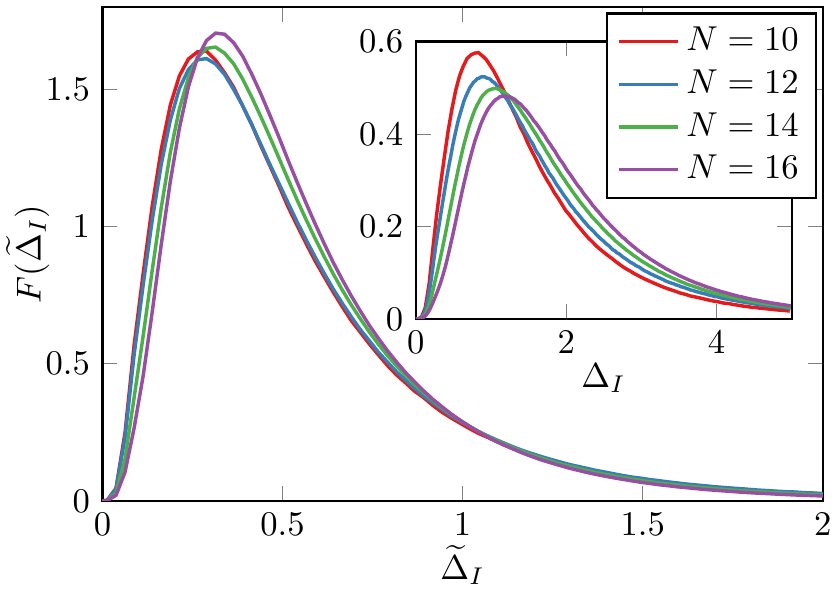}
\caption{\label{fig:fig11} 
Distributions $F(\tilde{\Delta}_{I})$ \emph{vs} $\tilde{\Delta}_{I} =\Delta_{I}/\mu_{E}$ for $W/t =1$ and
system sizes $N=10$ (red), $12$ (blue), $14$ (green), $16$ (purple). \emph{Inset}: corresponding distributions for 
$\Delta_{I}$. See text for discussion. 
}
\end{figure}

In an extended regime it is as we have emphasized $\dtilIono = \dIono/\mu_{E}$ (with $\mu_{E}\propto \sqrt{N}$) which is
perforce  non-zero and finite in the thermodynamic limit, rather than $\dIono$ itself. For the mean values 
$\langle \dtilIono\rangle$ and $\langle \dIono\rangle$, fig.\ \ref{fig:fig1} shows the former to be well converged with system size for the modest $N$'s amenable to calculation; so in consequence $\langle \dIono\rangle$ increases $\propto \sqrt{N}$ 
(fig.\ \ref{fig:fig1} inset) for given $W/t$. An arithmetic mean is of course merely one reflection of a  distribution. 
The evolution with system size and disorder of the geometric mean, $\tildeltypo$, of $\dtilIono$  has been considered 
in fig.\ \ref{fig:fig4}; as found there, it is not yet converged within the available $N$-range even in the extended regime.
This behavior is naturally reflected in the full distribution $F(\dtilIono)$, as illustrated in fig.\ \ref{fig:fig11} 
for $W/t=1$, well inside the extended phase. $F(\dtilIono)$ appears quite well converged in $N$ for $\dtilIono \gtrsim 1$, 
but some $N$-dependence clearly remains for lower values of $\dtilIono$ (the distribution of $\dIono$ by contrast 
shows clear $N$-dependence for all $\dIono$, fig.\ \ref{fig:fig11} inset, and as discussed above cannot converge as $N\rightarrow \infty$).

\begin{figure}
\includegraphics{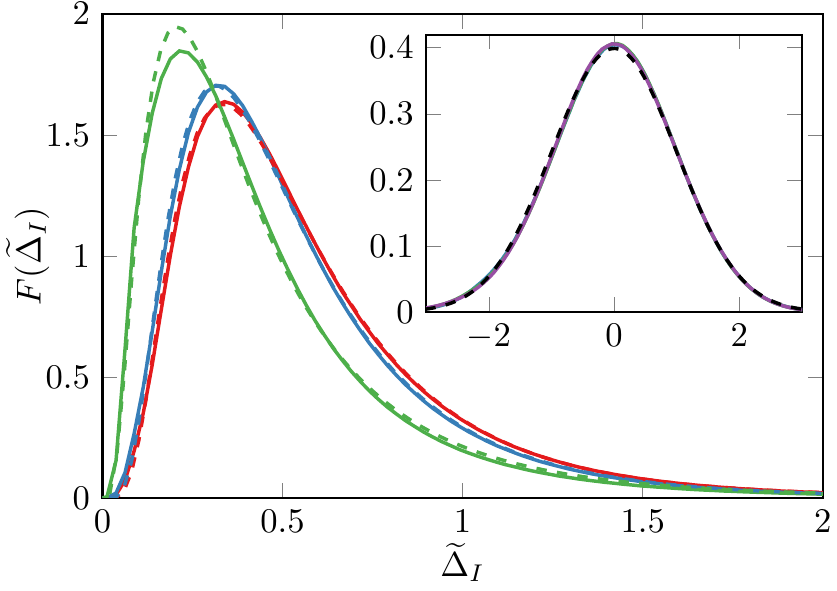}
\caption{\label{fig:fig12} 
Distributions $F(\tilde{\Delta}_{I})$ \emph{vs} $\tilde{\Delta}_{I}$
for $W/t =1$ (red), $2$ (blue) and $4$ (green), compared to fits (dashed lines) to the log-normal form 
eq.\ \ref{eq:56}. Data for $N=16$.
\emph{Inset}: For $W/t=2$, convergence with increasing $N$ of the distribution  ${\cal{F}}(x)$ of 
$x = [\ln(\dtilIono/\tildeltypo)]/\sigma$, to a standard normal distribution (dashed line), shown for 
$N=12$ (blue), $14$ (green),  $16$ (purple). Results for different $N$ are practically indistinguishable.  
}
\end{figure}

$F(\dtilIono)$ sufficiently deep in the extended phase is in fact found to be of log-normal (LN) form,
\begin{equation}
\label{eq:56}
F(\dtilIo)~=~\frac{1}{\sqrt{2\pi}\sigma} \frac{1}{\dtilIo} \exp\Big(
-\frac{[\ln(\dtilIono/\tildeltypo)]^{2}}{2\sigma^{2}}\Big)
\end{equation}
(with $\sigma^{2}$ the variance of $\ln \dtilIono$). This is  illustated in fig.\ \ref{fig:fig12} where comparison of the numerical  $F(\dtilIono)$ to eq.\ \ref{eq:56} is shown for $W/t =1,2,4$ (with $N=16$),  the data being well captured by the 
LN distribution for essentially all $\dtilIono$ (including deep in the tails). 

Further, given eq.\ \ref{eq:56} and  defining $x = [\ln(\dtilIono/\tildeltypo)]/\sigma$, its probability density
${\cal{F}}(x)$ should then be of standard normal form ${\cal{F}}(x) =[2\pi]^{-1/2}\exp(-x^{2}/2)$. That the numerics 
indeed converge to this form with increasing system size $N$ (and do so rapidly) is shown in the inset to 
fig.\ \ref{fig:fig12} for $W/t =2$. Data for $W/t =1$ likewise scale cleanly onto this common form, although by $W/t =4$ slight departures from it arise, reflecting the further evolution of the distribution with increasing disorder in the 
extended phase (see below). The occurrence of an $F(\dtilIono)$ of LN form for weak disorder seems physically intuitive; 
and although our simple mean-field approach yields a different form (eq.\ \ref{eq:43}), the latter does capture
qualitatively the long-tailed character of $F(\dtilIono)$, with a mode which diminishes with increasing disorder.

\begin{figure}
%
\includegraphics{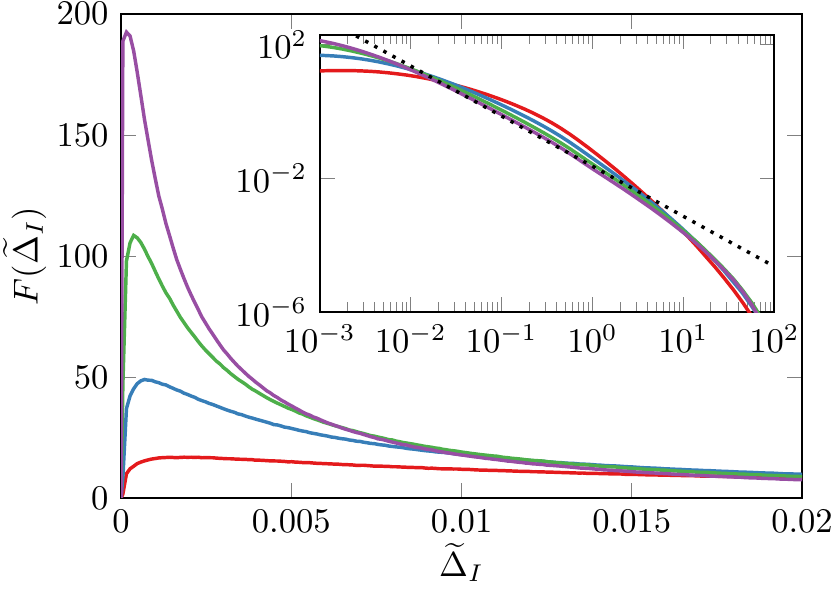}
\caption{\label{fig:fig13} 
Distributions $F(\tilde{\Delta}_{I})$ \emph{vs} $\tilde{\Delta}_{I}$ for $W/t =8$ (red), $10$ (blue), $12$ (green) 
and $14$ (purple), with $N=16$. 
\emph{Inset}: same data on log-log scale, with the dotted line showing an emergent 
$\tilde{\Delta}_{I}^{-3/2}$ component to the tail of $F(\tilde{\Delta}_{I})$.
}
\end{figure}

Consider now the evolution of $F(\dtilIono)$ with further increasing disorder towards $W/t =12$ (the transition occurring 
between
$W/t\sim 12$$-$$16$,~\cite{Pal+HusePRB2010,ReichmanPRB2010,DeLucaScardicchioEPL2013,LuitzAletPRB2015,BeraFHMBardarsonPRL2015}
probably closer to the upper part of it~\cite{LuitzAletPRB2015}). Fig.\ \ref{fig:fig13} shows the numerical $F(\dtilIono)$ 
for $W/t =8,10,12,14$. The distributions remain unimodal and long-tailed, and with increasing disorder naturally become increasingly strongly peaked at progressively lower $\dtilIono$. The inset shows the same results on a log-scale, with emphasis as such on the behavior of the tails in $F(\dtilIono)$. From this, an intermediate regime of L\'evy-like 
power-law behavior $\propto {\dtilIono}^{-3/2}$ (dotted line) is seen to emerge  with increasing disorder on approaching 
the transition from the extended side (as arises in the mean-field approach, sec.\ \ref {subsection:MBD1}); 
the range of which grows in extent with increasing disorder, before ultimately crossing over to a slower decay.

\begin{figure}
\includegraphics{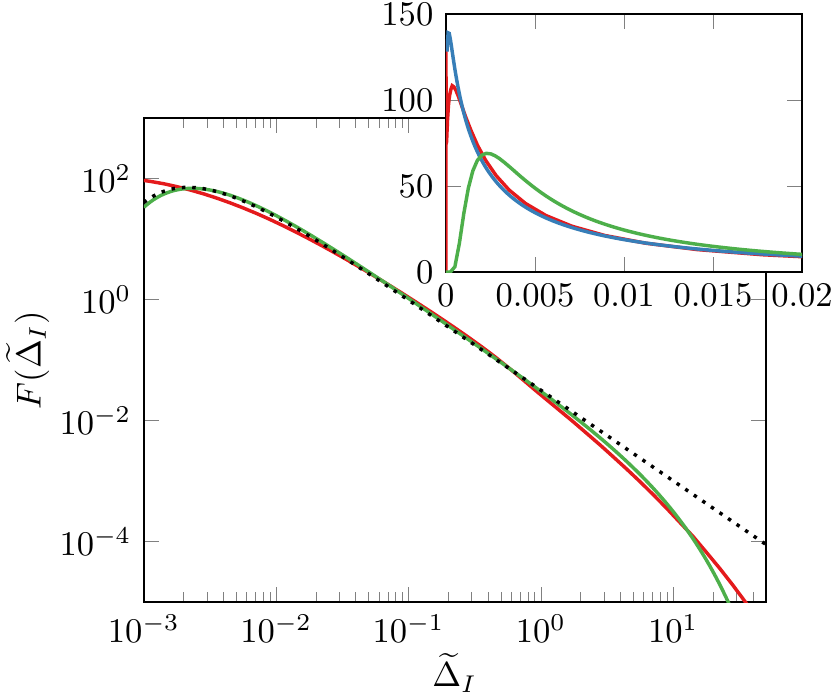}
\caption{\label{fig:fig14} 
For $W/t =12$, numerical $F(\tilde{\Delta}_{I})$ \emph{vs} $\tilde{\Delta}_{I}$ (red line) for $N=16$, on
a log-log scale. Fits to both an inverse Gaussian (green line), and a L\'evy distribution (dotted line) are shown.
\emph{Inset}: same data on a linear scale, showing also the fit to a LN distribution (blue line). 
Discussion in text.
}
\end{figure}

To pursue this further, fig.\ \ref{fig:fig14} shows $F(\dtilIono)$ for $W/t =12$, on a log-scale in the main figure. 
The natural distribution  to compare to the numerics is an inverse Gaussian for $F(\dtilIono)$ itself, i.e.\
\begin{equation}
\label{eq:57}
F(\dtilIo)~=~ \sqrt{\frac{\xi^{\prime}}{\pi}}~\tilde{\Delta}_{I}^{-3/2}
~\exp\left[
-\frac{\xi^{\prime}~(\tfrac{1}{\tilde{c}}\dtilIo-1)^{2}}{\dtilIo} \right]
\end{equation}
with $\tilde{c}=\langle \dtilIono\rangle$ again fixed by eq.\ \ref{eq:55} (for the same reasons as 
given there); and which generates characteristic  L\'evy tails $\propto {\dtilIono}^{-3/2}$, before becoming exponentially damped. The inverse Gaussian is compared to numerical results in fig.\ \ref{fig:fig14} and seen in particular to capture well 
the tails of the distribution, including the departure from the clearly visible power-law tail of the corresponding pure 
L\'evy distribution (eq.\ \ref{eq:57} with  $\tilde{c}\rightarrow \infty$), also shown (dotted line).

While the tail of $F(\dtilIono)$ appears to be described by inverse Gaussian/L\'evy, the bulk of the distribution resides 
at lower values  of $\dtilIono$, as shown in the inset to fig.\ \ref{fig:fig14} on a linear scale; and which, as seen,
is rather well captured by a LN distribution (blue line). The latter seems natural given the clear dominance of the LN form for lower disorder values (fig.\ \ref{fig:fig12}); although in the tail region of $F(\dtilIono)$ the LN distribution decays more rapidly than inverse Gaussian/L\'evy, such that the overall $\dtilIono$-dependence of the distribution thus appears to involve a crossover between LN and inverse Gaussian/L\'evy behaviors.


\section{Concluding remarks}
\label{section:summary}

In this paper we have considered many-body localization in the widely studied model of spinless 
fermions,~\cite{Oganesyan+HusePRB2007} on a lattice of $N \equiv L^{d}$ real-space sites.
Here the problem has been studied directly from the perspective of the underlying Fock-space lattice of many-body 
states; with FS sites coupled by single-fermion hoppings, such that a typical coordination number for a FS site 
is extensive in $N$. Such a perspective requires explanation of how an incipient divergence in the FS coordination 
number is effectively mitigated, such that an MBL phase even exists in the thermodynamic limit $N\rightarrow \infty$.
Central to that end, innocuous though it may seem at first sight, has been the rescaling of energy in terms of the 
standard deviation of the eigenvalue spectrum $\mu_{E} \propto \sqrt{N}$.

Exploiting the mapping to a tight-binding model in FS, we have focused specifically on local~\cite{PWA1958} FS 
propagators, via their associated (Feenberg) self-energies. The imaginary part  $\dtilIono$ of the rescaled 
self-energy is of primary importance, being finite in the former case with probability unity over an ensemble of 
disorder realizations, and vanishingly small for an MBL phase. We have thus focused on appropriate probability 
distributions for it; noting in particular that the geometric mean of the distribution can act as a suitable order 
parameter for the transition to the MBL phase.

A self-consistent, probabilistic mean-field approach was first developed. Despite its simplicity and 
natural limitations, this offers physical insight and yields quite a rich description of the problem; including  recovery 
of a stable MBL phase and an MBL transition in the thermodynamic limit, and the notable prediction that the appropriate 
self-energy distribution throughout the MBL phase should be characterized `universally' by a long-tailed L\'evy distribution.
Informed by the mean-field picture, but free from its assumptions, detailed numerical results from exact diagonalization in 
$1d$ have also been presented. As shown, these provide further detailed information about the underlying probability distributions in both phases, as well as broad support for the substantive predictions arising at mean-field level.


\begin{acknowledgments}
Many helpful discussions with John Chalker, H R Krishnamurthy, Sthitadhi Roy  and Peter Wolynes are gratefully 
acknowledged. We 
also 
thank the EPSRC for support, under grant EP/L015722/1 for the TMCS Centre for Doctoral Training, 
and grant EP/N01930X/1. The work is compliant with EPSRC Open Data requirements.
\end{acknowledgments}


\appendix

\section{}
\label{section:App1}

From the definition eq.\ \ref{eq:14} of the Feenberg self-energy, $S_{I}(\w) =\w +i\eta -\eIo -1/G_{I}(\w)$, 
with the local FS propagator $G_{I}(\w)$ given in terms of the squared eigenfunction amplitudes 
$|A_{mI}|^{2} = |\langle I|\Psi_{m}\rangle |^{2}$  (eq.\ \ref{eq:4}) by 
$G_{I}(\w) = \sum_{m} |A_{mI}|^{2}/(\w +i\eta -E_{m})$. From this, the imaginary part of the self-energy 
$\Delta_{I}(\w) =-\mathrm{Im}S_{I}(\w)$ is 
\begin{equation}
\label{eq:A1}
\Delta_{I}^{\pd}(\w)~=~\frac{\pi D_{I}(\w)}{ [G_{I}^{R}(\w)]^{2} +[\pi D_{I}(\w)]^{2}}~-~\eta
\end{equation}
with the real/imaginary parts of $G_{I}(\w) = G_{I}^{R}(\w) -i\pi D_{I}(\w)$ given by
\begin{equation}
\label{eq:A2}
\begin{split}
\pi D_{I}(\w)~=&~\sum_{m}  \frac{\eta ~|A_{mI}|^{2}}{(\w -E_{m})^{2} +\eta^{2}}
\\
G_{I}^{R}(\w) ~=&~\sum_{m}  \frac{|A_{mI}|^{2}}{(\w -E_{m})^{2} +\eta^{2}}(\w -E_{m}) .
\end{split} 
\end{equation}
Eqs.\ \ref{eq:A1},\ref{eq:A2} are used directly in sec.\ \ref{section:EDresults} to obtain 
the relevant self-energy distributions by exact diagonalization. 

As mentioned in sec.\ \ref{section:propagatorsformal}, our aim here is to sketch some simple 
arguments suggesting that, whether MBL or one-body localization (1BL) is considered, $\Delta_{I}(\w)$ 
is non-zero for extended states (and proportional to $\mu_{E}$ in the many-body case),
while by contrast $\Delta_{I}(\w) \propto \eta$ is vanishingly small for localized states. 
This behavior is also supported by the numerics of sec.\ \ref{section:EDresults}. In the following, the 
site label $I$ refers as usual to the FS site $|I\rangle \equiv|\{\n{i}\}\rangle$ in the MBL case, 
while for 1BL it implicitly refers to a single real-space site $i$ 
($|I\rangle \equiv |i\rangle = c_{i}^{\dagger}|\mathrm{vac}\rangle$).

\emph{Extended states}.  
To make the main points here, we consider the limit of extended states for weak disorder $W/t \ll 1$, on 
the standard assumption made in this regime that $|A_{mI}|^{2} \sim 1/\nh$ (for 1BL, $\nh \equiv {^{N}}C_{1} =N$).
With this, eq.\ \ref{eq:A2} for $D_{I}(\w)$ gives
\begin{equation}
\label{eq:A3}
\begin{split}
D_{I}(\w) ~=&~ \frac{1}{\nh} \sum_{m} \frac{\eta/\pi}{(\w -E_{m})^{2} +\eta^{2}}
\\
\equiv &~ \frac{1}{\nh} \sum_{m} \delta(\w -E_{m}) ~=~ D(\w),
\end{split}
\end{equation}
with $D(\w)$ the total density of states/eigenvalue spectrum. That $D_{I}(\w) \equiv D(\w)$ is physically clear, 
since $|A_{mI}|^{2} \sim 1/\nh$ is tantamount to treating all sites $I$ as equivalent in this regime.
From eq.\ \ref{eq:A1},  
 \begin{equation}
\label{eq:A4}
\Delta_{I}^{\pd}(\w)~=~\frac{\pi D(\w)}{ [G^{R}(\w)]^{2} +[\pi D(\w)]^{2}}
\end{equation}
with $G^{R}(\w) \equiv G_{I}^{R}(\w)$ the Hilbert transform of $D(\w)$ (and where $\eta \equiv 0+$ in eq.\ \ref{eq:A1} can clearly be neglected). 

For 1BL, $D(\w)$ is finite throughout the band of single-particle states, whence so too is $\Delta_{I}(\w)$. 
For the MBL case $D(\w)$ is the Gaussian eq.\ \ref{eq:10}, with standard deviation $\mu_{E}$.
To be concrete, consider the band center $\w =\overline{E}$, for which $G_{I}^{R}(\w =\overline{E}) =0$ by 
symmetry. With this, eq.\ \ref{eq:A4} gives $\Delta_{I}(\w =\overline{E}) = [\pi D(\overline{E})]^{-1}$, 
whence  (from eq.\ \ref{eq:10}) 
 \begin{equation}
\label{eq:A5}
\Delta_{I}^{\pd}(\w)~\propto ~\mu_{E}^{\pd}.
\end{equation}
This result is not confined to the band center; the Hilbert transform of
$D(\w)$ can be shown to be $G^{R}(\w) = \mu_{E}^{-1} \times \sqrt{2}F_{d}(\sqrt{2} \wtil)$ 
with $F_{d}$ the Dawson function, which from eq.\ \ref{eq:A4} guarantees eq.\ \ref{eq:A5} for 
arbitrary $\wtil = (\w -\overline{E})/\mu_{E}$.

As above, $\Delta_{I}(\w)$ is non-zero for both 1BL and MBL. In the former case it is finite, as above. 
For the many-body case (eq.\ \ref{eq:A5}), since $\mu_{E} \propto \sqrt{N}$, so too is  $\Delta_{I}(\w)$.
Hence $\tilde{\Delta}_{I}(\wtil) =\Delta_{I}(\w)/\mu_{E}$ remains finite in the thermodynamic limit
$N\rightarrow\infty$. It is thus $\tilde{\Delta}_{I}\propto\Delta_{I}/\sqrt{N}$ on which one must focus 
(this argument being complementary to that given in sec.\ \ref{section:propagatorsformal}).
This scaling behavior is also corroborated by the numerical results of fig.\ \ref{fig:fig1}, where the mean 
$\tilde{\Delta}_{I}$ is shown (and is very well converged with system size $N$ by $N \simeq 12$).

\emph{Localized states}. 
To be specific here, consider $\w =E_{n}$ for some particular localized state $|\Psi_{n}\rangle$ with 
eigenvalue $E_{n}$.  From eq.\ \ref{eq:A2}, 
\begin{equation}
\label{eq:A6}
\begin{split}
\pi D_{I}(E_{n})~=~&\frac{|A_{nI}|^{2}}{\eta} 
+\eta\sum_{m (\neq n)}  \frac{|A_{mI}|^{2}}{(E_{n} -E_{m})^{2} +\eta^{2}}
\\
G_{I}^{R}(E_{n}) ~=~&\sum_{m (\neq n)}  \frac{|A_{mI}|^{2}}{(E_{n} -E_{m})^{2} +\eta^{2}}(E_{n} -E_{m}) .
\end{split} 
\end{equation}
\comment{
\begin{subequations}
\label{eq:A6}
\begin{align}
\pi D_{I}(E_{n}) = &\frac{|A_{nI}|^{2}}{\eta} 
+\eta\sum_{m (\neq n)}  \frac{|A_{mI}|^{2}}{(E_{n} -E_{m})^{2} +\eta^{2}}
\\
G_{I}^{R}(E_{n}) =&\sum_{m (\neq n)}  \frac{|A_{mI}|^{2}}{(E_{n} -E_{m})^{2} +\eta^{2}}(E_{n} -E_{m}) 
\end{align} 
\end{subequations}
}
There are two potential categories of sites $I$ to be considered: (a) those for which 
$|A_{nI}|^{2}$ vanishes in the thermodynamic limit, and (b) those for which $|A_{nI}|^{2}$ 
remains non-zero in this limit. For the MBL case only the former category 
arises~\cite{LuitzAletPRB2015,DeLucaScardicchioEPL2013} (as discussed in sec.\ \ref{section:model}); 
since an MBL state has support on an exponentially large number $\sim \nh^{\alpha}$ ($\alpha <1$) of FS 
sites, with typical $|A_{mI}|^{2}$s for such sites of order $\nh^{-\alpha}$, vanishing in the thermodynamic limit.
For 1BL by contrast, both categories above are relevant, as here a finite number of real-space sites have finite
$|A_{mI}|^{2}$s in the thermodynamic limit.

For category (a), the $|A_{nI}|^{2}/\eta$ term in eq.\ \ref{eq:A6} vanishes, and we assume the remaining 
contribution to $\pi D_{I}(E_{n})$ is typically of order $\eta$ while $G_{I}^{R}(E_{n})$ is typically 
finite. From eq.\ \ref{eq:A1}, $\Delta_{I}(E_{n}) \sim [\pi D_{I}/(G_{I}^{R})^{2} -\eta]$ is then 
$\propto \eta$. For category (b) by contrast, where $|A_{nI}|^{2}$ remains non-zero in the thermodynamic limit, the 
$|A_{nI}|^{2}/\eta$ term completely dominates in eq.\ \ref{eq:A6}, with $\pi D_{I}(E_{n}) \sim |A_{nI}|^{2}/\eta$,
while $G_{I}^{R}(E_{n})$ is again typically finite. From eq.\ \ref{eq:A1}, $\Delta_{I}(E_{n}) \sim [1/\pi D_{I} -\eta]$,
which again is $\propto \eta$. Overall, $\Delta_{I} \propto \eta$ then arises for both categories (a) and (b).



\bibliography{paper}

\begin{thebibliography}{89}%
\makeatletter
\providecommand \@ifxundefined [1]{%
 \@ifx{#1\undefined}
}%
\providecommand \@ifnum [1]{%
 \ifnum #1\expandafter \@firstoftwo
 \else \expandafter \@secondoftwo
 \fi
}%
\providecommand \@ifx [1]{%
 \ifx #1\expandafter \@firstoftwo
 \else \expandafter \@secondoftwo
 \fi
}%
\providecommand \natexlab [1]{#1}%
\providecommand \enquote  [1]{``#1''}%
\providecommand \bibnamefont  [1]{#1}%
\providecommand \bibfnamefont [1]{#1}%
\providecommand \citenamefont [1]{#1}%
\providecommand \href@noop [0]{\@secondoftwo}%
\providecommand \href [0]{\begingroup \@sanitize@url \@href}%
\providecommand \@href[1]{\@@startlink{#1}\@@href}%
\providecommand \@@href[1]{\endgroup#1\@@endlink}%
\providecommand \@sanitize@url [0]{\catcode `\\12\catcode `\$12\catcode
  `\&12\catcode `\#12\catcode `\^12\catcode `\_12\catcode `\%12\relax}%
\providecommand \@@startlink[1]{}%
\providecommand \@@endlink[0]{}%
\providecommand \url  [0]{\begingroup\@sanitize@url \@url }%
\providecommand \@url [1]{\endgroup\@href {#1}{\urlprefix }}%
\providecommand \urlprefix  [0]{URL }%
\providecommand \Eprint [0]{\href }%
\providecommand \doibase [0]{http://dx.doi.org/}%
\providecommand \selectlanguage [0]{\@gobble}%
\providecommand \bibinfo  [0]{\@secondoftwo}%
\providecommand \bibfield  [0]{\@secondoftwo}%
\providecommand \translation [1]{[#1]}%
\providecommand \BibitemOpen [0]{}%
\providecommand \bibitemStop [0]{}%
\providecommand \bibitemNoStop [0]{.\EOS\space}%
\providecommand \EOS [0]{\spacefactor3000\relax}%
\providecommand \BibitemShut  [1]{\csname bibitem#1\endcsname}%
\let\auto@bib@innerbib\@empty
\bibitem [{\citenamefont {Anderson}(1958)}]{PWA1958}%
  \BibitemOpen
  \bibfield  {author} {\bibinfo {author} {\bibfnamefont {P.~W.}\ \bibnamefont
  {Anderson}},\ }\href@noop {} {\bibfield  {journal} {\bibinfo  {journal}
  {Phys. Rev.}\ }\textbf {\bibinfo {volume} {109}},\ \bibinfo {pages} {1492}
  (\bibinfo {year} {1958})}\BibitemShut {NoStop}%
\bibitem [{\citenamefont {Mott}(1990)}]{MottMIT}%
  \BibitemOpen
  \bibfield  {author} {\bibinfo {author} {\bibfnamefont {N.~F.}\ \bibnamefont
  {Mott}},\ }\href@noop {} {\emph {\bibinfo {title} {Metal-Insulator
  Transitions}}}\ (\bibinfo  {publisher} {Taylor and Francis},\ \bibinfo
  {address} {London},\ \bibinfo {year} {1990})\BibitemShut {NoStop}%
\bibitem [{\citenamefont {Gornyi}\ \emph {et~al.}(2005)\citenamefont {Gornyi},
  \citenamefont {Mirlin},\ and\ \citenamefont
  {Polyakov}}]{MirlinPolyakovlPRL2005}%
  \BibitemOpen
  \bibfield  {author} {\bibinfo {author} {\bibfnamefont {I.~V.}\ \bibnamefont
  {Gornyi}}, \bibinfo {author} {\bibfnamefont {A.~D.}\ \bibnamefont {Mirlin}},
  \ and\ \bibinfo {author} {\bibfnamefont {D.~G.}\ \bibnamefont {Polyakov}},\
  }\href@noop {} {\bibfield  {journal} {\bibinfo  {journal} {Phys. Rev. Lett.}\
  }\textbf {\bibinfo {volume} {95}},\ \bibinfo {pages} {206603} (\bibinfo
  {year} {2005})}\BibitemShut {NoStop}%
\bibitem [{\citenamefont {Basko}\ \emph {et~al.}(2006)\citenamefont {Basko},
  \citenamefont {Aleiner},\ and\ \citenamefont {Altshuler}}]{BAAAnnPhys2006}%
  \BibitemOpen
  \bibfield  {author} {\bibinfo {author} {\bibfnamefont {D.~M.}\ \bibnamefont
  {Basko}}, \bibinfo {author} {\bibfnamefont {I.~L.}\ \bibnamefont {Aleiner}},
  \ and\ \bibinfo {author} {\bibfnamefont {B.~L.}\ \bibnamefont {Altshuler}},\
  }\href@noop {} {\bibfield  {journal} {\bibinfo  {journal} {Ann. Phys.
  (N.Y.)}\ }\textbf {\bibinfo {volume} {321}},\ \bibinfo {pages} {1126}
  (\bibinfo {year} {2006})}\BibitemShut {NoStop}%
\bibitem [{\citenamefont {Basko}\ \emph {et~al.}(2007)\citenamefont {Basko},
  \citenamefont {Aleiner},\ and\ \citenamefont
  {Altshuler}}]{AltshuleretalPRB2007}%
  \BibitemOpen
  \bibfield  {author} {\bibinfo {author} {\bibfnamefont {D.~M.}\ \bibnamefont
  {Basko}}, \bibinfo {author} {\bibfnamefont {I.~L.}\ \bibnamefont {Aleiner}},
  \ and\ \bibinfo {author} {\bibfnamefont {B.~L.}\ \bibnamefont {Altshuler}},\
  }\href@noop {} {\bibfield  {journal} {\bibinfo  {journal} {Phys. Rev. B}\
  }\textbf {\bibinfo {volume} {76}},\ \bibinfo {pages} {052203} (\bibinfo
  {year} {2007})}\BibitemShut {NoStop}%
\bibitem [{\citenamefont {Oganesyan}\ and\ \citenamefont
  {Huse}(2007)}]{Oganesyan+HusePRB2007}%
  \BibitemOpen
  \bibfield  {author} {\bibinfo {author} {\bibfnamefont {V.}~\bibnamefont
  {Oganesyan}}\ and\ \bibinfo {author} {\bibfnamefont {D.~A.}\ \bibnamefont
  {Huse}},\ }\href@noop {} {\bibfield  {journal} {\bibinfo  {journal} {Phys.
  Rev. B}\ }\textbf {\bibinfo {volume} {75}},\ \bibinfo {pages} {155111}
  (\bibinfo {year} {2007})}\BibitemShut {NoStop}%
\bibitem [{\citenamefont {\v{Z}nidari\v{c}}\ \emph {et~al.}(2008)\citenamefont
  {\v{Z}nidari\v{c}}, \citenamefont {Prosen},\ and\ \citenamefont
  {Prelov\v{s}ek}}]{PeterP-PRB2008}%
  \BibitemOpen
  \bibfield  {author} {\bibinfo {author} {\bibfnamefont {M.}~\bibnamefont
  {\v{Z}nidari\v{c}}}, \bibinfo {author} {\bibfnamefont {T.}~\bibnamefont
  {Prosen}}, \ and\ \bibinfo {author} {\bibfnamefont {P.}~\bibnamefont
  {Prelov\v{s}ek}},\ }\href@noop {} {\bibfield  {journal} {\bibinfo  {journal}
  {Phys. Rev. B}\ }\textbf {\bibinfo {volume} {77}},\ \bibinfo {pages} {064426}
  (\bibinfo {year} {2008})}\BibitemShut {NoStop}%
\bibitem [{\citenamefont {Pal}\ and\ \citenamefont
  {Huse}(2010)}]{Pal+HusePRB2010}%
  \BibitemOpen
  \bibfield  {author} {\bibinfo {author} {\bibfnamefont {A.}~\bibnamefont
  {Pal}}\ and\ \bibinfo {author} {\bibfnamefont {D.~A.}\ \bibnamefont {Huse}},\
  }\href@noop {} {\bibfield  {journal} {\bibinfo  {journal} {Phys. Rev. B}\
  }\textbf {\bibinfo {volume} {82}},\ \bibinfo {pages} {174411} (\bibinfo
  {year} {2010})}\BibitemShut {NoStop}%
\bibitem [{\citenamefont {Berkelbach}\ and\ \citenamefont
  {Reichman}(2010)}]{ReichmanPRB2010}%
  \BibitemOpen
  \bibfield  {author} {\bibinfo {author} {\bibfnamefont {T.~C.}\ \bibnamefont
  {Berkelbach}}\ and\ \bibinfo {author} {\bibfnamefont {D.~R.}\ \bibnamefont
  {Reichman}},\ }\href@noop {} {\bibfield  {journal} {\bibinfo  {journal}
  {Phys. Rev. B}\ }\textbf {\bibinfo {volume} {81}},\ \bibinfo {pages} {224429}
  (\bibinfo {year} {2010})}\BibitemShut {NoStop}%
\bibitem [{\citenamefont {Monthus}\ and\ \citenamefont
  {Garel}(2010)}]{Monthus+GarelPRB2010}%
  \BibitemOpen
  \bibfield  {author} {\bibinfo {author} {\bibfnamefont {C.}~\bibnamefont
  {Monthus}}\ and\ \bibinfo {author} {\bibfnamefont {T.}~\bibnamefont
  {Garel}},\ }\href@noop {} {\bibfield  {journal} {\bibinfo  {journal} {Phys.
  Rev. B}\ }\textbf {\bibinfo {volume} {81}},\ \bibinfo {pages} {134202}
  (\bibinfo {year} {2010})}\BibitemShut {NoStop}%
\bibitem [{\citenamefont {Bardarson}\ \emph {et~al.}(2012)\citenamefont
  {Bardarson}, \citenamefont {Pollmann},\ and\ \citenamefont
  {Moore}}]{BardarsonPollmannMoorePRL2012}%
  \BibitemOpen
  \bibfield  {author} {\bibinfo {author} {\bibfnamefont {J.~H.}\ \bibnamefont
  {Bardarson}}, \bibinfo {author} {\bibfnamefont {F.}~\bibnamefont {Pollmann}},
  \ and\ \bibinfo {author} {\bibfnamefont {J.~E.}\ \bibnamefont {Moore}},\
  }\href@noop {} {\bibfield  {journal} {\bibinfo  {journal} {Phys. Rev. Lett.}\
  }\textbf {\bibinfo {volume} {109}},\ \bibinfo {pages} {017202} (\bibinfo
  {year} {2012})}\BibitemShut {NoStop}%
\bibitem [{\citenamefont {Serbyn}\ \emph
  {et~al.}(2013{\natexlab{a}})\citenamefont {Serbyn}, \citenamefont
  {Papi\v{c}},\ and\ \citenamefont {Abanin}}]{SerbynPapicAbaninPRL2013}%
  \BibitemOpen
  \bibfield  {author} {\bibinfo {author} {\bibfnamefont {M.}~\bibnamefont
  {Serbyn}}, \bibinfo {author} {\bibfnamefont {Z.}~\bibnamefont {Papi\v{c}}}, \
  and\ \bibinfo {author} {\bibfnamefont {D.~A.}\ \bibnamefont {Abanin}},\
  }\href@noop {} {\bibfield  {journal} {\bibinfo  {journal} {Phys. Rev. Lett.}\
  }\textbf {\bibinfo {volume} {110}},\ \bibinfo {pages} {260601} (\bibinfo
  {year} {2013}{\natexlab{a}})}\BibitemShut {NoStop}%
\bibitem [{\citenamefont {{De Luca}}\ and\ \citenamefont
  {Scardicchio}(2013)}]{DeLucaScardicchioEPL2013}%
  \BibitemOpen
  \bibfield  {author} {\bibinfo {author} {\bibfnamefont {A.}~\bibnamefont {{De
  Luca}}}\ and\ \bibinfo {author} {\bibfnamefont {A.}~\bibnamefont
  {Scardicchio}},\ }\href@noop {} {\bibfield  {journal} {\bibinfo  {journal}
  {Europhys. Lett.}\ }\textbf {\bibinfo {volume} {101}},\ \bibinfo {pages}
  {37003} (\bibinfo {year} {2013})}\BibitemShut {NoStop}%
\bibitem [{\citenamefont {Iyer}\ \emph {et~al.}(2013)\citenamefont {Iyer},
  \citenamefont {Oganesyan}, \citenamefont {Refael},\ and\ \citenamefont
  {Huse}}]{IyerHusePRB2013}%
  \BibitemOpen
  \bibfield  {author} {\bibinfo {author} {\bibfnamefont {S.}~\bibnamefont
  {Iyer}}, \bibinfo {author} {\bibfnamefont {V.}~\bibnamefont {Oganesyan}},
  \bibinfo {author} {\bibfnamefont {G.}~\bibnamefont {Refael}}, \ and\ \bibinfo
  {author} {\bibfnamefont {D.~A.}\ \bibnamefont {Huse}},\ }\href@noop {}
  {\bibfield  {journal} {\bibinfo  {journal} {Phys. Rev. B}\ }\textbf {\bibinfo
  {volume} {87}},\ \bibinfo {pages} {134202} (\bibinfo {year}
  {2013})}\BibitemShut {NoStop}%
\bibitem [{\citenamefont {Serbyn}\ \emph
  {et~al.}(2013{\natexlab{b}})\citenamefont {Serbyn}, \citenamefont
  {Papi\v{c}},\ and\ \citenamefont {Abanin}}]{SerbynPapicAbaninLIOMPRL2013}%
  \BibitemOpen
  \bibfield  {author} {\bibinfo {author} {\bibfnamefont {M.}~\bibnamefont
  {Serbyn}}, \bibinfo {author} {\bibfnamefont {Z.}~\bibnamefont {Papi\v{c}}}, \
  and\ \bibinfo {author} {\bibfnamefont {D.~A.}\ \bibnamefont {Abanin}},\
  }\href@noop {} {\bibfield  {journal} {\bibinfo  {journal} {Phys. Rev. Lett.}\
  }\textbf {\bibinfo {volume} {111}},\ \bibinfo {pages} {127201} (\bibinfo
  {year} {2013}{\natexlab{b}})}\BibitemShut {NoStop}%
\bibitem [{\citenamefont {Vosk}\ and\ \citenamefont
  {Altman}(2013)}]{VoskAltmanPRL2013}%
  \BibitemOpen
  \bibfield  {author} {\bibinfo {author} {\bibfnamefont {R.}~\bibnamefont
  {Vosk}}\ and\ \bibinfo {author} {\bibfnamefont {E.}~\bibnamefont {Altman}},\
  }\href@noop {} {\bibfield  {journal} {\bibinfo  {journal} {Phys. Rev. Lett.}\
  }\textbf {\bibinfo {volume} {110}},\ \bibinfo {pages} {067204} (\bibinfo
  {year} {2013})}\BibitemShut {NoStop}%
\bibitem [{\citenamefont {Huse}\ \emph {et~al.}(2013)\citenamefont {Huse},
  \citenamefont {Nandkishore}, \citenamefont {Oganesyan}, \citenamefont {Pal},\
  and\ \citenamefont {Sondhi}}]{Huse+Pal+SondhiPRB2013}%
  \BibitemOpen
  \bibfield  {author} {\bibinfo {author} {\bibfnamefont {D.~A.}\ \bibnamefont
  {Huse}}, \bibinfo {author} {\bibfnamefont {R.}~\bibnamefont {Nandkishore}},
  \bibinfo {author} {\bibfnamefont {V.}~\bibnamefont {Oganesyan}}, \bibinfo
  {author} {\bibfnamefont {A.}~\bibnamefont {Pal}}, \ and\ \bibinfo {author}
  {\bibfnamefont {S.~L.}\ \bibnamefont {Sondhi}},\ }\href@noop {} {\bibfield
  {journal} {\bibinfo  {journal} {Phys. Rev. B}\ }\textbf {\bibinfo {volume}
  {88}},\ \bibinfo {pages} {014206} (\bibinfo {year} {2013})}\BibitemShut
  {NoStop}%
\bibitem [{\citenamefont {Vosk}\ and\ \citenamefont
  {Altman}(2014)}]{VoskAltmanPRL2014}%
  \BibitemOpen
  \bibfield  {author} {\bibinfo {author} {\bibfnamefont {R.}~\bibnamefont
  {Vosk}}\ and\ \bibinfo {author} {\bibfnamefont {E.}~\bibnamefont {Altman}},\
  }\href@noop {} {\bibfield  {journal} {\bibinfo  {journal} {Phys. Rev. Lett.}\
  }\textbf {\bibinfo {volume} {112}},\ \bibinfo {pages} {217204} (\bibinfo
  {year} {2014})}\BibitemShut {NoStop}%
\bibitem [{\citenamefont {Pekker}\ \emph {et~al.}(2014)\citenamefont {Pekker},
  \citenamefont {Refael}, \citenamefont {Altman}, \citenamefont {Demler},\ and\
  \citenamefont {Oganesyan}}]{AltmanDemlerOganesyanPRX2014}%
  \BibitemOpen
  \bibfield  {author} {\bibinfo {author} {\bibfnamefont {D.}~\bibnamefont
  {Pekker}}, \bibinfo {author} {\bibfnamefont {G.}~\bibnamefont {Refael}},
  \bibinfo {author} {\bibfnamefont {E.}~\bibnamefont {Altman}}, \bibinfo
  {author} {\bibfnamefont {E.}~\bibnamefont {Demler}}, \ and\ \bibinfo {author}
  {\bibfnamefont {V.}~\bibnamefont {Oganesyan}},\ }\href@noop {} {\bibfield
  {journal} {\bibinfo  {journal} {Phys. Rev. X}\ }\textbf {\bibinfo {volume}
  {4}},\ \bibinfo {pages} {011052} (\bibinfo {year} {2014})}\BibitemShut
  {NoStop}%
\bibitem [{\citenamefont {Kj\"all}\ \emph {et~al.}(2014)\citenamefont
  {Kj\"all}, \citenamefont {Bardarson},\ and\ \citenamefont
  {Pollmann}}]{BardarsonPollmannPRL2014}%
  \BibitemOpen
  \bibfield  {author} {\bibinfo {author} {\bibfnamefont {J.~A.}\ \bibnamefont
  {Kj\"all}}, \bibinfo {author} {\bibfnamefont {J.~H.}\ \bibnamefont
  {Bardarson}}, \ and\ \bibinfo {author} {\bibfnamefont {F.}~\bibnamefont
  {Pollmann}},\ }\href@noop {} {\bibfield  {journal} {\bibinfo  {journal}
  {Phys. Rev. Lett.}\ }\textbf {\bibinfo {volume} {113}},\ \bibinfo {pages}
  {107204} (\bibinfo {year} {2014})}\BibitemShut {NoStop}%
\bibitem [{\citenamefont {Chandran}\ \emph {et~al.}(2014)\citenamefont
  {Chandran}, \citenamefont {Khemani}, \citenamefont {Laumann},\ and\
  \citenamefont {Sondhi}}]{LaumannSondhiPRB2014}%
  \BibitemOpen
  \bibfield  {author} {\bibinfo {author} {\bibfnamefont {A.}~\bibnamefont
  {Chandran}}, \bibinfo {author} {\bibfnamefont {V.}~\bibnamefont {Khemani}},
  \bibinfo {author} {\bibfnamefont {C.~R.}\ \bibnamefont {Laumann}}, \ and\
  \bibinfo {author} {\bibfnamefont {S.~L.}\ \bibnamefont {Sondhi}},\
  }\href@noop {} {\bibfield  {journal} {\bibinfo  {journal} {Phys. Rev. B}\
  }\textbf {\bibinfo {volume} {89}},\ \bibinfo {pages} {144201} (\bibinfo
  {year} {2014})}\BibitemShut {NoStop}%
\bibitem [{\citenamefont {Huse}\ \emph {et~al.}(2014)\citenamefont {Huse},
  \citenamefont {Nandkishore},\ and\ \citenamefont
  {Oganesyan}}]{Huse+Nand+OganPRB2014}%
  \BibitemOpen
  \bibfield  {author} {\bibinfo {author} {\bibfnamefont {D.~A.}\ \bibnamefont
  {Huse}}, \bibinfo {author} {\bibfnamefont {R.}~\bibnamefont {Nandkishore}}, \
  and\ \bibinfo {author} {\bibfnamefont {V.}~\bibnamefont {Oganesyan}},\
  }\href@noop {} {\bibfield  {journal} {\bibinfo  {journal} {Phys. Rev. B}\
  }\textbf {\bibinfo {volume} {90}},\ \bibinfo {pages} {174202} (\bibinfo
  {year} {2014})}\BibitemShut {NoStop}%
\bibitem [{\citenamefont {Nandkishore}\ and\ \citenamefont
  {Huse}(2015)}]{HuseReview2015}%
  \BibitemOpen
  \bibfield  {author} {\bibinfo {author} {\bibfnamefont {R.}~\bibnamefont
  {Nandkishore}}\ and\ \bibinfo {author} {\bibfnamefont {D.~A.}\ \bibnamefont
  {Huse}},\ }\href@noop {} {\bibfield  {journal} {\bibinfo  {journal} {Annu.
  Rev. Condens. Matter Phys.}\ }\textbf {\bibinfo {volume} {6}},\ \bibinfo
  {pages} {15} (\bibinfo {year} {2015})}\BibitemShut {NoStop}%
\bibitem [{\citenamefont {Lazarides}\ \emph {et~al.}(2015)\citenamefont
  {Lazarides}, \citenamefont {Das},\ and\ \citenamefont
  {Moessner}}]{DasMoessner2015}%
  \BibitemOpen
  \bibfield  {author} {\bibinfo {author} {\bibfnamefont {A.}~\bibnamefont
  {Lazarides}}, \bibinfo {author} {\bibfnamefont {A.}~\bibnamefont {Das}}, \
  and\ \bibinfo {author} {\bibfnamefont {R.}~\bibnamefont {Moessner}},\
  }\href@noop {} {\bibfield  {journal} {\bibinfo  {journal} {Phys. Rev. Lett.}\
  }\textbf {\bibinfo {volume} {115}},\ \bibinfo {pages} {030402} (\bibinfo
  {year} {2015})}\BibitemShut {NoStop}%
\bibitem [{\citenamefont {Ponte}\ \emph {et~al.}(2015)\citenamefont {Ponte},
  \citenamefont {Papi\'{c}}, \citenamefont {Huveneers},\ and\ \citenamefont
  {Abanin}}]{PapicAbaninPRL2015}%
  \BibitemOpen
  \bibfield  {author} {\bibinfo {author} {\bibfnamefont {P.}~\bibnamefont
  {Ponte}}, \bibinfo {author} {\bibfnamefont {Z.}~\bibnamefont {Papi\'{c}}},
  \bibinfo {author} {\bibfnamefont {F.}~\bibnamefont {Huveneers}}, \ and\
  \bibinfo {author} {\bibfnamefont {D.~A.}\ \bibnamefont {Abanin}},\
  }\href@noop {} {\bibfield  {journal} {\bibinfo  {journal} {Phys. Rev. Lett.}\
  }\textbf {\bibinfo {volume} {114}},\ \bibinfo {pages} {140401} (\bibinfo
  {year} {2015})}\BibitemShut {NoStop}%
\bibitem [{\citenamefont {Agarwal}\ \emph {et~al.}(2015)\citenamefont
  {Agarwal}, \citenamefont {Gopalakrishnan}, \citenamefont {Knap},
  \citenamefont {M\"{u}ller},\ and\ \citenamefont
  {Demler}}]{GopalKDemlerPRL2015}%
  \BibitemOpen
  \bibfield  {author} {\bibinfo {author} {\bibfnamefont {K.}~\bibnamefont
  {Agarwal}}, \bibinfo {author} {\bibfnamefont {S.}~\bibnamefont
  {Gopalakrishnan}}, \bibinfo {author} {\bibfnamefont {M.}~\bibnamefont
  {Knap}}, \bibinfo {author} {\bibfnamefont {M.}~\bibnamefont {M\"{u}ller}}, \
  and\ \bibinfo {author} {\bibfnamefont {E.}~\bibnamefont {Demler}},\
  }\href@noop {} {\bibfield  {journal} {\bibinfo  {journal} {Phys. Rev. Lett.}\
  }\textbf {\bibinfo {volume} {114}},\ \bibinfo {pages} {160401} (\bibinfo
  {year} {2015})}\BibitemShut {NoStop}%
\bibitem [{\citenamefont {{Bar Lev}}\ \emph {et~al.}(2015)\citenamefont {{Bar
  Lev}}, \citenamefont {Cohen},\ and\ \citenamefont
  {Reichman}}]{ReichmanPRL2015}%
  \BibitemOpen
  \bibfield  {author} {\bibinfo {author} {\bibfnamefont {Y.}~\bibnamefont {{Bar
  Lev}}}, \bibinfo {author} {\bibfnamefont {G.}~\bibnamefont {Cohen}}, \ and\
  \bibinfo {author} {\bibfnamefont {D.~R.}\ \bibnamefont {Reichman}},\
  }\href@noop {} {\bibfield  {journal} {\bibinfo  {journal} {Phys. Rev. Lett.}\
  }\textbf {\bibinfo {volume} {114}},\ \bibinfo {pages} {100601} (\bibinfo
  {year} {2015})}\BibitemShut {NoStop}%
\bibitem [{\citenamefont {Chandran}\ \emph {et~al.}(2015)\citenamefont
  {Chandran}, \citenamefont {Kim}, \citenamefont {Vidal},\ and\ \citenamefont
  {Abanin}}]{AbaninLIOMs2015}%
  \BibitemOpen
  \bibfield  {author} {\bibinfo {author} {\bibfnamefont {A.}~\bibnamefont
  {Chandran}}, \bibinfo {author} {\bibfnamefont {I.~H.}\ \bibnamefont {Kim}},
  \bibinfo {author} {\bibfnamefont {G.}~\bibnamefont {Vidal}}, \ and\ \bibinfo
  {author} {\bibfnamefont {D.~A.}\ \bibnamefont {Abanin}},\ }\href@noop {}
  {\bibfield  {journal} {\bibinfo  {journal} {Phys. Rev. B}\ }\textbf {\bibinfo
  {volume} {91}},\ \bibinfo {pages} {085425} (\bibinfo {year}
  {2015})}\BibitemShut {NoStop}%
\bibitem [{\citenamefont {Bera}\ \emph {et~al.}(2015)\citenamefont {Bera},
  \citenamefont {Schomerus}, \citenamefont {Heidrich-Meisner},\ and\
  \citenamefont {Bardarson}}]{BeraFHMBardarsonPRL2015}%
  \BibitemOpen
  \bibfield  {author} {\bibinfo {author} {\bibfnamefont {S.}~\bibnamefont
  {Bera}}, \bibinfo {author} {\bibfnamefont {H.}~\bibnamefont {Schomerus}},
  \bibinfo {author} {\bibfnamefont {F.}~\bibnamefont {Heidrich-Meisner}}, \
  and\ \bibinfo {author} {\bibfnamefont {J.~H.}\ \bibnamefont {Bardarson}},\
  }\href@noop {} {\bibfield  {journal} {\bibinfo  {journal} {Phys. Rev. Lett.}\
  }\textbf {\bibinfo {volume} {115}},\ \bibinfo {pages} {046603} (\bibinfo
  {year} {2015})}\BibitemShut {NoStop}%
\bibitem [{\citenamefont {Modak}\ and\ \citenamefont
  {Mukerjee}(2015)}]{SubrotoPRL2015}%
  \BibitemOpen
  \bibfield  {author} {\bibinfo {author} {\bibfnamefont {R.}~\bibnamefont
  {Modak}}\ and\ \bibinfo {author} {\bibfnamefont {S.}~\bibnamefont
  {Mukerjee}},\ }\href@noop {} {\bibfield  {journal} {\bibinfo  {journal}
  {Phys. Rev. Lett.}\ }\textbf {\bibinfo {volume} {115}},\ \bibinfo {pages}
  {230401} (\bibinfo {year} {2015})}\BibitemShut {NoStop}%
\bibitem [{\citenamefont {Li}\ \emph {et~al.}(2015)\citenamefont {Li},
  \citenamefont {Ganeshan}, \citenamefont {Pixley},\ and\ \citenamefont
  {DasSarma}}]{PixleyDasSarmaPRL2015}%
  \BibitemOpen
  \bibfield  {author} {\bibinfo {author} {\bibfnamefont {X.}~\bibnamefont
  {Li}}, \bibinfo {author} {\bibfnamefont {S.}~\bibnamefont {Ganeshan}},
  \bibinfo {author} {\bibfnamefont {J.~H.}\ \bibnamefont {Pixley}}, \ and\
  \bibinfo {author} {\bibfnamefont {S.}~\bibnamefont {DasSarma}},\ }\href@noop
  {} {\bibfield  {journal} {\bibinfo  {journal} {Phys. Rev. Lett.}\ }\textbf
  {\bibinfo {volume} {115}},\ \bibinfo {pages} {186601} (\bibinfo {year}
  {2015})}\BibitemShut {NoStop}%
\bibitem [{\citenamefont {Mondragon-Shem}\ \emph {et~al.}(2015)\citenamefont
  {Mondragon-Shem}, \citenamefont {Pal}, \citenamefont {Hughes},\ and\
  \citenamefont {Laumann}}]{Mondragon-ShemME_PRB2015}%
  \BibitemOpen
  \bibfield  {author} {\bibinfo {author} {\bibfnamefont {I.}~\bibnamefont
  {Mondragon-Shem}}, \bibinfo {author} {\bibfnamefont {A.}~\bibnamefont {Pal}},
  \bibinfo {author} {\bibfnamefont {T.~L.}\ \bibnamefont {Hughes}}, \ and\
  \bibinfo {author} {\bibfnamefont {C.~R.}\ \bibnamefont {Laumann}},\
  }\href@noop {} {\bibfield  {journal} {\bibinfo  {journal} {Phys. Rev. B}\
  }\textbf {\bibinfo {volume} {92}},\ \bibinfo {pages} {064203} (\bibinfo
  {year} {2015})}\BibitemShut {NoStop}%
\bibitem [{\citenamefont {Luitz}\ \emph {et~al.}(2015)\citenamefont {Luitz},
  \citenamefont {Laflorencie},\ and\ \citenamefont {Alet}}]{LuitzAletPRB2015}%
  \BibitemOpen
  \bibfield  {author} {\bibinfo {author} {\bibfnamefont {D.~J.}\ \bibnamefont
  {Luitz}}, \bibinfo {author} {\bibfnamefont {N.}~\bibnamefont {Laflorencie}},
  \ and\ \bibinfo {author} {\bibfnamefont {F.}~\bibnamefont {Alet}},\
  }\href@noop {} {\bibfield  {journal} {\bibinfo  {journal} {Phys. Rev. B}\
  }\textbf {\bibinfo {volume} {91}},\ \bibinfo {pages} {081103} (\bibinfo
  {year} {2015})}\BibitemShut {NoStop}%
\bibitem [{\citenamefont {Devakul}\ and\ \citenamefont
  {Singh}(2015)}]{Devakul15}%
  \BibitemOpen
  \bibfield  {author} {\bibinfo {author} {\bibfnamefont {T.}~\bibnamefont
  {Devakul}}\ and\ \bibinfo {author} {\bibfnamefont {R.~R.~P.}\ \bibnamefont
  {Singh}},\ }\href {\doibase 10.1103/PhysRevLett.115.187201} {\bibfield
  {journal} {\bibinfo  {journal} {Phys. Rev. Lett.}\ }\textbf {\bibinfo
  {volume} {115}},\ \bibinfo {pages} {187201} (\bibinfo {year}
  {2015})}\BibitemShut {NoStop}%
\bibitem [{\citenamefont {Johri}\ \emph {et~al.}(2015)\citenamefont {Johri},
  \citenamefont {Nandkishore},\ and\ \citenamefont
  {Bhatt}}]{NandkishoreBhattPRL2015}%
  \BibitemOpen
  \bibfield  {author} {\bibinfo {author} {\bibfnamefont {S.}~\bibnamefont
  {Johri}}, \bibinfo {author} {\bibfnamefont {R.}~\bibnamefont {Nandkishore}},
  \ and\ \bibinfo {author} {\bibfnamefont {R.~N.}\ \bibnamefont {Bhatt}},\
  }\href@noop {} {\bibfield  {journal} {\bibinfo  {journal} {Phys. Rev. Lett.}\
  }\textbf {\bibinfo {volume} {114}},\ \bibinfo {pages} {117401} (\bibinfo
  {year} {2015})}\BibitemShut {NoStop}%
\bibitem [{\citenamefont {Vosk}\ \emph {et~al.}(2015)\citenamefont {Vosk},
  \citenamefont {Huse},\ and\ \citenamefont {Altman}}]{VoskHuseAltmanPRX2015}%
  \BibitemOpen
  \bibfield  {author} {\bibinfo {author} {\bibfnamefont {R.}~\bibnamefont
  {Vosk}}, \bibinfo {author} {\bibfnamefont {D.~A.}\ \bibnamefont {Huse}}, \
  and\ \bibinfo {author} {\bibfnamefont {E.}~\bibnamefont {Altman}},\
  }\href@noop {} {\bibfield  {journal} {\bibinfo  {journal} {Phys. Rev. X}\
  }\textbf {\bibinfo {volume} {5}},\ \bibinfo {pages} {031032} (\bibinfo {year}
  {2015})}\BibitemShut {NoStop}%
\bibitem [{\citenamefont {Vasseur}\ \emph {et~al.}(2015)\citenamefont
  {Vasseur}, \citenamefont {Potter},\ and\ \citenamefont
  {Parameswaran}}]{SIDVaasseurPotterPRL2015}%
  \BibitemOpen
  \bibfield  {author} {\bibinfo {author} {\bibfnamefont {R.}~\bibnamefont
  {Vasseur}}, \bibinfo {author} {\bibfnamefont {A.~C.}\ \bibnamefont {Potter}},
  \ and\ \bibinfo {author} {\bibfnamefont {S.~A.}\ \bibnamefont
  {Parameswaran}},\ }\href@noop {} {\bibfield  {journal} {\bibinfo  {journal}
  {Phys. Rev. Lett.}\ }\textbf {\bibinfo {volume} {114}},\ \bibinfo {pages}
  {217201} (\bibinfo {year} {2015})}\BibitemShut {NoStop}%
\bibitem [{\citenamefont {Potter}\ \emph {et~al.}(2015)\citenamefont {Potter},
  \citenamefont {Vasseur},\ and\ \citenamefont
  {Parameswaran}}]{SIDPotterVasseurPRX2015}%
  \BibitemOpen
  \bibfield  {author} {\bibinfo {author} {\bibfnamefont {A.~C.}\ \bibnamefont
  {Potter}}, \bibinfo {author} {\bibfnamefont {R.}~\bibnamefont {Vasseur}}, \
  and\ \bibinfo {author} {\bibfnamefont {S.~A.}\ \bibnamefont {Parameswaran}},\
  }\href@noop {} {\bibfield  {journal} {\bibinfo  {journal} {Phys. Rev. X}\
  }\textbf {\bibinfo {volume} {5}},\ \bibinfo {pages} {031033} (\bibinfo {year}
  {2015})}\BibitemShut {NoStop}%
\bibitem [{\citenamefont {Serbyn}\ \emph {et~al.}(2015)\citenamefont {Serbyn},
  \citenamefont {Papi{\'c}},\ and\ \citenamefont
  {Abanin}}]{Serbyn+P+AbaninPRX2015}%
  \BibitemOpen
  \bibfield  {author} {\bibinfo {author} {\bibfnamefont {M.}~\bibnamefont
  {Serbyn}}, \bibinfo {author} {\bibfnamefont {Z.}~\bibnamefont {Papi{\'c}}}, \
  and\ \bibinfo {author} {\bibfnamefont {D.~A.}\ \bibnamefont {Abanin}},\
  }\href@noop {} {\bibfield  {journal} {\bibinfo  {journal} {Phys. Rev. X}\
  }\textbf {\bibinfo {volume} {5}},\ \bibinfo {pages} {041047} (\bibinfo {year}
  {2015})}\BibitemShut {NoStop}%
\bibitem [{\citenamefont {Imbrie}(2016{\natexlab{a}})}]{ImbriePRL2016}%
  \BibitemOpen
  \bibfield  {author} {\bibinfo {author} {\bibfnamefont {J.~Z.}\ \bibnamefont
  {Imbrie}},\ }\href@noop {} {\bibfield  {journal} {\bibinfo  {journal} {Phys.
  Rev. Lett.}\ }\textbf {\bibinfo {volume} {117}},\ \bibinfo {pages} {027201}
  (\bibinfo {year} {2016}{\natexlab{a}})}\BibitemShut {NoStop}%
\bibitem [{\citenamefont {Imbrie}(2016{\natexlab{b}})}]{Imbrie16}%
  \BibitemOpen
  \bibfield  {author} {\bibinfo {author} {\bibfnamefont {J.~Z.}\ \bibnamefont
  {Imbrie}},\ }\href@noop {} {\bibfield  {journal} {\bibinfo  {journal} {J.
  Stat. Phys}\ }\textbf {\bibinfo {volume} {163}},\ \bibinfo {pages} {998}
  (\bibinfo {year} {2016}{\natexlab{b}})}\BibitemShut {NoStop}%
\bibitem [{\citenamefont {Pietracaprina}\ \emph {et~al.}(2016)\citenamefont
  {Pietracaprina}, \citenamefont {Ros},\ and\ \citenamefont
  {Scardicchio}}]{ScardicchioFAppPRB2016}%
  \BibitemOpen
  \bibfield  {author} {\bibinfo {author} {\bibfnamefont {F.}~\bibnamefont
  {Pietracaprina}}, \bibinfo {author} {\bibfnamefont {V.}~\bibnamefont {Ros}},
  \ and\ \bibinfo {author} {\bibfnamefont {A.}~\bibnamefont {Scardicchio}},\
  }\href@noop {} {\bibfield  {journal} {\bibinfo  {journal} {Phys. Rev. B}\
  }\textbf {\bibinfo {volume} {93}},\ \bibinfo {pages} {054201} (\bibinfo
  {year} {2016})}\BibitemShut {NoStop}%
\bibitem [{\citenamefont {{De Roeck}}\ \emph {et~al.}(2016)\citenamefont {{De
  Roeck}}, \citenamefont {Huveneers}, \citenamefont {M{\"u}ller},\ and\
  \citenamefont {Schiulaz}}]{DeRoeckNoMEsPRB2016}%
  \BibitemOpen
  \bibfield  {author} {\bibinfo {author} {\bibfnamefont {W.}~\bibnamefont {{De
  Roeck}}}, \bibinfo {author} {\bibfnamefont {F.}~\bibnamefont {Huveneers}},
  \bibinfo {author} {\bibfnamefont {M.}~\bibnamefont {M{\"u}ller}}, \ and\
  \bibinfo {author} {\bibfnamefont {M.}~\bibnamefont {Schiulaz}},\ }\href@noop
  {} {\bibfield  {journal} {\bibinfo  {journal} {Phys. Rev. B}\ }\textbf
  {\bibinfo {volume} {93}},\ \bibinfo {pages} {014203} (\bibinfo {year}
  {2016})}\BibitemShut {NoStop}%
\bibitem [{\citenamefont {Geraedts}\ \emph {et~al.}(2016)\citenamefont
  {Geraedts}, \citenamefont {Nandkishore},\ and\ \citenamefont
  {Regnault}}]{Geraedts16}%
  \BibitemOpen
  \bibfield  {author} {\bibinfo {author} {\bibfnamefont {S.~D.}\ \bibnamefont
  {Geraedts}}, \bibinfo {author} {\bibfnamefont {R.}~\bibnamefont
  {Nandkishore}}, \ and\ \bibinfo {author} {\bibfnamefont {N.}~\bibnamefont
  {Regnault}},\ }\href {\doibase 10.1103/PhysRevB.93.174202} {\bibfield
  {journal} {\bibinfo  {journal} {Phys. Rev. B}\ }\textbf {\bibinfo {volume}
  {93}},\ \bibinfo {pages} {174202} (\bibinfo {year} {2016})}\BibitemShut
  {NoStop}%
\bibitem [{\citenamefont {Khemani}\ \emph {et~al.}(2016)\citenamefont
  {Khemani}, \citenamefont {Lazarides}, \citenamefont {Moessner},\ and\
  \citenamefont {Sondhi}}]{Khemani2016}%
  \BibitemOpen
  \bibfield  {author} {\bibinfo {author} {\bibfnamefont {V.}~\bibnamefont
  {Khemani}}, \bibinfo {author} {\bibfnamefont {A.}~\bibnamefont {Lazarides}},
  \bibinfo {author} {\bibfnamefont {R.}~\bibnamefont {Moessner}}, \ and\
  \bibinfo {author} {\bibfnamefont {S.~L.}\ \bibnamefont {Sondhi}},\
  }\href@noop {} {\bibfield  {journal} {\bibinfo  {journal} {Phys. Rev. Lett.}\
  }\textbf {\bibinfo {volume} {116}},\ \bibinfo {pages} {250401} (\bibinfo
  {year} {2016})}\BibitemShut {NoStop}%
\bibitem [{\citenamefont {Serbyn}\ and\ \citenamefont
  {Moore}(2016)}]{Serbyn16}%
  \BibitemOpen
  \bibfield  {author} {\bibinfo {author} {\bibfnamefont {M.}~\bibnamefont
  {Serbyn}}\ and\ \bibinfo {author} {\bibfnamefont {J.~E.}\ \bibnamefont
  {Moore}},\ }\href {\doibase 10.1103/PhysRevB.93.041424} {\bibfield  {journal}
  {\bibinfo  {journal} {Phys. Rev. B}\ }\textbf {\bibinfo {volume} {93}},\
  \bibinfo {pages} {041424} (\bibinfo {year} {2016})}\BibitemShut {NoStop}%
\bibitem [{\citenamefont {Serbyn}\ \emph {et~al.}(2016)\citenamefont {Serbyn},
  \citenamefont {Michailidis}, \citenamefont {Abanin},\ and\ \citenamefont
  {Papi\ifmmode~\acute{c}\else \'{c}\fi{}}}]{Serbyn16b}%
  \BibitemOpen
  \bibfield  {author} {\bibinfo {author} {\bibfnamefont {M.}~\bibnamefont
  {Serbyn}}, \bibinfo {author} {\bibfnamefont {A.~A.}\ \bibnamefont
  {Michailidis}}, \bibinfo {author} {\bibfnamefont {D.~A.}\ \bibnamefont
  {Abanin}}, \ and\ \bibinfo {author} {\bibfnamefont {Z.}~\bibnamefont
  {Papi\ifmmode~\acute{c}\else \'{c}\fi{}}},\ }\href {\doibase
  10.1103/PhysRevLett.117.160601} {\bibfield  {journal} {\bibinfo  {journal}
  {Phys. Rev. Lett.}\ }\textbf {\bibinfo {volume} {117}},\ \bibinfo {pages}
  {160601} (\bibinfo {year} {2016})}\BibitemShut {NoStop}%
\bibitem [{\citenamefont {Baldwin}\ \emph {et~al.}(2016)\citenamefont
  {Baldwin}, \citenamefont {Laumann}, \citenamefont {Pal},\ and\ \citenamefont
  {Scardicchio}}]{BaldwinQREM2016}%
  \BibitemOpen
  \bibfield  {author} {\bibinfo {author} {\bibfnamefont {C.~L.}\ \bibnamefont
  {Baldwin}}, \bibinfo {author} {\bibfnamefont {C.~R.}\ \bibnamefont
  {Laumann}}, \bibinfo {author} {\bibfnamefont {A.}~\bibnamefont {Pal}}, \ and\
  \bibinfo {author} {\bibfnamefont {A.}~\bibnamefont {Scardicchio}},\
  }\href@noop {} {\bibfield  {journal} {\bibinfo  {journal} {Phys. Rev. B}\
  }\textbf {\bibinfo {volume} {93}},\ \bibinfo {pages} {024202} (\bibinfo
  {year} {2016})}\BibitemShut {NoStop}%
\bibitem [{\citenamefont {Prelov\v{s}ek}\ \emph {et~al.}(2016)\citenamefont
  {Prelov\v{s}ek}, \citenamefont {Bari\v{s}i{\'c}},\ and\ \citenamefont
  {\v{Z}nidari\v{c}}}]{PeterPHubbard2016}%
  \BibitemOpen
  \bibfield  {author} {\bibinfo {author} {\bibfnamefont {P.}~\bibnamefont
  {Prelov\v{s}ek}}, \bibinfo {author} {\bibfnamefont {O.~S.}\ \bibnamefont
  {Bari\v{s}i{\'c}}}, \ and\ \bibinfo {author} {\bibfnamefont {M.}~\bibnamefont
  {\v{Z}nidari\v{c}}},\ }\href@noop {} {\bibfield  {journal} {\bibinfo
  {journal} {Phys. Rev. B}\ }\textbf {\bibinfo {volume} {94}},\ \bibinfo
  {pages} {241104} (\bibinfo {year} {2016})}\BibitemShut {NoStop}%
\bibitem [{\citenamefont {{ De Roeck}}\ and\ \citenamefont
  {Huveneers}(2017)}]{DeRoeckNoMBLinHigherDimPRB2017}%
  \BibitemOpen
  \bibfield  {author} {\bibinfo {author} {\bibfnamefont {W.}~\bibnamefont {{ De
  Roeck}}}\ and\ \bibinfo {author} {\bibfnamefont {F.}~\bibnamefont
  {Huveneers}},\ }\href@noop {} {\bibfield  {journal} {\bibinfo  {journal}
  {Phys. Rev. B}\ }\textbf {\bibinfo {volume} {95}},\ \bibinfo {pages} {155129}
  (\bibinfo {year} {2017})}\BibitemShut {NoStop}%
\bibitem [{\citenamefont {Rademaker}\ and\ \citenamefont
  {Ortu{\~n}o}(2016)}]{RademakerPRL2016}%
  \BibitemOpen
  \bibfield  {author} {\bibinfo {author} {\bibfnamefont {L.}~\bibnamefont
  {Rademaker}}\ and\ \bibinfo {author} {\bibfnamefont {M.}~\bibnamefont
  {Ortu{\~n}o}},\ }\href@noop {} {\bibfield  {journal} {\bibinfo  {journal}
  {Phys. Rev. Lett.}\ }\textbf {\bibinfo {volume} {116}},\ \bibinfo {pages}
  {010404} (\bibinfo {year} {2016})}\BibitemShut {NoStop}%
\bibitem [{\citenamefont {Parameswaran}\ \emph {et~al.}(2017)\citenamefont
  {Parameswaran}, \citenamefont {Potter},\ and\ \citenamefont
  {Vasseur}}]{Parameswaran17}%
  \BibitemOpen
  \bibfield  {author} {\bibinfo {author} {\bibfnamefont {S.~A.}\ \bibnamefont
  {Parameswaran}}, \bibinfo {author} {\bibfnamefont {A.~C.}\ \bibnamefont
  {Potter}}, \ and\ \bibinfo {author} {\bibfnamefont {R.}~\bibnamefont
  {Vasseur}},\ }\href {\doibase 10.1002/andp.201600302} {\bibfield  {journal}
  {\bibinfo  {journal} {Annalen der Physik}\ }\textbf {\bibinfo {volume}
  {529}},\ \bibinfo {pages} {1600302} (\bibinfo {year} {2017})},\ \bibinfo
  {note} {1600302}\BibitemShut {NoStop}%
\bibitem [{\citenamefont {Lezama}\ \emph {et~al.}(2017)\citenamefont {Lezama},
  \citenamefont {Bera}, \citenamefont {Schomerus}, \citenamefont
  {Heidrich-Meisner},\ and\ \citenamefont {Bardarson}}]{Lezama17}%
  \BibitemOpen
  \bibfield  {author} {\bibinfo {author} {\bibfnamefont {T.~L.~M.}\
  \bibnamefont {Lezama}}, \bibinfo {author} {\bibfnamefont {S.}~\bibnamefont
  {Bera}}, \bibinfo {author} {\bibfnamefont {H.}~\bibnamefont {Schomerus}},
  \bibinfo {author} {\bibfnamefont {F.}~\bibnamefont {Heidrich-Meisner}}, \
  and\ \bibinfo {author} {\bibfnamefont {J.~H.}\ \bibnamefont {Bardarson}},\
  }\href {\doibase 10.1103/PhysRevB.96.060202} {\bibfield  {journal} {\bibinfo
  {journal} {Phys. Rev. B}\ }\textbf {\bibinfo {volume} {96}},\ \bibinfo
  {pages} {060202} (\bibinfo {year} {2017})}\BibitemShut {NoStop}%
\bibitem [{\citenamefont {Nag}\ and\ \citenamefont {Garg}(2017)}]{Nag17}%
  \BibitemOpen
  \bibfield  {author} {\bibinfo {author} {\bibfnamefont {S.}~\bibnamefont
  {Nag}}\ and\ \bibinfo {author} {\bibfnamefont {A.}~\bibnamefont {Garg}},\
  }\href {\doibase 10.1103/PhysRevB.96.060203} {\bibfield  {journal} {\bibinfo
  {journal} {Phys. Rev. B}\ }\textbf {\bibinfo {volume} {96}},\ \bibinfo
  {pages} {060203} (\bibinfo {year} {2017})}\BibitemShut {NoStop}%
\bibitem [{\citenamefont {Moessner}\ and\ \citenamefont
  {Sondhi}(2017)}]{MoessnerSondhi2017}%
  \BibitemOpen
  \bibfield  {author} {\bibinfo {author} {\bibfnamefont {R.}~\bibnamefont
  {Moessner}}\ and\ \bibinfo {author} {\bibfnamefont {S.~L.}\ \bibnamefont
  {Sondhi}},\ }\href@noop {} {\bibfield  {journal} {\bibinfo  {journal} {Nature
  Physics}\ }\textbf {\bibinfo {volume} {13}},\ \bibinfo {pages} {424}
  (\bibinfo {year} {2017})}\BibitemShut {NoStop}%
\bibitem [{\citenamefont {Dumitrescu}\ \emph {et~al.}(2017)\citenamefont
  {Dumitrescu}, \citenamefont {Vasseur},\ and\ \citenamefont
  {Potter}}]{Dumitrescu17}%
  \BibitemOpen
  \bibfield  {author} {\bibinfo {author} {\bibfnamefont {P.~T.}\ \bibnamefont
  {Dumitrescu}}, \bibinfo {author} {\bibfnamefont {R.}~\bibnamefont {Vasseur}},
  \ and\ \bibinfo {author} {\bibfnamefont {A.~C.}\ \bibnamefont {Potter}},\
  }\href {\doibase 10.1103/PhysRevLett.119.110604} {\bibfield  {journal}
  {\bibinfo  {journal} {Phys. Rev. Lett.}\ }\textbf {\bibinfo {volume} {119}},\
  \bibinfo {pages} {110604} (\bibinfo {year} {2017})}\BibitemShut {NoStop}%
\bibitem [{\citenamefont {De~Tomasi}\ \emph {et~al.}(2017)\citenamefont
  {De~Tomasi}, \citenamefont {Bera}, \citenamefont {Bardarson},\ and\
  \citenamefont {Pollmann}}]{DeTomasi17}%
  \BibitemOpen
  \bibfield  {author} {\bibinfo {author} {\bibfnamefont {G.}~\bibnamefont
  {De~Tomasi}}, \bibinfo {author} {\bibfnamefont {S.}~\bibnamefont {Bera}},
  \bibinfo {author} {\bibfnamefont {J.~H.}\ \bibnamefont {Bardarson}}, \ and\
  \bibinfo {author} {\bibfnamefont {F.}~\bibnamefont {Pollmann}},\ }\href
  {\doibase 10.1103/PhysRevLett.118.016804} {\bibfield  {journal} {\bibinfo
  {journal} {Phys. Rev. Lett.}\ }\textbf {\bibinfo {volume} {118}},\ \bibinfo
  {pages} {016804} (\bibinfo {year} {2017})}\BibitemShut {NoStop}%
\bibitem [{\citenamefont {Imbrie}\ \emph {et~al.}(2017)\citenamefont {Imbrie},
  \citenamefont {Ros},\ and\ \citenamefont {Scardicchio}}]{Imbrie17}%
  \BibitemOpen
  \bibfield  {author} {\bibinfo {author} {\bibfnamefont {J.~Z.}\ \bibnamefont
  {Imbrie}}, \bibinfo {author} {\bibfnamefont {V.}~\bibnamefont {Ros}}, \ and\
  \bibinfo {author} {\bibfnamefont {A.}~\bibnamefont {Scardicchio}},\
  }\href@noop {} {\bibfield  {journal} {\bibinfo  {journal} {Ann. Phys.
  (Berlin)}\ }\textbf {\bibinfo {volume} {529}},\ \bibinfo {pages} {1600278}
  (\bibinfo {year} {2017})}\BibitemShut {NoStop}%
\bibitem [{\citenamefont {Wahl}\ \emph {et~al.}()\citenamefont {Wahl},
  \citenamefont {Pal},\ and\ \citenamefont {Simon}}]{Wahl17}%
  \BibitemOpen
  \bibfield  {author} {\bibinfo {author} {\bibfnamefont {T.~B.}\ \bibnamefont
  {Wahl}}, \bibinfo {author} {\bibfnamefont {A.}~\bibnamefont {Pal}}, \ and\
  \bibinfo {author} {\bibfnamefont {S.~H.}\ \bibnamefont {Simon}},\ }\href@noop
  {} {\bibinfo  {journal} {arXiv:1711.02678}\ }\BibitemShut {NoStop}%
\bibitem [{\citenamefont {Mierzejewski}\ \emph {et~al.}(2018)\citenamefont
  {Mierzejewski}, \citenamefont {Kozarzewski},\ and\ \citenamefont
  {Prelov\v{s}ek}}]{MarcinM+PeterPPTB2018}%
  \BibitemOpen
\bibfield  {journal} {  }\bibfield  {author} {\bibinfo {author} {\bibfnamefont
  {M.}~\bibnamefont {Mierzejewski}}, \bibinfo {author} {\bibfnamefont
  {M.}~\bibnamefont {Kozarzewski}}, \ and\ \bibinfo {author} {\bibfnamefont
  {P.}~\bibnamefont {Prelov\v{s}ek}},\ }\href@noop {} {\bibfield  {journal}
  {\bibinfo  {journal} {Phys. Rev. B}\ }\textbf {\bibinfo {volume} {97}},\
  \bibinfo {pages} {064204} (\bibinfo {year} {2018})}\BibitemShut {NoStop}%
\bibitem [{\citenamefont {Welsh}\ and\ \citenamefont {Logan}(2018)}]{SWDEL1}%
  \BibitemOpen
  \bibfield  {author} {\bibinfo {author} {\bibfnamefont {S.}~\bibnamefont
  {Welsh}}\ and\ \bibinfo {author} {\bibfnamefont {D.~E.}\ \bibnamefont
  {Logan}},\ }\href@noop {} {\bibfield  {journal} {\bibinfo  {journal} {J.
  Phys.: Condens. Matter}\ }\textbf {\bibinfo {volume} {30}},\ \bibinfo {pages}
  {405601} (\bibinfo {year} {2018})}\BibitemShut {NoStop}%
\bibitem [{\citenamefont {Feenberg}(1948)}]{Feenberg1948}%
  \BibitemOpen
  \bibfield  {author} {\bibinfo {author} {\bibfnamefont {E.}~\bibnamefont
  {Feenberg}},\ }\href@noop {} {\bibfield  {journal} {\bibinfo  {journal}
  {Phys. Rev.}\ }\textbf {\bibinfo {volume} {74}},\ \bibinfo {pages} {206}
  (\bibinfo {year} {1948})}\BibitemShut {NoStop}%
\bibitem [{\citenamefont {Economou}(2006)}]{Economoubook}%
  \BibitemOpen
  \bibfield  {author} {\bibinfo {author} {\bibfnamefont {E.~N.}\ \bibnamefont
  {Economou}},\ }\href@noop {} {\emph {\bibinfo {title} {Green's Functions in
  Quantum Physics}}}\ (\bibinfo  {publisher} {Springer},\ \bibinfo {address}
  {Berlin},\ \bibinfo {year} {2006})\BibitemShut {NoStop}%
\bibitem [{\citenamefont {Economou}\ and\ \citenamefont
  {Cohen}(1972)}]{EconomousCohen1972}%
  \BibitemOpen
  \bibfield  {author} {\bibinfo {author} {\bibfnamefont {E.~N.}\ \bibnamefont
  {Economou}}\ and\ \bibinfo {author} {\bibfnamefont {M.~H.}\ \bibnamefont
  {Cohen}},\ }\href@noop {} {\bibfield  {journal} {\bibinfo  {journal} {Phys.
  Rev. B}\ }\textbf {\bibinfo {volume} {5}},\ \bibinfo {pages} {2931} (\bibinfo
  {year} {1972})}\BibitemShut {NoStop}%
\bibitem [{\citenamefont {Abou-Chacra}\ \emph {et~al.}(1973)\citenamefont
  {Abou-Chacra}, \citenamefont {Anderson},\ and\ \citenamefont
  {Thouless}}]{AAT1973}%
  \BibitemOpen
  \bibfield  {author} {\bibinfo {author} {\bibfnamefont {R.}~\bibnamefont
  {Abou-Chacra}}, \bibinfo {author} {\bibfnamefont {P.~W.}\ \bibnamefont
  {Anderson}}, \ and\ \bibinfo {author} {\bibfnamefont {D.~J.}\ \bibnamefont
  {Thouless}},\ }\href@noop {} {\bibfield  {journal} {\bibinfo  {journal} {J.
  Phys. C}\ }\textbf {\bibinfo {volume} {6}},\ \bibinfo {pages} {1734}
  (\bibinfo {year} {1973})}\BibitemShut {NoStop}%
\bibitem [{\citenamefont {Thouless}(1973)}]{ThoulessReview1974}%
  \BibitemOpen
  \bibfield  {author} {\bibinfo {author} {\bibfnamefont {D.~J.}\ \bibnamefont
  {Thouless}},\ }\href@noop {} {\bibfield  {journal} {\bibinfo  {journal}
  {Phys. Rep.}\ }\textbf {\bibinfo {volume} {13}},\ \bibinfo {pages} {93}
  (\bibinfo {year} {1973})}\BibitemShut {NoStop}%
\bibitem [{\citenamefont {Licciardello}\ and\ \citenamefont
  {Economou}(1975)}]{LicciardelloEconomou1975}%
  \BibitemOpen
  \bibfield  {author} {\bibinfo {author} {\bibfnamefont {D.~C.}\ \bibnamefont
  {Licciardello}}\ and\ \bibinfo {author} {\bibfnamefont {E.~N.}\ \bibnamefont
  {Economou}},\ }\href@noop {} {\bibfield  {journal} {\bibinfo  {journal}
  {Phys. Rev. B}\ }\textbf {\bibinfo {volume} {11}},\ \bibinfo {pages} {3697}
  (\bibinfo {year} {1975})}\BibitemShut {NoStop}%
\bibitem [{\citenamefont {Heinrichs}(1977)}]{Heinrichs1977}%
  \BibitemOpen
  \bibfield  {author} {\bibinfo {author} {\bibfnamefont {J.}~\bibnamefont
  {Heinrichs}},\ }\href@noop {} {\bibfield  {journal} {\bibinfo  {journal}
  {Phys. Rev. B}\ }\textbf {\bibinfo {volume} {16}},\ \bibinfo {pages} {4365}
  (\bibinfo {year} {1977})}\BibitemShut {NoStop}%
\bibitem [{\citenamefont {Fleishman}\ and\ \citenamefont
  {Stein}(1979)}]{Stein1979}%
  \BibitemOpen
  \bibfield  {author} {\bibinfo {author} {\bibfnamefont {L.}~\bibnamefont
  {Fleishman}}\ and\ \bibinfo {author} {\bibfnamefont {D.~L.}\ \bibnamefont
  {Stein}},\ }\href@noop {} {\bibfield  {journal} {\bibinfo  {journal} {J.
  Phys. C}\ }\textbf {\bibinfo {volume} {12}},\ \bibinfo {pages} {4817}
  (\bibinfo {year} {1979})}\BibitemShut {NoStop}%
\bibitem [{\citenamefont {Logan}\ and\ \citenamefont
  {Wolynes}(1987{\natexlab{a}})}]{DELPGWPRB1987}%
  \BibitemOpen
  \bibfield  {author} {\bibinfo {author} {\bibfnamefont {D.~E.}\ \bibnamefont
  {Logan}}\ and\ \bibinfo {author} {\bibfnamefont {P.~G.}\ \bibnamefont
  {Wolynes}},\ }\href@noop {} {\bibfield  {journal} {\bibinfo  {journal} {Phys.
  Rev. B}\ }\textbf {\bibinfo {volume} {36}},\ \bibinfo {pages} {4135}
  (\bibinfo {year} {1987}{\natexlab{a}})}\BibitemShut {NoStop}%
\bibitem [{\citenamefont {Logan}\ and\ \citenamefont
  {Wolynes}(1987{\natexlab{b}})}]{DELPGWdipolar1987}%
  \BibitemOpen
  \bibfield  {author} {\bibinfo {author} {\bibfnamefont {D.~E.}\ \bibnamefont
  {Logan}}\ and\ \bibinfo {author} {\bibfnamefont {P.~G.}\ \bibnamefont
  {Wolynes}},\ }\href@noop {} {\bibfield  {journal} {\bibinfo  {journal} {J.
  Chem. Phys.}\ }\textbf {\bibinfo {volume} {87}},\ \bibinfo {pages} {7199}
  (\bibinfo {year} {1987}{\natexlab{b}})}\BibitemShut {NoStop}%
\bibitem [{\citenamefont {Logan}\ and\ \citenamefont
  {Wolynes}(1986)}]{DELPGWJCP1986}%
  \BibitemOpen
  \bibfield  {author} {\bibinfo {author} {\bibfnamefont {D.~E.}\ \bibnamefont
  {Logan}}\ and\ \bibinfo {author} {\bibfnamefont {P.~G.}\ \bibnamefont
  {Wolynes}},\ }\href@noop {} {\bibfield  {journal} {\bibinfo  {journal} {J.
  Chem. Phys.}\ }\textbf {\bibinfo {volume} {85}},\ \bibinfo {pages} {937}
  (\bibinfo {year} {1986})}\BibitemShut {NoStop}%
\bibitem [{\citenamefont {Logan}\ and\ \citenamefont
  {Wolynes}(1985)}]{DELPGWPRB1985}%
  \BibitemOpen
  \bibfield  {author} {\bibinfo {author} {\bibfnamefont {D.~E.}\ \bibnamefont
  {Logan}}\ and\ \bibinfo {author} {\bibfnamefont {P.~G.}\ \bibnamefont
  {Wolynes}},\ }\href@noop {} {\bibfield  {journal} {\bibinfo  {journal} {Phys.
  Rev. B}\ }\textbf {\bibinfo {volume} {31}},\ \bibinfo {pages} {2437}
  (\bibinfo {year} {1985})}\BibitemShut {NoStop}%
\bibitem [{\citenamefont {Logan}\ and\ \citenamefont
  {Wolynes}(1984)}]{DELPGWPRB1984}%
  \BibitemOpen
  \bibfield  {author} {\bibinfo {author} {\bibfnamefont {D.~E.}\ \bibnamefont
  {Logan}}\ and\ \bibinfo {author} {\bibfnamefont {P.~G.}\ \bibnamefont
  {Wolynes}},\ }\href@noop {} {\bibfield  {journal} {\bibinfo  {journal} {Phys.
  Rev. B}\ }\textbf {\bibinfo {volume} {29}},\ \bibinfo {pages} {6560}
  (\bibinfo {year} {1984})}\BibitemShut {NoStop}%
\bibitem [{\citenamefont {Logan}\ and\ \citenamefont
  {Wolynes}(1990)}]{DELPGWJCP1990}%
  \BibitemOpen
  \bibfield  {author} {\bibinfo {author} {\bibfnamefont {D.~E.}\ \bibnamefont
  {Logan}}\ and\ \bibinfo {author} {\bibfnamefont {P.~G.}\ \bibnamefont
  {Wolynes}},\ }\href@noop {} {\bibfield  {journal} {\bibinfo  {journal} {J.
  Chem. Phys.}\ }\textbf {\bibinfo {volume} {93}},\ \bibinfo {pages} {4994}
  (\bibinfo {year} {1990})}\BibitemShut {NoStop}%
\bibitem [{\citenamefont {Leitner}(2015)}]{LeitnerReview2015}%
  \BibitemOpen
  \bibfield  {author} {\bibinfo {author} {\bibfnamefont {D.~M.}\ \bibnamefont
  {Leitner}},\ }\href@noop {} {\bibfield  {journal} {\bibinfo  {journal} {Adv.
  Phys.}\ }\textbf {\bibinfo {volume} {64}},\ \bibinfo {pages} {445} (\bibinfo
  {year} {2015})}\BibitemShut {NoStop}%
\bibitem [{\citenamefont {Altshuler}\ and\ \citenamefont
  {Prigodin}(1989)}]{Altshuler+Prigodin1989}%
  \BibitemOpen
  \bibfield  {author} {\bibinfo {author} {\bibfnamefont {B.~L.}\ \bibnamefont
  {Altshuler}}\ and\ \bibinfo {author} {\bibfnamefont {V.~N.}\ \bibnamefont
  {Prigodin}},\ }\href@noop {} {\bibfield  {journal} {\bibinfo  {journal} {Sov.
  Phys. JETP}\ }\textbf {\bibinfo {volume} {68}},\ \bibinfo {pages} {198}
  (\bibinfo {year} {1989})}\BibitemShut {NoStop}%
\bibitem [{FN1()}]{FN1}%
  \BibitemOpen
  \href@noop {} {}\bibinfo {note} {Means are denoted generally by $\overline{O}
  =\langle\mathrm{Tr}\hat{O}\rangle_{\epsilon}$ with $\mathrm{Tr}\hat{O}
  =\tfrac{1}{N_{s}}\sum_{I}\langle I|\hat{O}|I\rangle$ an average over FS
  sites, and $\langle ....\rangle_{\epsilon}$ a disorder average; and variances
  by $\mu_{O}^{2}=\langle
  \mathrm{Tr}\big([\hat{O}-\mathrm{Tr}\hat{O}]^{2}\big)\rangle_{\epsilon}$.}\BibitemShut
  {Stop}%
\bibitem [{FN2()}]{FN2}%
  \BibitemOpen
  \href@noop {} {}\bibinfo {note} {Strictly speaking the $\gtilJo$ in eq.\
  \ref{eq:21} are propagators for sites $J$ with site $I$ removed from the
  problem; but the difference is readily shown to lead to
  ${\cal{O}}(1/\mu_{E}^{2}) \propto 1/N$ corrections to $\stilIo$, which are
  irrelevant in the thermodynamic limit.}\BibitemShut {Stop}%
\bibitem [{FN3()}]{FN3}%
  \BibitemOpen
  \href@noop {} {}\bibinfo {note} {With $N_{n}$ denoting the number of terms
  contributing in $n^{\mathrm{th}}$-order to the RPS, arguments can be given to
  show that higher-order contributions to $\stilIo$ do not dominate if
  $N_{n}/\mu_{E}^{n}$ remains finite as $N\rightarrow \infty$, and that this
  condition is satisfied at least for finite $n$.}\BibitemShut {Stop}%
\bibitem [{\citenamefont {Georges}\ \emph {et~al.}(1996)\citenamefont
  {Georges}, \citenamefont {Kotliar}, \citenamefont {Krauth},\ and\
  \citenamefont {Rosenberg}}]{dmftgeorgeskotliar}%
  \BibitemOpen
  \bibfield  {author} {\bibinfo {author} {\bibfnamefont {A.}~\bibnamefont
  {Georges}}, \bibinfo {author} {\bibfnamefont {G.}~\bibnamefont {Kotliar}},
  \bibinfo {author} {\bibfnamefont {W.}~\bibnamefont {Krauth}}, \ and\ \bibinfo
  {author} {\bibfnamefont {M.~J.}\ \bibnamefont {Rosenberg}},\ }\href@noop {}
  {\bibfield  {journal} {\bibinfo  {journal} {Rev. Mod. Phys.}\ }\textbf
  {\bibinfo {volume} {68}},\ \bibinfo {pages} {13} (\bibinfo {year}
  {1996})}\BibitemShut {NoStop}%
\bibitem [{\citenamefont {Hewson}(1993)}]{hewsonbook}%
  \BibitemOpen
  \bibfield  {author} {\bibinfo {author} {\bibfnamefont {A.~C.}\ \bibnamefont
  {Hewson}},\ }\href@noop {} {\emph {\bibinfo {title} {The {K}ondo Problem to
  Heavy Fermions}}}\ (\bibinfo  {publisher} {Cambridge University Press},\
  \bibinfo {address} {Cambridge},\ \bibinfo {year} {1993})\BibitemShut
  {NoStop}%
\bibitem [{FN4()}]{FN4}%
  \BibitemOpen
  \href@noop {} {}\bibinfo {note} {Further, since the energy ${\cal{E}}_{J}$ of
  \emph{all} states $J$ under $H_{0}$ differs from ${\cal{E}}_{I}$ by a finite
  amount ${\cal{O}}(W,V)$ (eq.\ \ref{eq:25}), but $Z_{I} \propto N\rightarrow
  \infty$, the remaining states of the Wilson chain decouple completely (the
  so-called zero-bandwidth limit of the Anderson model~\cite{hewsonbook}). The
  $|0\rangle$-orbital itself thus couples only to the `impurity' $|I\rangle$.
  There are then only two eigenstates in which $|I\rangle$ participates, namely
  the symmetric and antisymmetric combinations of $|I\rangle$ and
  $|0\rangle$.}\BibitemShut {Stop}%
\bibitem [{FN5()}]{FN5}%
  \BibitemOpen
  \href@noop {} {}\bibinfo {note} {Its origins within the present description
  are easily seen. For low enough $\lambda$ (where $\tilde{{\cal{E}}}_{I}
  =(\tilde{{\cal{E}}}_{I}-\overline{\e})/\mu_{E}$ may be neglected), the local
  FS propagators and self-energies are essentially site-independent,
  $\tilde{G}_{I}(\wtil) \equiv \tilde{G}(\wtil)$ and $\tilde{S}_{I}(\wtil)
  \equiv \tilde{S}(\wtil)$; and follow using eqs.\
  \ref{eq:15},\ref{eq:21},\ref{eq:28} as $\tilde{S}(\wtil)
  =\Gamma\tilde{G}(\wtil)$ and hence $\tilde{G}(\wtil) = [\wtil^{+}-\Gamma
  \tilde{G}(\wtil)]^{-1}$. The spectrum $\tilde{D}(\wtil)$ follows trivially,
  $\pi\sqrt{\Gamma}\tilde{D}(\wtil) =[1-({\wtil}^{2}/4\Gamma)]^{1/2}$. Hence
  $\tilde{\Delta}_{\mathrm{t}}(\wtil)$ has the golden rule form of eq.\
  \ref{eq:19}, $\tildeltyp =\pi\Gamma \tilde{D}(\wtil)$, recovering the result
  quoted for the band center $\wtil =0$. Note that $\tilde{D}(\wtil)$ here is
  semi-elliptic rather than the Gaussian eq.\ \ref{eq:12}, because the problem
  is treated at second-order RPS level.}\BibitemShut {Stop}%
\bibitem [{FNl()}]{FNlater}%
  \BibitemOpen
  \href@noop {} {}\bibinfo {note} {Reflecting the configurational disorder
  arising from distributing $N_{e}$ fermions over $N$ sites.}\BibitemShut
  {Stop}%
\bibitem [{FN6()}]{FN6}%
  \BibitemOpen
  \href@noop {} {}\bibinfo {note} {For a finite-size system, the band center is
  taken as the center of gravity $\mathrm{Tr}H =N_{s}^{-1}\sum_{n} E_{n}$ for
  each disorder realization, for the reasons explained in
  [\onlinecite{SWDEL1}].}\BibitemShut {Stop}%
\bibitem [{\citenamefont {Atas}\ \emph {et~al.}(2013)\citenamefont {Atas},
  \citenamefont {Bogomolny}, \citenamefont {Giraud},\ and\ \citenamefont
  {Roux}}]{AtasPRL2013}%
  \BibitemOpen
  \bibfield  {author} {\bibinfo {author} {\bibfnamefont {Y.~Y.}\ \bibnamefont
  {Atas}}, \bibinfo {author} {\bibfnamefont {E.}~\bibnamefont {Bogomolny}},
  \bibinfo {author} {\bibfnamefont {O.}~\bibnamefont {Giraud}}, \ and\ \bibinfo
  {author} {\bibfnamefont {G.}~\bibnamefont {Roux}},\ }\href@noop {} {\bibfield
   {journal} {\bibinfo  {journal} {Phys. Rev. Lett.}\ }\textbf {\bibinfo
  {volume} {110}},\ \bibinfo {pages} {084101} (\bibinfo {year}
  {2013})}\BibitemShut {NoStop}%
\bibitem [{FN7()}]{FN7}%
  \BibitemOpen
  \href@noop {} {}\bibinfo {note} {So $\eta \sim 4t/N$ for $W\ll 4t$ (with $4t$
  the full width of the non-disordered band), while $\eta \sim W/N$ for $ W \gg
  4t$.}\BibitemShut {Stop}%
\bibitem [{\citenamefont {J{\o}rgensen}(1982)}]{Jorgensenbook}%
  \BibitemOpen
  \bibfield  {author} {\bibinfo {author} {\bibfnamefont {B.}~\bibnamefont
  {J{\o}rgensen}},\ }\href@noop {} {\emph {\bibinfo {title} {Statistical
  Properties of the Generalized Inverse Gaussian Distribution}}}\ (\bibinfo
  {publisher} {Springer-Verlag, Lecture Notes in Statistics, Vol. 9},\ \bibinfo
  {address} {New York},\ \bibinfo {year} {1982})\BibitemShut {NoStop}%
\end{thebibliography}%

\end{document}